\documentclass[resetfootnote ]{aastex702}

\usepackage{amssymb,amsfonts}
\usepackage{amsthm}
\usepackage{booktabs}
\usepackage{graphicx}
\usepackage{subcaption}
\usepackage{longtable}
\usepackage{multirow}
\usepackage{xspace}
\usepackage[font=small,labelfont=bf]{caption}
\usepackage{textcomp}
\usepackage[misc]{ifsym}
\usepackage{CJK}
\usepackage[whole]{bxcjkjatype}
\usepackage{CJKutf8}
\usepackage[utf8]{inputenc}

\usepackage{lmodern}
\usepackage{amsmath}
\usepackage{hyperref}
\usepackage{caption}
\usepackage{upgreek}
\usepackage{ulem}

\graphicspath{{figures/}}

\providecommand{\araa}{Annu. Rev. Astron. Astrophys.} 
\providecommand{\aj}{Astron. J.} 
\providecommand{\apj}{Astrophys. J.} 
\providecommand{\apjl}{Astrophys. J. Lett.} 
\providecommand{\apjs}{Astrophys. J. Suppl. Ser.} 
\providecommand{\ao}{Appl. Opt.} 
\providecommand{\aap}{Astron. Astrophys.} 
\providecommand{\aaps}{Astron. Astrophys. Suppl.} 
\providecommand{\mnras}{Mon. Not. R. Astron. Soc.} 
\providecommand{\nat}{Nature} 
\providecommand{\nar}{New Astron. Rev.} 
\providecommand{\pra}{Phys. Rev. A} 
\providecommand{\prl}{Phys. Rev. Lett.} 
\providecommand{\ssr}{Space Sci. Rev.} 

\begin{document}
\title{The 91T/99aa-like Type Ia Supernova 2019vrq, Part II: 3D, Non-LTE, Low-Amplitude, Pulsating Delayed-Detonation Models of a Promising Standard Candle}

\author[0000-0002-4338-6586]{Peter Hoeflich*}
\affiliation{Department of Physics, Florida State University, Tallahassee, Florida 32306-4350, USA}
\email[show]{*phoeflich@fsu.edu} 

\author[orcid=0000-0002-6535-8500]{Yi Yang\begin{CJK*}{UTF8}{gbsn}
(杨轶)\end{CJK*}}
\affiliation{Department of Physics, Tsinghua University, Qinghua Yuan, Beijing 100084, China}
\email[show]{yi\_yang@mail.tsinghua.edu.cn} 

\author[0000-0003-1349-6538]{J. Craig Wheeler}
\affiliation{Department of Astronomy, University of Texas, 2515 Speedway, Stop C1400, Austin, TX 78712-1205, USA}
\email[unshow]{wheel@astro.as.utexas.edu}

\author[0000-0003-1637-9679]{Dietrich Baade}
\affiliation{European Organisation for Astronomical Research in the Southern Hemisphere (ESO), Karl-Schwarzschild-Str.\ 2, 85748 Garching b.\ M{\"u}nchen, Germany}
\email[unshow]{dbaade@eso.org}

\author[0000-0001-7101-9831]{Aleksandar Cikota}
\affiliation{Gemini Observatory,NSFs NOIRLab, Casilla 603, La Serena, Chile}
\email{aleksandar.cikota@noirlab.edu}

\author[0009-0001-9148-8421]{E.~Fereidouni}
\affiliation{Department of Physics, Florida State University, Tallahassee, Florida 32306-4350, USA}
\email{ef22g@fsu.edu}

\author[0000-0001-7092-9374]{D. Mishra}
\affiliation{George P. and Cynthia Woods Mitchell
    Institute for Fundamental Physics and Astronomy,
    Department of Physics and Astronomy, Texas 
             A\&M University, College Station, TX 77843, USA}
             \affiliation{Department of Physics and Astronomy, Texas A\&M University, 4242 TAMU, College Station, TX 77843, USA}
\email{mdivya@tamu.edu}



\author[0000-0001-6107-0887]{S. Shiber}
\affiliation{Department of Physics, Florida State University, Tallahassee, Florida 32306-4350, USA}
\email{sshiber@fsu.edu} 

\author[0000-0001-9910-9230]{M.~Marengo}
\affiliation{Department of Physics, Florida State University, Tallahassee, Florida 32306-4350, USA}
\email{mmarengo@fsu.edu}

\author[0000-0001-5888-2542]{T.~Mera}
\affiliation{Department of Physics, Florida State University, Tallahassee, Florida 32306-4350, USA}\affiliation{Institute for Astronomy, University of Hawai’i at Manoa, 2680 Woodlawn Dr., Hawai’i, HI 96822, USA}
\email{tycomera@gmail.com}


\author[0000-0002-5221-7557]{C. Ashall}
\affiliation{Institute for Astronomy, University of Hawai’i at Manoa, 2680 Woodlawn Dr., Hawai’i, HI 96822, USA}
\email{cashall@hawaii.edu}

\author[0000-0001-7186-105X]{K. Medler}
\affiliation{Institute for Astronomy, University of Hawai’i at Manoa, 2680 Woodlawn Dr., Hawai’i, HI 96822, USA}
\email{kyle.medler@sky.com}

\author[0000-0002-7305-8321]{C.~M.~Pfeffer}
\affiliation{Institute for Astronomy, University of Hawai’i at Manoa, 2680 Woodlawn Dr., Hawai’i, HI 96822, USA}
\email{cpfeffer@hawaii.edu}

\author[0000-0003-0183-451X]{Lauren~Aldoroty}
\affiliation{University of Maryland Baltimore County, 1000 Hilltop Cir, Baltimore, MD 21250, USA}
\affiliation{NASA GSFC, 8800 Greenbelt Rd, Greenbelt, MD 20771, USA}
\email{laldoroty@umbc.edu}



\author[0000-0002-1296-6887]{Lluís Galbany}
\affiliation{Institute of Space Sciences (ICE-CSIC), Campus UAB, Carrer de Can Magrans, s/n, E-08193 Barcelona, Spain}
\affiliation{Institut d'Estudis Espacials de Catalunya (IEEC), 08860 Castelldefels (Barcelona), Spain}
\email{luisgalbany@gmail.com}




\author[0000-0003-2734-0796]{M.~M.~Phillips}
\affiliation{Las Campanas Observatory, Carnegie
    Observatories, Casilla 601, La Serena, Chile}
\email{mmp@lco.cl}





\begin{abstract}
We analyze the overluminous, 91T/99aa-like Type~Ia SN\,2019vrq, based on light curves(LCs) and spectropolarimetric time-series.
We employ
3D-radiation-hydrodynamical, full non-LTE simulations of low-amplitude, radially-pulsating-off-center-delayed-detonations(PDD) of a possibly rotating near-$M_{\rm Ch}$ white dwarf (WD) to reproduce the LCs and spectra.  The progenitor originates from a $7\,M_\odot$ main-sequence star of solar metallicity. The explosion yields $0.86\,M_\odot$ of $^{56}$Ni and $0.023M_\odot$ of $^{58}$Ni. The latter falls a factor of $\approx 10$
below that of `classical' delayed-detonations for 91T/99aa-like SNe, 
a diagnostic that is directly testable with JWST. The slow deflagration leaves a bound, pulsating WD. The detonation is triggered at $0.8\,M_\odot$. 
LCs and spectra require an outer $\approx0.11\,M_\odot$ of unburned material with twice-solar Fe, plausibly the ashes of an earlier, unsuccessful explosion, and low-level mixing of nuclear-statistical-equilibrium(NSE) elements. The spectra reflect early high ionization followed by recombination, with the photosphere shifting from intermediate-mass-element- to NSE-dominated layers about a week before maximum.
The early high-velocity (HV) Ca{\sc ii}~IR3 line (likely produced by an aspherical density shell at $\approx$24,000km/s of $\lesssim 10^{-2}\,M_\odot$) arises from an ionization sandwich rather than a double structure in abundances. After Ca recombines, the Ca\,{\sc ii}-IR3 wing reaches $\approx$33{,}000km/s well beyond the HV component. The low polarization is due to low scattering in an iron-group-dominated photosphere, consistent with asphericities $\lesssim 20\%$ and resulting in a directional luminosity dependence $\lesssim10$--$15\%$ from the outer layers, and a dispersion of 35\% in total. The LC-shape provides a further probe of asphericity, consistent with the locally tested polarimetry limits and relevant for high-z cosmology.
\end{abstract}


\section{Introduction}
\label{sec:intro}

Thermonuclear explosions of white dwarfs (WDs), so-called Type~Ia supernovae
(SNe~Ia), are a critical step on the cosmic distance ladder and led to the
discovery of the accelerating expansion of the Universe
\citep{1998AJ....116.1009R, 1999ApJ...517..565P, 2016ApJ...826...56R,
2022ApJ...934L...7R}. This is remarkable given their photometric diversity,
spanning some 3~mag in peak brightness, and their spectral diversity, both of
which suggest a rather inhomogeneous population of progenitor WDs (see the
reviews by \citealt{2011NatCo...2..350H, 2013FrPhy...8..116H,
2014ARA&A..52..107M, 2017suex.book.....B, 2017hsn..book.1151H,
2017hsn..book..317T, 2023RAA....23h2001L}). The empirical relation between
light-curve (LC) shape and peak brightness allows individual  SNe~Ia to be used as
quasi-standard candles with an accuracy of $\approx0.1$--$0.2$ mag out to
redshifts of $z\sim 1 $--$2$ whereas an accuracy of $\approx 0.01{-}~ 0.02$ mag is required to probe
the nature of the acceleration.

The error budget in cosmology can be reduced by using the statistics of large samples of SNe~Ia
and by eliminating outliers. Recent results have
suggested the first evidence for a deviation from the standard cosmological model
\citep{AbdulKarim2025DESIDR2II, AbdulKarim2025DESIDR2LyA}, at a level that
depends on controlling the diversity of SNe~Ia and compensating for their
possible evolution with redshift.

Among the outliers that are carefully culled are the SN\,1991T/SN\,1999aa-like
(91T/99aa-like) SNe~Ia \citep{1992ApJ...384L..15F, 1992AJ....103.1632P,
1992ApJ...387L..33R, 2001ApJ...546..734L}. They are defined by weak Si\,{\sc ii}
features with Doppler shifts of the absorption minimum  $\approx25\%$ larger and 
brighter by $\approx0.3$~mag compared to `typical' bright SNe~Ia. They have
emerged as a bright, homogeneous group powered by the radioactive decay of
$\gtrsim0.8\,M_\odot$ of $^{56}$Ni ($^{56}$Ni~$\rightarrow$~$^{56}$Co~$\rightarrow$~$^{56}$Fe).
A small spread in peak luminosity, $\Delta M(B)\leq0.35$~mag, together with a
slow rise and decline and blue colors, makes them potential candidates as true
standard candles for high-precision cosmology \citep{2022ApJ...938...47P,
2022ApJ...938...83Y, 2024ApJS..273...16P}. Moreover, their association with
galaxies of high star formation implies a short delay time between star
formation and explosion \citep{2000AJ....120.1479H,2017hsn..book..317T,2022ApJ...938...47P}. Light curve analyses suggest progenitors with main-sequence masses close to
$7\,M_\odot$. Consequently, they are expected to evolve from rare objects
locally ($\approx5\%$ of all SNe~Ia) to common ones at high redshift (up to
$\approx30\%$; \citealt{2024ApJ...969...80C}), making them
potential tracers of star formation at high redshift.

Modeling efforts under the assumption of local thermodynamic equilibrium (LTE)
suggest higher temperatures and an outermost layer enriched in iron-group elements (IGE) in 91T/99aa-like SNe than those inferred for normal events.
Suggested scenarios for 91T-like SNe include (1) classical and (2) pulsating delayed-detonation explosions with ejecta masses $M$ near the Chandrasekhar mass ${\rm M_{\rm Ch}}$, with a deflagration to detonation transition (DDT) of the nuclear flame; (3) dynamical and violent mergers ($M\approx 1.2$--$1.6~M_\odot$); or (4) He-triggered detonations (HeD) of sub-${\rm M_{\rm Ch}}$ close to the high mass end for C/O-WDs ($M \approx 1.2 M_\odot$) e.g. \citet{1992ApJ...387L..33R,1992ApJ...397..304J,1995A&A...297..509M,2014MNRAS.445..711S}. The
high $^{56}$Ni mass points to explosions at the upper end of the WD mass range
in all SN~Ia scenarios currently considered. The scenarios differ, however, in
the mass of the exploding WD, the amount of electron-capture (EC) elements
produced, and the directional dependence of the luminosity.

Significant progress in modeling has been presented recently. The classical delayed-detonation (DD) scenario \citep{khok89} of a near $M_{\rm Ch}$ WD in which a deflagration front transitions to a detonation front in an expanding WD, has been suggested for 91T/99aa-like SNe based on high-resolution, full 3D hydrodynamical simulations starting from a static WD, close to $M_{\rm Ch}$~that is centrally ignited but neglects pre-existing velocity and magnetic fields \citep{2026arXiv260521575P}. It uses advanced hydrodynamics, non-LTE ionization, and source functions, although the level populations are treated in LTE. 
To first order, these simulations predict an overall spectral evolution consistent with observations of 91T/99aa-like SNe\,Ia. However, the resulting spectra systematically show  too large Doppler shifts of Si{\sc\,ii} and fall short in explaining the slow rise and decline of the LCs. In principle, a change of the initial WD properties and a pulsational phase prior to the DDT can solve these problems \citep{1993A&A...270..223K,1993A&AS...97..221H}.
We therefore use PDDs in this study.
`Classical' DD and PDD models are very similar but differ by the DDT taking place in a WD unbound and bound at the time of the DDT, respectively, and in the physics of the instability leading to the DDT. The differences have a profound impact on the details of the spectral evolution and luminosity, as we show here.


To make 91T/99aa-like SNe better standard candles, a better understanding of the
underlying physics and the explosion scenario is needed. This is practical
because their homogeneity allows for detailed analysis of individual objects as
representatives for the whole class. In turn, the quantitative properties derived may allow the
diversity of progenitor scenarios, each with its specific underlying physics and
progenitor system, to be narrowed down. This defines the motivation for the
detailed analysis presented in this paper.

Here we use a hybrid approach, combining constraints from high-quality
observations with detailed 3D radiation-hydrodynamical simulations performed
with our code HYDRA, which includes a detailed nuclear network and full non-LTE
treatment, i.e.\ departures from equilibrium in the ionization balance and level
populations, and the transport of high-energy photons and particles (Sect.~\ref{sec:num}).
This code has been widely applied to other SNe~Ia to compute
high-precision light curves, flux, and polarization spectra from the
optical to the mid-infrared. Throughout this paper, we present consistency checks
and estimates of the uncertainties.

The analysis is applied to SN\,2019vrq, which occurred in a low-surface-brightness emission-line spiral galaxy (see Paper I) at a
redshift of $z=0.01308$  \citep{2009MNRAS.399..683J}. The former is consistent with SN\,2019vrq originating from a young population in a star-burst galaxy.

Classification spectra
obtained a day after discovery match 91T/99aa-like events well
\citep{2019TNSCR2483....1I, 2019TNSCR2898....1Z}, and the LCs and time series of
spectra are comparable to the best data sets in this class, with the addition of
spectropolarimetric observations \citep{2022ApJ...938...83Y}. What sets
SN\,2019vrq apart are the high-precision data obtained with the Very Large Telescope  (VLT), including
flux and polarization spectra at $-9$, $+2$, and $+12$~days relative to maximum
light, at the high signal-to-noise ratio required for a detailed analysis.

In Paper~I \citep{PaperI} we presented the details of the data reduction needed to achieve this
accuracy, provided model-independent estimates of the overall light-curve and
spectral properties of SN\,2019vrq, and demonstrated that its flux spectra and
LCs are consistent with those of 91T/99aa-like events. Specifically:

\begin{itemize}
\item A detailed comparison of SN\,2019vrq in the context of 91T/99aa-like SNe
      established that it shows all the characteristics of the class and is not
      an outlier.
\item A detailed multi-wavelength analysis of the broadband colors enabled 
      the reconstruction of the bolometric LC and the evaluation of its
      uncertainties.
\item The analysis of Paper I provided the basic observational properties. In addition, estimates were derived for the absolute peak brightnesses and the mass of the radioactive $^{56}$Ni using empirical tools such as \textsc{SNooPy} \citep{2014ApJ...789...32B} and semi-analytical relations from the literature.
\item  The observed polarization P is low (typically $\lesssim$ 0.3 \% ), providing a tight constraint on SN~Ia simulations \citep{2001ApJ...556..302H,2006NewAR..50..470H,
2010ApJ...725L.167M,2012A&A...545A...7P,2016MNRAS.455.1060B,2023MNRAS.520..560H}.
\end{itemize}

A better understanding of the physics of 91T/99aa-like SNe is required to
improve the current accuracy to the percent level in luminosity, both for these locally rare
events and in light of the few well-observed objects expected at high
redshift. In this paper (Paper~II), a key question is the limit on the
anisotropy of the luminosity, which for typical SNe~Ia is known to reach
$\approx20\%$, with a statistical distribution that depends on the explosion
mechanism \citep{2026ApJ...996L..10C}. Specifically, we are guided by and  address the following questions:

\begin{itemize}
\item Can the light curves, flux, and polarization spectra be understood
      within the framework of low-amplitude pulsational delayed-detonation
      models? Are double features in Ca and Si, namely high-velocity (HV) and photospheric components, related to separate burning layers,
      and what is the physical origin?
      [Q1]
\item What is the physical basis for understanding the spectra and line profiles
      of SN\,2019vrq? [Q2]
\item Is the low polarization a consequence of the physics of the envelope, or
      is SN\,2019vrq particularly round, with nearly isotropic emission? [Q3]
\item Can asphericities be probed by the LCs, and are the results consistent with
      the upper limit from polarimetry? [Q4]
\item What is the evolutionary pathway that produces 91T/99aa-like SNe, and can
      they serve as true standard candles? [Q5]
\item Can we estimate the level of uncertainty in their use as true standard
      candles, and what are the future prospects using JWST and the Roman Space Telescope? [Q6]
\end{itemize}

The paper is organized as follows. In Sect.~\ref{sec:model_explosion}, we discuss the
difference between `classical' DD and PDD explosions
in the context of 91T/99aa-like SNe. Both are near-$M_{\rm Ch}$ explosions, but the main differences are the amounts of unburned matter in the outer layers and of electron-capture elements (EC). In Sect. \ref{sec:model_construct}, we discuss the setup of our simulations and the free parameters that characterize the initial WD, flame physics, and the approximations employed. In Sect. \ref{sec:model}, the synthetic and observed LCs are compared, including our LC-based method to probe asphericity.
In Sect. \ref{sec:asplum}, limits on the geometrical asphericity and the directional dependence of the luminosity are derived. In Sect. \ref{sec:HVCa}, the potential imprint of the progenitor system is discussed within the framework of low-amplitude PDDs.
In Sect. \ref{sec:alternative}, alternative explosion scenarios are discussed in light of the analysis of SN\,2019vrq.
In Sect. \ref{summary}, the main results are summarized, including possible implications for cosmology. Open questions with respect
to the progenitor system and the prospects for probing the initial conditions of the explosion 
with JWST are discussed.

\begin{figure}[ht!]
    \centering
    \includegraphics[trim={0.0cm 0.0cm 0.0cm 0.0cm},clip,width=0.9\textwidth]{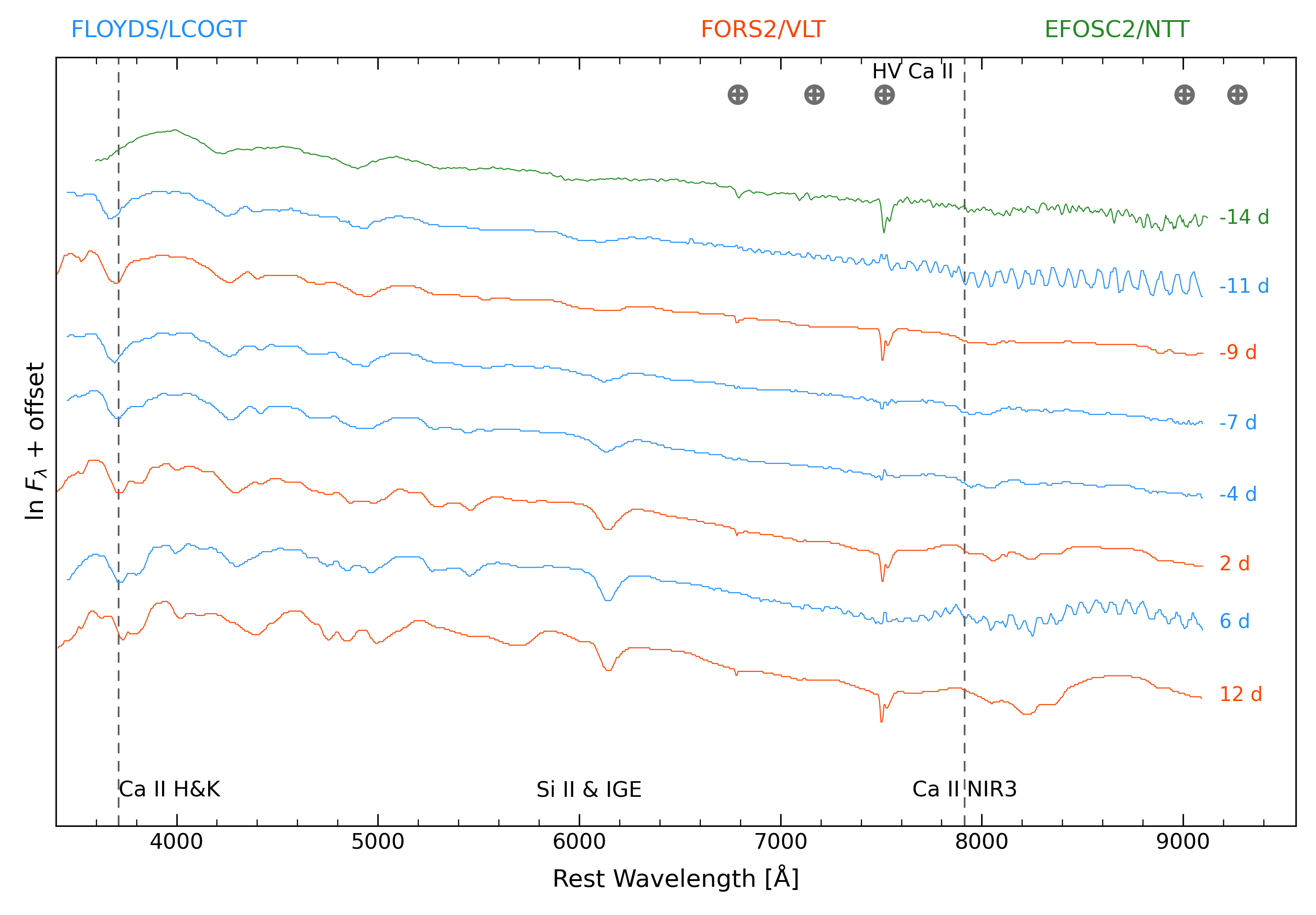}
    \vspace{-0.0 cm}
    \caption{{\bf Spectral time series of SN\,2019vrq.} Phases and instruments are labeled on the right and named on the top, respectively. 
    Ca{\sc\,ii}\,H\&K and Si{\sc\,ii}\,$\lambda$6355 + IGE blends, and Ca{\sc\,ii} NIR3 features are indicated. Note the high S/N data in the FORS2/VLT spectra (red) (Paper I), which, for the first time, allow a quantitative analysis of weak lines and their evolution (see Sect.~\ref{sec:spec_model}), the classification spectrum (green) obtained by Ihanec \& Wevers \footnote{IAU Transient Name Server} and FLOYDS/LCOGT (blue) spectra \citep{2022ApJ...938...83Y}. The wiggles in the FLOYDS spectra are due to fringing in the detector. Additional FLOYDS/LCOGT spectra can be found in Paper I.
    }\label{Fig_spec} 
\end{figure}

\section{Explosion Models and Geometry of 91T/99\lowercase{aa}-like SNe}~\label{sec:model_explosion}
A comprehensive exploration of all possible explosion scenarios lies beyond the scope of this work. We present our PDD model in detail, followed by a comparison between observations and synthetic LCs and flux spectra that require the use of PDD. The M$_{\rm Ch}$ large-amplitude pulsating delayed-detonation (PDD) scenario was previously invoked to explain the relatively slow rise and decline LCs of SN\,1990N~\citep{1992A&A...253L...9K} which resembles spectroscopically a normal-bright SN~Ia with a strong Si{\sc\,ii}\,$\lambda$6355 line like a normal SNe\,Ia, and it has been applied to SN\,1991T \citep{1992A&A...259..549H}, the prototype of the class considered here.
 Subsequently, PDDs have been employed to subluminous SNe~Ia to evaluate
possible molecule formation \citep{1995ApJ...444..831H}. Further systematic studies of the
pulsational mechanism, explosion, and LCs have been presented by \citet{1994ApJ...427..330A}, \citet{2009ApJ...695.1244B}, and \citet{1996ApJ...457..500H}, with models being applied to study the bifurcation among normal SNe~Ia \citep{2007ApJ...666.1083Q}. 
Here, the need for PDDs for SN\,2019vrq is discussed due to its similarity to 91T/99aa-like SNe, and we study the physics that separates SN\,1991T/SN\,1999aa from other subclasses.   Our quantitative results are employed for a qualitative discussion of alternative explosions and their issues in explaining SN\,2019vrq.
Our discussion and modeling focus on near-M$_{\rm Ch}$ explosions, particularly the `classical' DD and PDD scenarios. We identify both the similarities and the key physical differences between the `classical' DD and PDD models that serve to explain why 91T/99aa-like events form a distinct subclass associated with the physics of the DDT. 
 Following the descriptions in previous works~\citep{1991A&A...246..481H, 2023MNRAS.520..560H}, we employ full non-LTE radiative-transfer simulations for ionization balance, level populations, and non-thermal energy deposition by positrons and $\gamma$-rays. This level of sophistication is required because polarization is highly sensitive to the thermalization depth set by Thomson and Rayleigh scattering.

Both `classical' delayed-detonation and PDD models share the essential feature of a transition from the explosive burning of a deflagration to the detonation phase (DDT). 
The density regime where this transition occurs corresponds to conditions where the effective flame speed approaches the Chapman–Jouguet limit, which is $\approx$40\% of the sound speed for an electron-degenerate mixture with a mass ratio of C/O$=$1 (see, e.g., Figures 3 and 14 of \citealp{1995ApJ...449..695K}). In this regime, turbulent mixing of burned and unburned material dramatically enhances the burning rate and promotes the formation of the detonation by the Zeldovich gradient mechanism in the regime of distributed burning~\citep{1970JAMTP..11..264Z, 1995ApJ...449..695K}. 
The principal distinction between DD and PDD lies in the physical origin of the turbulence that leads to shocks that trigger the DDT. In the DD framework, a subsonic turbulent deflagration propagates through the expanding WD. Numerous mechanisms for turbulence generation and mixing have been proposed (see Sect.~3.1.2 of \citealp{2006NuPhA.777..579H} and references therein). 
Previous studies demonstrated that a DDT can arise naturally from turbulence produced by Rayleigh-Taylor (RT) unstable burning fronts in the distributed regime during the phase of peak turbulence~\citep{2011PhRvL.107e4501P, poludnenko2016b}.
In recent work on 91T/99aa-like SNe and classical  DD models~\citep{2026arXiv260521575P}, RT instabilities with slowly rising plumes close to the center produce turbulence and mix on small scales, causing an early DDT after $\approx$2.5 seconds. However, simulations that account for pre-existing turbulent fields have shown that, near the center, pockets of unburned material are burned away within a few tenths of a second~\citep{2026ApJ..1003L..37S}, long before the slowly rising RT can induce mixing.

We note that the details of the initial thermonuclear runaway of the WD ignition still remain unresolved in any simulation. 
Previous studies suggest that non-explosive burning prior to ignition, when the burning timescales become shorter than the hydrodynamical timescales, can eliminate steep temperature and density gradients, leading to a successful deflagration of the explosive phase of burning rather than a detonation~\citep{1986ApJ...307..619M, 1993ApJ...419L..77K, 1994MsT..........1N, 1995ApJ...452..779N}. 
In contrast, in the PDD scenario, the initial deflagration is too weak to unbind the WD. The star undergoes partial expansion followed by the fallback of material onto the slowly expanding inner layers~\citep{1991A&A...245..114K, 1996ApJ...457..500H}. The interaction between the material falling back and the inner expanding material can create the turbulence that can lead to a DDT. 
For pulsation to occur, the energy release must be lower than that corresponding to a classical  DD explosion by $\approx$70\%~\citep{1996ApJ...457..500H}. This, in turn, requires a reduced burning rate during the early deflagration phase~\citep{1996ApJ...457..500H, 2005ApJ...632..443L}, a condition that places important constraints on the progenitor structure and ignition conditions. 
The amplitude of the pulsation is expected to span a continuum from strong contraction to nearly unbound configurations approaching the classical  DD limit~\citep{1996ApJ...457..500H, 2018ApJ...861..119D}.

Classical  DD models tend to produce LCs that rise to maximum about one day faster than observed for normal SNe\,Ia~\citep{2017ApJ...846...58H}. 
In contrast, most 91T/99aa-like SNe rise $\gtrsim$ two days longer than normal SNe\,Ia~\citep{2011MNRAS.416.2607G}. This slower rise naturally favors PDD-like explosions for this subclass of events.
Moreover, a fundamental limitation of classical  DD models is that their burning products typically follow an abundance profile that is exponential with radius and extends to the outer ejecta~\citep{1998ApJ...495..617H, 2006ApJ...636..400Q}. This property is incompatible with the stronger IGE absorption features observed in 91T/99aa-like SNe (see Fig. 5 of \citet{2026arXiv260521575P}.
Earlier spherical simulations of large-amplitude PDDs, as well as models including extended envelopes as proxies for merger or core-degenerate scenarios, successfully reproduced slow-rising LCs and high luminosities and were applied to SN\,1991T~\citep{1993A&A...270..223K,1993A&AS...97..221H}. 
These large-amplitude PDD models produced 0.2–-0.4\,M$_{\odot}$ of unburned C/O-rich material that violates spectral constraints, namely the early appearance of products related to explosive C and O burning. 
To first order, the mass of unburned C/O-rich material in models of bright SNe\,Ia scales with the mass of the outer density enhancement formed by the infall of the outer layers during pulsation (Fig.~10 of \citealp{2007ApJ...666.1083Q}). The mass of the density enhancement correlates with pulsation amplitude~\citep{2017ApJ...846...58H, 2018ApJ...861..119D}.

Our goal is therefore to identify the regime of low-amplitude PDDs capable of reproducing the observed properties of SN\,2019vrq, namely the appearance of IME elements in early spectra at $\approx -9 $ days (Fig.\ref{Fig_spec}), while clarifying the physical origin of the low polarization observed in 91T/99aa-like events. Over time, deeper layers are exposed, and thus the mass observed at a given time provides the link between the model structure and observables (Fig. \ref{fig:mass}). The strongest constraints are given by the monochromatic LCs (Fig. \ref{fig:lc_prof}) and the flux spectra (Fig. \ref{fig:model_spec}).
The intrinsically weak polarization may also provide critical limits on luminosity anisotropy and hence on the reliability of 91T/99aa-like objects as standard candles (see \citealp{2024A&A...686A.227P} and references therein). 
In this work, we follow the pulsational phase under spherical symmetry while imprinting the three-dimensional (3D) structures from the early deflagration and detonation phases onto the ejecta and compute full 3D flux and polarization spectra.

\section{Hydrodynamics and Setup of the PDD Model for SN\,2019\lowercase{vrq}}~\label{sec:model_construct}
\subsection{Numerical Method and Calibration} \label{sec:num}
Our simulations for SN\,2019vrq employ the HYDrodynamical RAdiation code (HYDRA) \citep{2021ApJ...923..210H,2025arXiv250107654H}  with a resolution of $300^{3}$ in a Cartesian grid. For the initial radial pulsational phase, a spherical 1D grid in the comoving frame with 6 times higher resolution than the 3D grid is adopted. We use a network of 218 isotopes \textcolor{black}{with rates based on the REACLIB (\cite{cy10}, and references therein). For the initial structure of the WD, an equation of state (EOS) is used for a partially degenerate and partially relativistic Bose- and Fermi-gas. Interactions are included due to effects of Coulomb corrections, quantum-relativistic effects on the electron component, and electron-positron pair production \citep{1931ApJ....74...81C,vanHorn69,slattery82}. Electron screenings are taken from \citet{graboske73} for the weak, intermediate, and intermediate-strong regimes, and \citet{itoh79} for the strong regime. For the atomic model we consider He, C, O, N, Ne, Mg, Si, P, S, Ar, Ca, Ti, Cr, Mn, Fe, Co, and Ni if present in a given zone within ionization levels I-IV, with a total of $\approx$ 2000 - 3000 superlevels. In the simulations here, we employ sparse level merging of atomic levels with strong optical transitions to optimize accuracy. The elements, ions, and levels are adjusted dynamically, depending on the occupation numbers in each computational zone.} 
HYDRA employs detailed hydrodynamics, nuclear networks, magnetic fields, and transport for high-energy photons and non-thermal leptons, and detailed non-LTE atomic models  \citep{1993A&A...268..570H, 1995ApJ...440..821H, 1995ApJ...444..831H, 2003ASPC..288..185H, 2003LNP...635..203H, 2014ApJ...795...84P, 2021ApJ...923..210H, 2024JPhCS2742a2024H, 2025arXiv250107654H}.
This allows us to calculate the thermonuclear explosion for given initial conditions and to compute the LC, flux, and polarization spectra with high precision. The simulations adopt some parametrized features and hence are not ``first-principles''.


The simulations have limited resolution, which hinders detailed flame-physics analysis. We parameterize the effective rate of deflagration burning~\citep{2000ApJ...528..854D}, as calibrated by high-resolution, full 3D simulations done with the adaptive mesh code ALLA~\citep{1998JCoPh.143..519K, 2000astro.ph..8463K}.
Guided by the time series of spectra of SN\,2019vrq (Fig. \ref{Fig_spec}) and using spherical PDDs, the Euler number and initial central density of the WD ($\rho_{c}$) have been evaluated to find a low-amplitude PDD with $\approx$0.1\,M$_{\odot}$ in unburned layers. 
We also carried out separate simulations to characterize the thermonuclear runaway, the effects of WD rotation, and the sensitivity to physical conditions of the progenitor star, such as its main-sequence mass ($M_{\rm MS}$) and initial metallicity, as constrained by observations. 
\begin{figure*}[h]
\centering
     \includegraphics[width=0.46\linewidth]{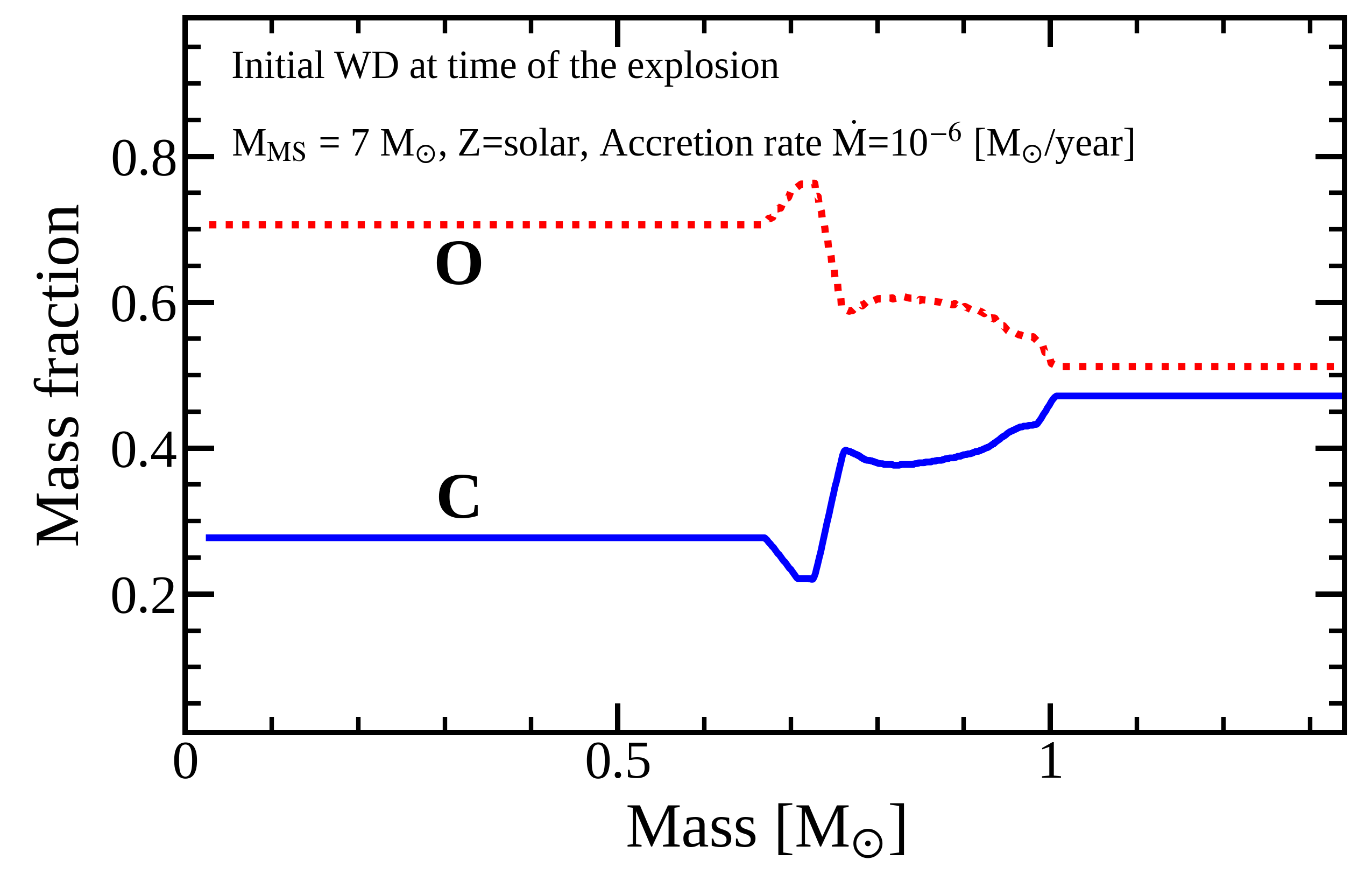}  \includegraphics[width=0.46\linewidth]{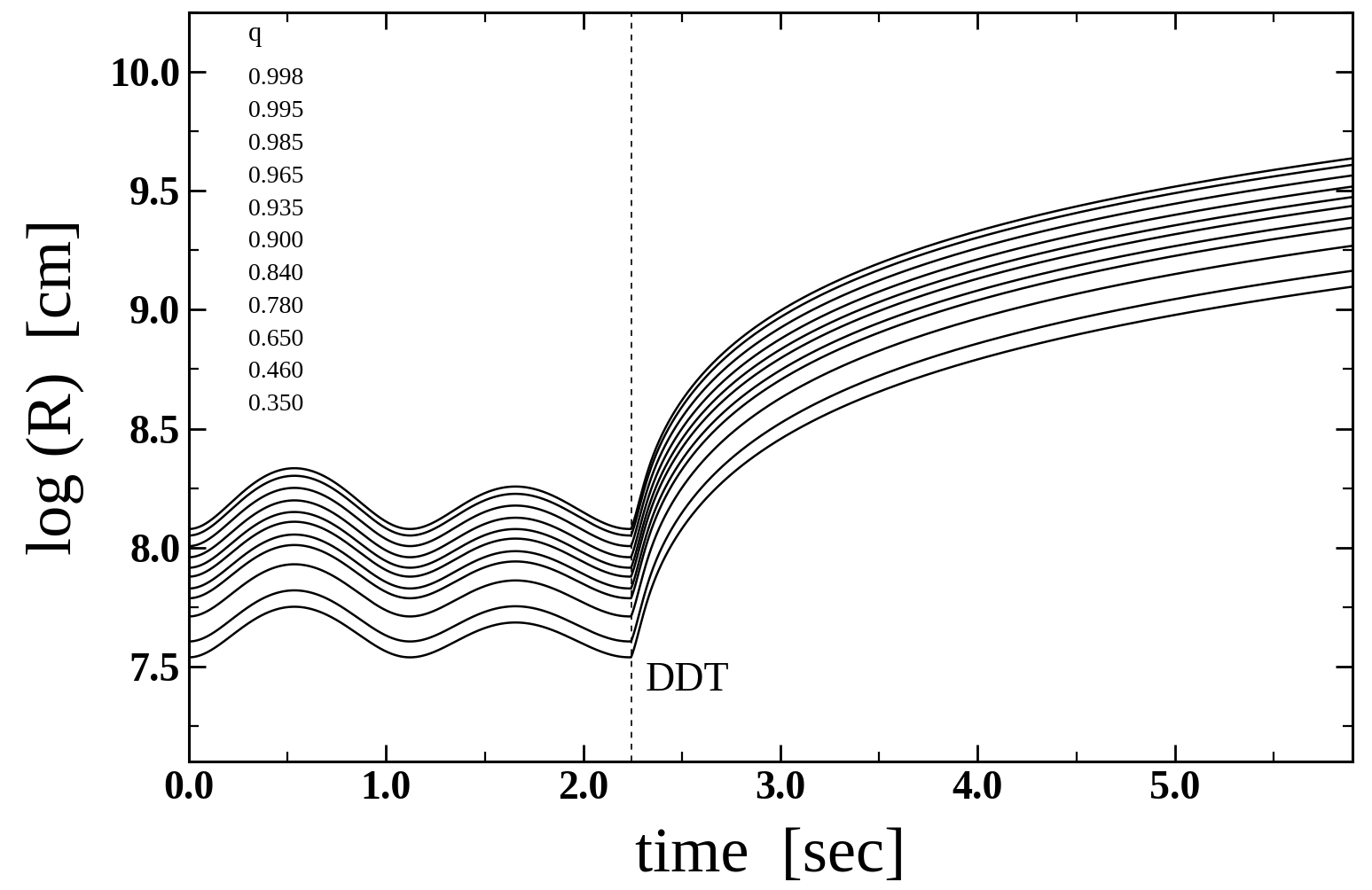}
     \caption{{ Initial abundance structure of the WD, and the evolution of mass elements q of the near $M_{\rm Ch}$ PDD. {Left:}  The mass fraction of C and O as a function of mass in the initial near $M_{\rm Ch}$ mass WD originating from a progenitor of 7\,M$_{\odot}$ with solar metallicity (left), and the evolution of the mass zones as a function of time starting with t=0 sec being the time of the explosive burning (right).} 
      {\sl Left:} The results were obtained based on detailed stellar evolution models~\citep{2001ApJ...557..279D} followed by accretion of H- or He-rich matter at a rate of $10^{-6}$\,M$_{\odot}$\,year$^{-1}$~\citep{2019nuco.conf..187H}. The layers up to $\approx$0.7\,M$_{\odot}$ are a result of central, convective helium burning with overshooting, followed by stellar shell flashes close to $\approx$1.1 $M_\odot$ and the pycnonuclear burning atop an accreting WD. Note that the outer $\approx 10^{-4}$\,M$_{\odot}$ layers are He-rich. {\sl Right:} The evolution of the mass elements is shown for $q \geq 0.35 ~M_{\rm Ch}$, which are all burned during the detonation phase. Note that instabilities in the region ($q \gtrsim$ 0.84 $M_{\rm Ch}$) later visible as QSE(Si) region (Fig.\ref{fig:Model}) are stretched by a factor of $\approx 10$ during the acceleration phase by the detonation. Note that after the runaway, the amplitude of the WD pulsation decreases with time. 
     }\label{fig:progenitor}
\end{figure*}

\subsection{The Pulsational Phase, DDT, and 3D Mapping}
\label{sec:DDT}
For a sufficiently slow deflagration, the initial pulsation phase lasts longer than the WD sound-crossing time. The resulting dynamical response is insufficient to unbind the WD, causing a radial pulsation of finite amplitude. For the time evolution of the mass elements, see Fig. \ref{fig:progenitor}.  The DDT is triggered at about 2.2 sec after the thermonuclear runaway in the layers of maximum deceleration at $q \approx  0.8~ M_{\odot} \approx 0.6 ~M_{\rm Ch} $. During the deceleration phase of mass layers, and based on 3D simulations for deflagration instabilities (see text), chemical mixing is assumed over two pressure scale heights. 
     The first pulsation mixes some burned and unburned material on the largest RT scale \citep{taylor1950,Chandrasekhar1961,1995ApJ...449..695K} and develops over the pulsational period, whereas the second pulsation triggers the DDT.  The deflagration depends sensitively on the initial conditions of the WD such as pre-existing turbulence and magnetic fields  \citep{2003Sci...299...77G,2026ApJ..1003L..37S,2026arXiv260813432S}, possibly WD rotation, and details of the ignition process in the WD \citep{2002ApJ...568..779H,2005ApJ...632..443L,2011ApJ...740....8Z}. Although detailed information about the deflagration burning will be wiped out in most of the ejecta by the detonation, our treatment of mixing is a limitation.  

Our simulations of the hydrodynamics of the deflagration phase implement spherical symmetry. At the end of the deflagration phase, the radial structure is then mapped onto the 3D grid to follow the detonation, which involves the triggering of the DDT. 
The approximation of the spherical density structure of models with WD mass near $M_{\rm Ch}$ is supported by all full 3D simulations~\citep{2001AIPC..556..301K,2003Sci...299...77G, 2005ApJ...623..337G, 2007ApJ...668.1132R,2021ApJ...923..210H} of both DD and PDD. We employ 3D hydrodynamic simulations for the phase starting at the DDT because the underlying physics is multi-dimensional. 
Despite evidence of spherical symmetry in the density structure, the DDT is likely to occur at an off-center point. This will lead to an axisymmetric component in the abundance distribution, for which evidence is found in the structure of SN remnants and in the polarization~\citep{2015ApJ...804..140F,2006NewAR..50..470H, 2018ApJ...854...55Y}.

According to high-resolution 3D explosion simulations, Rayleigh-Taylor (RT) instabilities during the deflagration phase and detonation-cellular instabilities during the detonation phase \citep{2025ApJ...982..204K} would imprint clumpy structures at the abundance interfaces on a range of small scales, with random orientations. 
Both instabilities produce corresponding abundance imprints at the end of the deflagration and detonation phases, respectively. The abundance instabilities `freeze out' in mass in an expanding, but still accelerating ejecta~\citep{2001AIPC..556..301K, 2003Sci...299...77G, 2005ApJ...623..337G}.
At the end of the deflagration phase, the inhomogeneous abundance structures fill the entire WD. However,
the detonation phase burns away the inhomogeneities inside the NSE-region, and subsequently, the remaining structure is accelerated (Figs. 1 and 3 in \citet{2005ApJ...623..337G}).
The average expansion rate is $\approx$2,000\,km\,s$^{-1}$ at the end of the acceleration phase compared to $\approx$10,000\,km\,s$^{-1}$ after the subsequent acceleration during the detonation. For layers resulting from explosive O burning in normal SNe\,Ia, the RT instabilities have typical scales that transform to $\approx$1,000--2,000\,km\,s$^{-1}$ in the envelope.
However, in 91T-like SNe, most of the partially burned layers are burned to NSE, with only a small amount of Si remaining in the outer, high-velocity layers of the envelope at the interface between NSE and Si. The largest eddy sizes of the remaining RT fingers are $\approx$6,000--7,000\,km\,s$^{-1}$ (Figs. \ref{fig:progenitor} \& \ref{fig:Model}), corresponding to a width of $\approx $140\,\AA. 
The size of the largest eddies is comparable to the thickness of the Si-rich layer. From the outside and in 91T/99aa-like SNe, the abundance pattern appears like a picket-fence pattern of intermediate-mass elements (IME) and iron-group elements (IGE) in angular space~\citep{2023ApJ...948...10A}. The increase in scales for the RT instabilities is caused by the high-velocity tail of $^{56}$Ni (Fig.~\ref{fig:Model}). 
Here, we imprint those abundance structures at the abundance interfaces during the homologous expansion phase. Note that the RT and cellular instabilities are well resolved in our subsequent radiation hydro simulations down to scales of $\approx$30--40\,\AA. 

Accreting WDs may be expected to rotate. Following \citet{1985A&A...146..260E}, an axisymmetric ellipsoidal density structure with axis ratio B/A is imprinted on the PDD model as a proxy for rotation. 
We do not include the origin of the high-velocity (HV) Ca{\sc\,ii} feature~\citep{2004ApJ...607..391G, 2023MNRAS.520..560H} in this section but discuss it separately (Sect. \ref{sec:HVCa}) because complex initial conditions would add the progenitor system to the parameter space. 


\subsection{Reference Model for SN 2019vrq}
\label{sec:refM}
As guided by observations and previous studies, our simulations adopt the scenario of a low-amplitude, off-centered PDD in a near-M$_{\rm Ch}$ WD. The low-amplitude pulsation models are chosen, and the pulsational amplitude has been tuned such that the mass of the unburned layers of the WD is $\approx$0.1\,M$_{\odot}$, consistent with the LCs and spectral evolution of SN\,2019vrq and 99aa-like SNe~\citep{2024ApJS..273...16P} (Sect.~\ref{sec:model}). 
Due to the small pre-expansion, the WD density remains high, thus producing a large amount of elements formed in nuclear statistical equilibrium (NSE) (Tab.~\ref{table_abundances} \& Fig. \ref{fig:Model2}). Based on our analysis of the LC and flux spectra of SN\,2019vrq, together with the LC properties of a wide sample of SNe including 91T/99aa-like events~\citep{2024ApJ...969...80C}, our PDD simulations adopt a WD originating from a main-sequence mass of M$_{\rm MS} = 7$\,M$_{\odot}$ with solar metallicity and a WD central density of $\rho_{\rm c}=10^{9}$\,g\,cm$^{-3}$. 
The actual total mass of our near $M_{\rm Ch}$ mass progenitor is 1.356\,M$_{\odot}$. 
For the flux spectra, twice-solar metallicity for Fe was used for the unburned layers to improve the spectral fit in Fig.~\ref{fig:model_spec}. It may hint at a `failed explosion', i.e., a thermonuclear runaway in a $M_{\rm Ch}$ WD that does not unbind the WD, with a $^{56}$Ni plume rising to the surface, and time to decay from $^{56}$Ni to $^{56}$Fe ~\citep{1994ApJS...92..501H}. 
We note that different values of $\rho_{c}$ hardly affect the amount of NSE (iron-group) production. However, they do influence the production of $^{56}$Ni and $^{58}$Ni, leading to a dispersion of $\approx$0.3\,mag in the SN peak brightness. The ratio of EC capture elements to $^{56}$Ni can be investigated through late-time nebular spectroscopy in the near and mid-infrared (NIR and MIR). 

Large M$_{\rm MS}$ results in an extended region of low C-abundance in the progenitor WD (Fig. \ref{fig:progenitor}), and thus reduces the total nuclear energy generation by $\approx$70\% during the deflagration phase~\citep{2001ApJ...557..279D, 1998ApJ...495..617H} compared to the mass ratio C/O$=$1 commonly used in 3D models. 
The slow deflagration burning resulting from the reduced carbon mass fraction is needed to allow for a pulsation rather than a prompt total disruption of the WD. Note that using the same description for the burning front but a large C/O ratio would result in a 'classical'  DD. $\rho_{c}$ was adopted from the LC analysis of 91T/99aa-like SNe~\citep{2024ApJ...969...80C}. 
For our reference model, we find a maximum deceleration at $M_{\rm DDT}\approx$ 0.8\,M$_{\odot}$ which, as discussed above, is the most likely location of the DDT. 



The local flame speed is sensitive to the local relative abundance of the WD matter, in particular, the C/O ratio. A low-amplitude PDD leads to a reduction in the effective deflagration speed by $\approx$70\% compared to the canonical values for normal events~\citep{1996ApJ...457..500H}. The average C/O ratio over the deflagration phase determines whether a WD becomes bound or unbound,
and the average C/O ratio over the entire WD determines the total explosion energy. 
We calibrate the deflagration speed using 3D simulations in which the deflagration front during the RT phase propagates mostly in a C/O$=$0.5 mixture~\citep{2001AIPC..556..301K, 2003Sci...299...77G, 2000ApJ...528..854D}. 
For the innermost layers, our resolution is insufficient for a detailed description of the early deflagration phase. Thus, we use the calibrated effective burning speed based on 3D simulations with adaptive mesh refinement (AMR)~\citep{2001ApJ...557..279D}. More details of the early hydrodynamical phase can be found in \citet{2023MNRAS.520..560H}. 

The speed of the deflagration flame also depends on several other quantities related to the conditions at the time of the thermonuclear runaway, including the strengths of the turbulence and the magnetic field, the number of ignition spots, and their spatial distribution~\citep{2002ApJ...568..779H, 2012ApJ...750L..19R, 2026ApJ..1003L..37S}. The slow deflagration in a PDD scheme also results in a smaller amount of WD mass that undergoes deflagration burning, namely M$_{\rm Defl} \approx$ 0.2\,M$_{\odot}$ over $\approx$two seconds during a rather slow expansion of the envelope.
The outer layers of the envelope remain bound, forming a temporary shell at $r\lesssim 10^{9}$\,cm, compared to shell formation at $r\approx 3\times 10^{10}$\,cm for large-amplitude pulsations, which started at twice the initial $\rho_{c}$ and with C/O $\approx 0.5$~\citep{1993A&A...270..223K}. The pre-expansion determines the mass of the outer layers that remain unburned after the explosion. 
For a pulsation to take place, low energy production during the deflagration phase~\citep{1992A&A...253L...9K} is necessary. Moreover, the low pulsation amplitude requires that the progenitor WD has a low C/O mass ratio (i.e., high $M_{\rm MS}$) and relatively low $\rho_{c}$. Both parameters were tuned in the current models. 

The pulsation decreases the EC production by $\approx$30\% compared to classical  DD models  because, in PDDs, the front propagates slowly, leading to little deflagration burning. The longer phase of deflagration (2.2 sec, see Fig.\ref{fig:progenitor}) does not compensate because the initial density has decreased by a factor of $\approx$3. 
Note that \citet{2026arXiv260521575P} found an EC production higher by a factor of 10 because high-density material is burned by a subsequent detonation (see above).

The timescale for the radial pulsation is given by the WD sound-crossing time, namely $ t _ {s} \approx 1$ seconds. The interaction layer between the expanding inner ejecta and the outer layers, which are slowing down during the pulsation, is RT-unstable. As a result, strong mixing and subsequent DDT would occur in this region~\citep{1995ApJ...444..831H}, coinciding with the region of maximum turbulence~\citep{2019AAS...23311307P}. 
In our models, the flame transition occurs at M$_{\rm DDT} \approx 0.8$\,M$_{\odot}$ in mass coordinates. The off-center DDT produces one-sided abundance distributions of $\approx$3,000\,km\,s$^{-1}$, very similar to normal SNe\,Ia. 
Our simulations imprinted the large-scale RT and cellular instabilities and the picket-fence structure suggested by detonation cells~(see Sect.\ref{sec:model_construct}). 

The direct consequence of low-amplitude pulsation is burning at higher densities than in the classical DD model. Therefore, a large $^{56}$Ni mass is expected at the expense of the products of explosive oxygen burning (Si/S), carbon burning (O/Ne/Mg), and unburned layers. 

\subsection{Final Overall Abundance Distribution as Signature of the Progenitor and Explosion}~\label{sec:model_abundance}
\begin{figure*}[h]
     \includegraphics[width=0.47\linewidth]{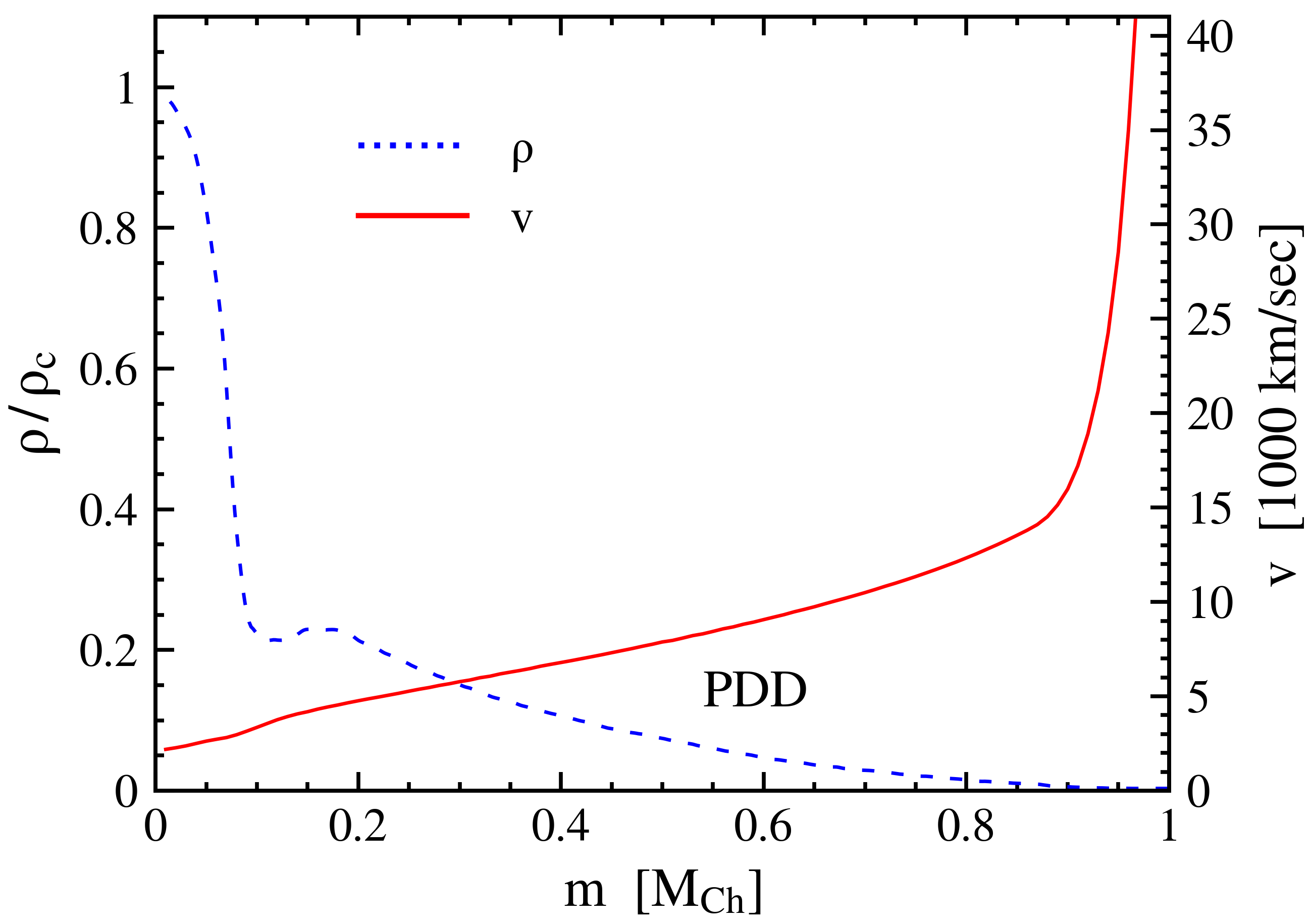}
      \includegraphics[width=0.52\linewidth]{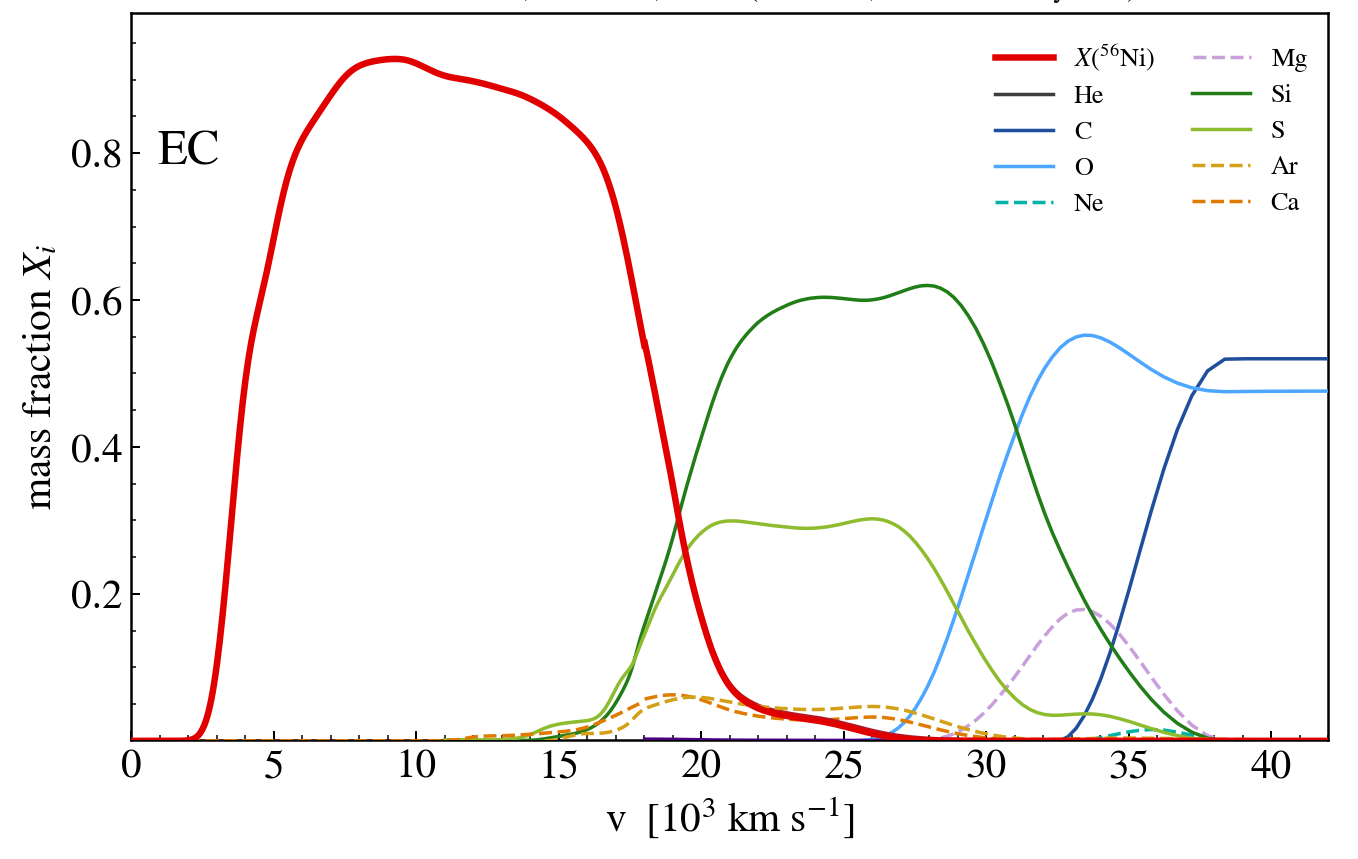}
     \caption{{\bf The pulsational delayed detonation model of SN\,2019vrq during the phase of homologous expansion.} The deflagration-to-detonation flame transition takes place at M$_{\rm DDT}=0.8$\,M$_{\odot}\approx 0.6 M_{\rm Ch}$. The left panel presents the angle-averaged density ($\rho$, left ordinate) and the expansion velocity ($v$, right ordinate) as a function of the enclosed WD mass in units of M$_{\rm Ch}$. The right panel displays the abundances of the elements relevant for the spectral  analysis and the locus of EC  as a function of the expansion velocity. }\label{fig:Model}
\end{figure*}

The abundance distribution of the low-amplitude PDD (Fig.~\ref{fig:Model}) differs significantly from that of the classical DD. Due to the slow deflagration flame speed in PDD, the EC element production is $\approx 1/3$ of that of a classical DD.

\begin{figure*}[ht]
\begin{center}
     \includegraphics[width=0.7\linewidth]{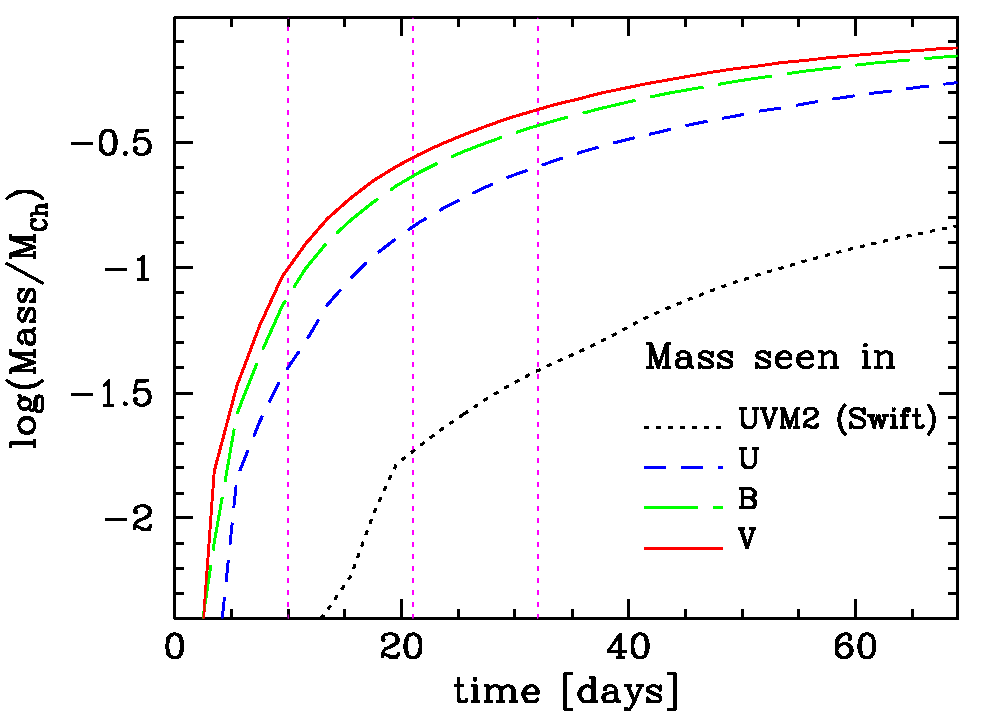}
\caption{{\bf Mass above the photosphere as a function of time for the PDD model of SN\,2019vrq.} Different curves present the estimates as measured in the {\it Swift} $uvm2$ and the Johnson $UBV$ bands as labeled. The vertical lines mark the epochs when the VLT spectropolarimetry was carried out.}\label{fig:mass}
\end{center}
\end{figure*}

The low-amplitude PDD yields a slow  pre-expansion of the outer unburned layers of the WD compared to a classical DD model. Consequently, the subsequent detonation burns most of the WD under high densities sufficient to reach NSE, providing a natural explanation of the large amount of $^{56}$Ni synthesized in the SN ejecta. Moreover, the small pre-expansion and pulsation lead to a wide transition zone between NSE and QSE elements on the level of $\approx1-2$\% between 11,000 and 18,000\,km\,s$^{-1}$. The long $^{56}$Ni tail at higher velocities is a result of instabilities and the off-center DDT when angle-averaged (see Fig.~\ref{fig:Model}). 
Note that the exact mass of the unburned material depends sensitively on the amplitude of the pulsation. In contrast, in bright classical  DD models, hardly any unburned C/O layers would remain, and the outgoing C/O detonation front in classical  DD models triggers a detonation in the outer $10^{-4}$\,M$_{\odot}$ helium layers~\citep{2019nuco.conf..187H}. A pulsation avoids such helium surface burning, which could be detected $\approx1-3$ days after the explosion (Fig.~\ref{fig:mass}). 

\begin{minipage}{0.44\textwidth}
    \centering
    \includegraphics[width=0.85\linewidth]{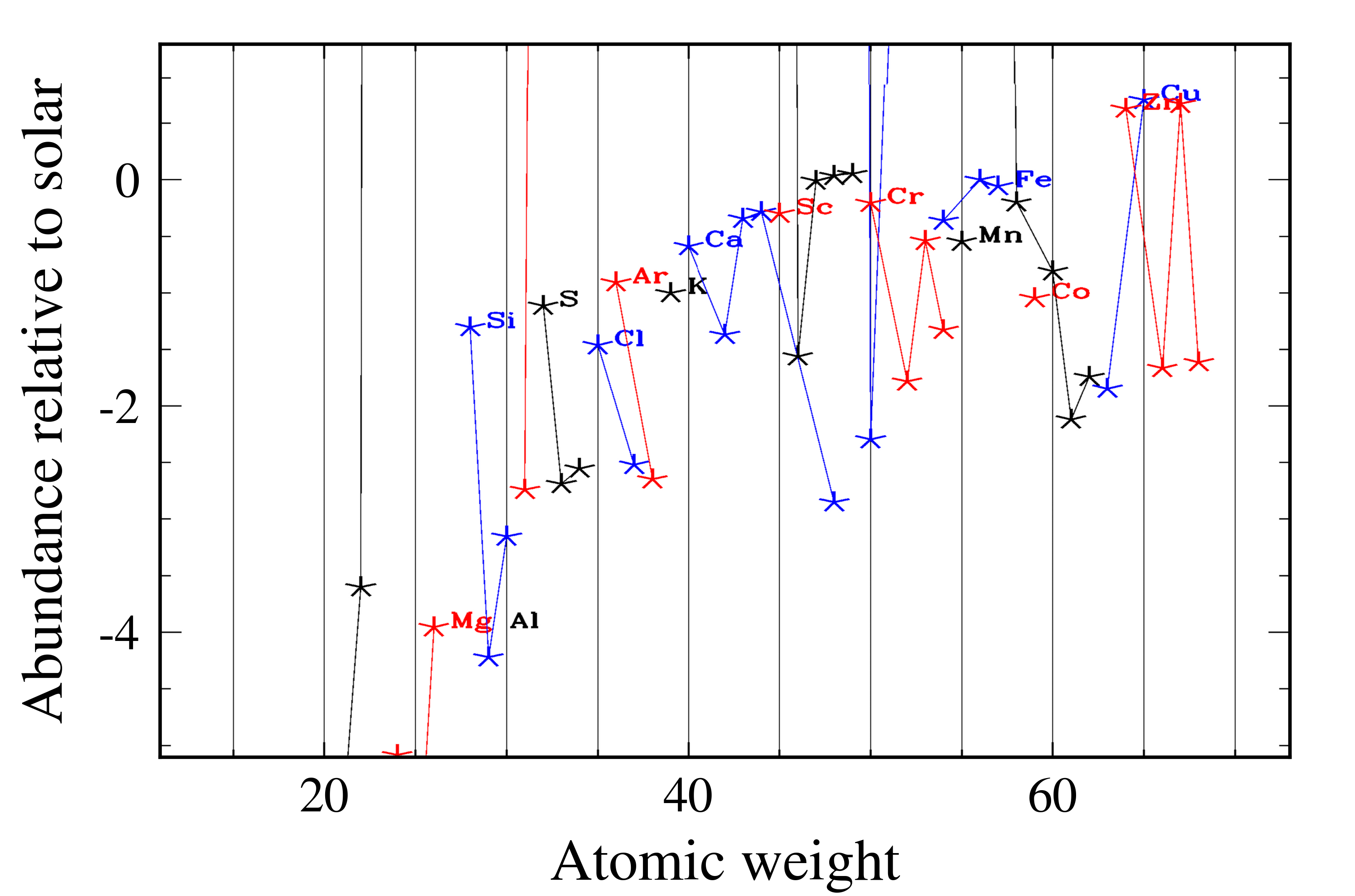}
    \captionof{figure}{Final isotopic abundances relative to solar iron for the model shown in Fig.~\ref{fig:Model}. The labels mark the elements and their isotopes, connected alternately by red and blue lines. The electron-capture production of our PDD is typical for SNe\,Ia, whereas the `classical' DD scenario for 91T/99aa-like events is likely to cause an excess of stable Mn, Ni, Co, and Cr. Note the implications for the chemical evolution of the Universe.}\label{fig:Model2}
  \end{minipage}
  \hfill
  \begin{minipage}[t]{0.52\textwidth}
    \centering
    \footnotesize
    \vskip -112pt
       \captionof{table}[b]{{\bf Normalized mass fractions $X_i$ of abundant stable isotopes after decay}.} The total M$(^{56}{\rm Ni})$ yields 0.86\,M$_\odot$.
\begin{tabular}{|ll|ll|ll|}
\hline
Isotope & $X_i$ & Isotope & $X_i$ &Isotope & $X_i$ \\
\hline
    $^{4}$He    &  1.321E-04&    $^{12}$C    &  3.562E-02    &  $^{16}$O     &  7.577E-02 \\
   $^{20}$Ne    &  3.289E-06&    $^{21}$Ne    &  3.300E-08   &   $^{22}$Ne    &  1.523E-03\\
       $^{23}$Na    &  8.371E-10   &   $^{24}$Mg    &  1.359E-03 &
   $^{25}$Mg    &  4.436E-08 \\   $^{26}$Mg    &  2.393E-05   &   $^{27}$Al    &  4.964E-07 &
   $^{28}$Si    &  7.802E-02\\  $^{29}$Si    &  4.621E-06   &   $^{30}$Si    &  3.423E-05 &
    $^{31}$P    &  1.805E-05    \\   $^{32}$S    &  5.435E-02 &
    $^{33}$S    &  1.117E-05&    $^{34}$S    &  8.222E-05    \\  $^{35}$Cl    &  1.680E-04 &
   $^{37}$Cl    &  4.491E-06&   $^{36}$Ar    &  1.464E-02   \\   $^{38}$Ar    &  4.677E-05 &
    $^{39}$K    &  3.950E-04&   $^{40}$Ca    &  1.868E-02   \\   $^{42}$Ca    &  1.931E-05 &
   $^{43}$Ca    &  4.565E-05&   $^{44}$Ca    &  7.252E-04   \\   $^{48}$Ca    &  1.613E-07 &
   $^{45}$Sc    &  1.648E-05  & 
   $^{46}$Ti    &  5.410E-06 \\   $^{47}$Ti    &  1.733E-04   &   $^{48}$Ti    &  1.902E-03 &
   $^{49}$Ti    &  1.438E-04 \\
    $^{50}$V    &  2.933E-09&    $^{51}$V    &  1.561E-04  &   $^{50}$Cr    &  3.245E-04 \\
   $^{52}$Cr    &  1.597E-04&   $^{53}$Cr    &  3.130E-04   &  $^{54}$Cr    &  1.254E-05 \\
   $^{55}$Mn    &  2.282E-03&   $^{54}$Fe    &  2.002E-02   &  $^{56}$Fe    &  6.397E-01 \\
   $^{57}$Fe    &  1.442E-02 &   $^{59}$Co    &  1.599E-04 &
   $^{58}$Ni    &  1.678E-02   \\ $^{60}$Ni    &  1.542E-03   &   $^{61}$Ni    &  3.353E-06 & & \\
  \hline
  \end{tabular}
\label{table_abundances}
\end{minipage}

\section{Synthetic Light Curves and Spectra Compared to Observations of SN\,2019\lowercase{vrq}}~\label{sec:model}

\subsection{Light Curve Properties and Limits on Asymmetry}~\label{sec:lc_model}
In Paper I and based on semi-analytical estimates for normal SNe~Ia, the empirical LC properties, such as the rise-time ($t_{B}=19.03 \pm 0.36 \pm 1.08$ days), the absolute brightness using the distance to the host ($M_{max}(B)=-19.44\pm 0.08$ mag), and the semi-bolometric LC have been derived.
The estimated $^{56}$Ni mass was estimated  to be $M(^{\rm 56}{\rm Ni})=1.02 \pm 0.14 $ and $0.77\pm 0.16 M_\odot  $ from the peak and tail, respectively.

From the simulations, the time of the explosion and the absolute brightness are given by the simulation and are not free parameters, allowing a physics-based interpretation
depending on the validity of the synthetic LCs. 
 The simulations use global energy conservation, resulting in the stability  of the simulated $L_{\rm bol}$ with respect to numerical errors and physical models. 
However, the  spatial resolution of $300^{3}$ is insufficient to resolve the optically thick X-ray/UV photosphere and hence to adequately describe the frequency redistribution between X-ray and UV fluxes. Monochromatic X-ray and UV fluxes are very uncertain. 
Note that for the bolometric and optical monochromatic LCs, the time-dependent diffusion of hard radiation is not relevant because the redistribution of energy to the optical and IR is governed by the rate equations with fast, allowed transitions via the incomplete Rosseland Cycle well below the UV and X-ray photosphere~\citep{2021ApJ...922..186H} (see also Fig.~\ref{fig:mass}). Therefore, stability in $L_{bol}$ translates into stability in B and V. 
Our numerical scheme has  been validated by a direct comparison  with both normal-bright and subluminous SNe~Ia \citep{2017ApJ...846...58H}.
Our calculations are also consistent with other full non-LTE codes, such as CMFGEN~\citep{2012MNRAS.424..252H,2022A&A...668A.163B}.



%
%

Synthetic LCs allow us to differentiate between 'classical' DD models and PDDs. Using the HYDRA code and the same setup of the exploding WD, simulations of a similarly bright `classical' DD model yield a $\approx$2.4 days shorter rise time than the PDD model~\citep{2017ApJ...846...58H}. 
The latter also produces bluer colors, higher luminosity, and less mixing of $^{56}$Ni to the surface layers, thus accounting for the longer rise times in general. The shape of the early LCs is heavily dependent on the effective photon diffusion timescale, which, to first order, is inversely proportional to the square of the expansion velocity, $v$, of the inner layers of the WD. Pulsation does effectively slow down the expansion of such inner layers and leads to high velocities in the outer layers (see Fig.~\ref{fig:Model}, and \citet{1993A&A...270..223K}). The mass of the unburned layers is directly proportional to the increase in the 
rise-time to maximum light and, thus, provides a scale to measure the mass of unburned ejecta via momentum conservation for a wide range of explosion scenarios \citep{1996ApJ...457..500H,2006ApJ...636..400Q}, and is independent of flux spectra.

\begin{figure*}[ht]
\centering
     \includegraphics[width=0.8\linewidth]{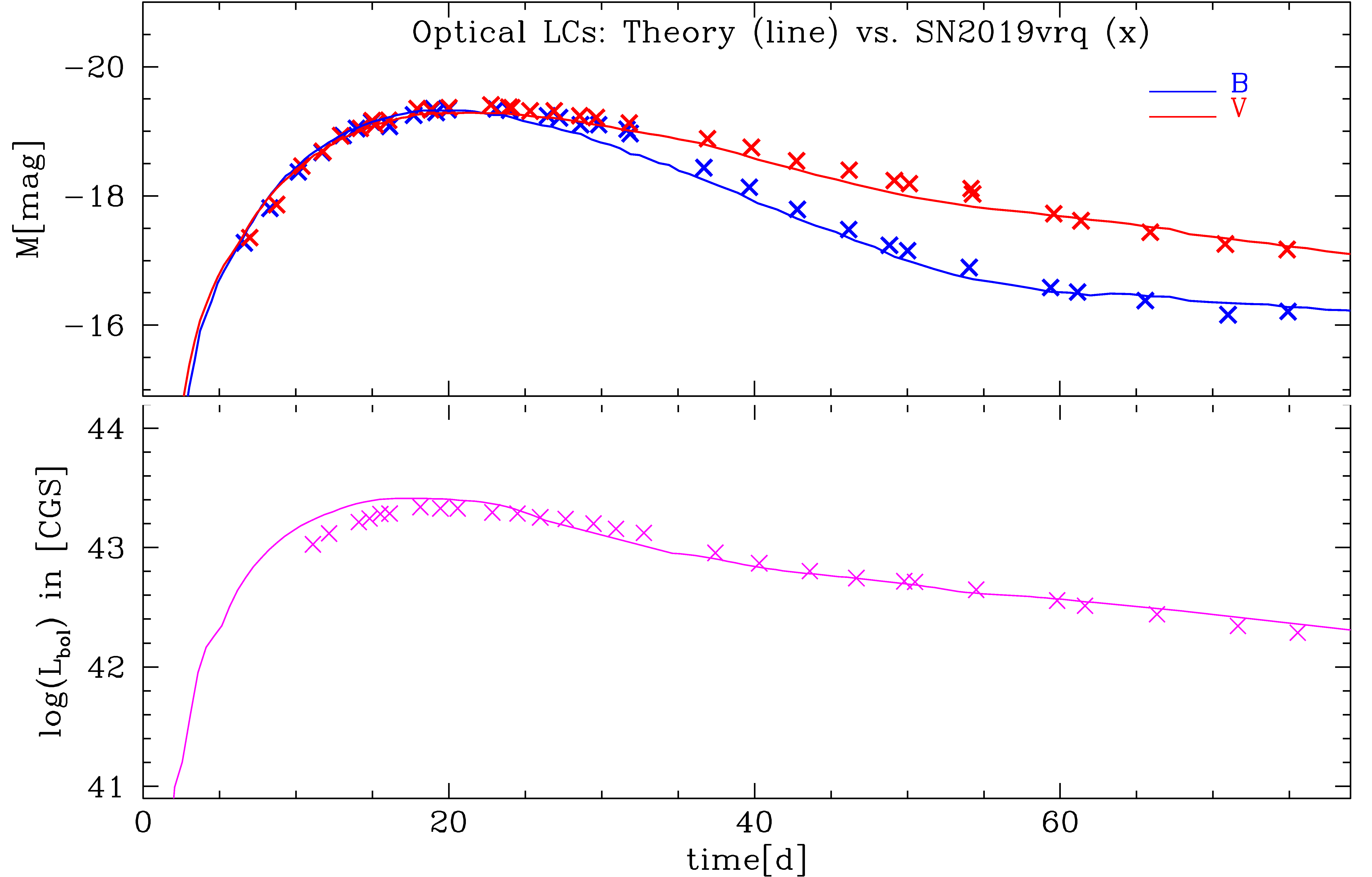}
     \caption{{\bf Comparison of synthetic and observed LCs using LC shapes in $B$ and $V$ with the zero time and absolute brightness and luminosity of the simulations.} The model has rise times in $L_{bol}$, B and V of $17.5\pm 1$, $19.7\pm 1$ and  $21.3 \pm 1.5$ days, respectively with uncertainties due to the flat maxima.  For the observations, the offset in rise time is $-$0.5 days, and the common 0.08\,mag difference in peak brightness of all LCs is well within the uncertainties using the empirical approach of Paper I, thus validating the reference model. The directional dependence of $L$ is $\approx$0.05\,dex, which is comparable to the error imposed by the grid.
     The model has been tested against normal SNe with NIR and MIR data (e.g. Fig.~11 in \citet{2018A&A...611A..58G}), and in \citep{2015ApJ...798...93T, 2024ApJ...975..203A}, which may be useful for high-precision cosmology (see Sect.~\ref{summary}). Note that the maximum-likelihood fit allows a shift of the observational zero-point in time within the uncertainties. 
    }\label{fig:lc_prof}
\end{figure*}

The best fit of the LC profiles is shown in Fig. \ref{fig:lc_prof}.
The good agreement in B and V validates the PDD models for SN\,2019vrq. The effects of inhomogeneity in the ejecta structures (`picket-fence') are well within the error margins for the optical wavelength range but result in a boost of the UV flux. The X-ray and UV flux combined contribute some 25 to 15\% to the luminosity a week before and at about maximum light, respectively. This is a factor $\approx 2$ larger than 
inferred in Paper I. This is possibly related to comparing synthetic bolometric LCs with empirical pseudo-bolometric LCs (Paper I). Post maximum, the increase in the redistribution to the IR and MIR starts at about maximum light and a few days past maximum, respectively, leveling off at $\approx 30\% $ some 60 to 70 days past the explosion in both the bolometric and $BV$ LCs. This is consistent with the MIR observations of SN\,2014J~\citep{2015ApJ...798...93T} and SNe~Ia observed in the same galaxy; see Fig.~11 in \citet{2015ApJ...798...93T} and \citet{2018A&A...611A..58G}. 

Note in Fig. \ref{fig:lc_prof} that the post-maximum decline ($\approx 25-50 $ days past explosion) in B and V is slightly too steep, possibly indicating that the model photosphere recedes slightly too fast past maximum because the density bump at 0.3 $M_\odot$ (see Fig. \ref{fig:Model}) should be at $\approx 0.5 ~M_\odot$   (Fig. \ref{fig:mass}). This interpretation is also supported by the spectra discussed below.
Agreement in  rise time of $B$ and $V$ measures the diffusion timescales $t_{\rm diff}(t)$ in the corresponding bands. To first order, $t_{\rm diff}(t_{\rm max}) \approx t_{\rm max}$~\citep{1982ApJ...253..785A}, and the entire ejecta contribute to $L$. 
The agreement between $B$ and $V$ in the rise supports the structure in the outer $0.3-0.4$\,M$_{\odot}$ that contributes to the LCs by day\,$-$9, when the first VLT spectrum was obtained. In addition, the post-maximum decline  $L_{\rm bol}$ is  also slightly too fast, suggesting that the stored energy is released too fast, consistent with the too-low Doppler shifts at day\,$+$12 (see below). Past day 50 to 60, the slopes in $L_{\rm bol}$, $B$ and $V$ agree with the observations because the redistribution to the NIR and MIR flattens out (see above). The actual decline rate is governed by radioactive decay, modified by the increasing escape probability of $\gamma$-ray photons. 


LC shapes may provide a new measure for asphericity.
 The overall agreement between synthetic and observed $B$ and $V$ across all phases of the LCs places additional limits on the directional dependence of the luminosity beyond what polarization can provide, which observationally is limited to local objects.
At late times, the entire envelope becomes optically thin and the brightness and colors become isotropic, whereas at early times, strong directional dependence is produced by axisymmetric density and abundance distributions expected for mergers or for explosions of a rapidly rotating WD. 
For SN\,2019vrq, the difference between observation and theory at early and late times $B$- and $V$-LCs is $\lesssim 0.3$\,mag. Assuming axisymmetric spheroids with a minor-to-major axis ratio of $\approx 0.8$ is consistent with axis ratios previously found for normal-bright SNe\,Ia~\citep{2001ApJ...556..302H, 2006NewAR..50..470H, 2012A&A...545A...7P, 2020ApJ...902...46Y, 2023MNRAS.520..560H}. This suggests that the size of the asphericity of 91T/99aa-likes is similar to the rest of SNe\,Ia.


\subsection{Flux Spectra}~\label{sec:spec_model}

\begin{figure*}[ht]
\includegraphics[width=1.0\linewidth]{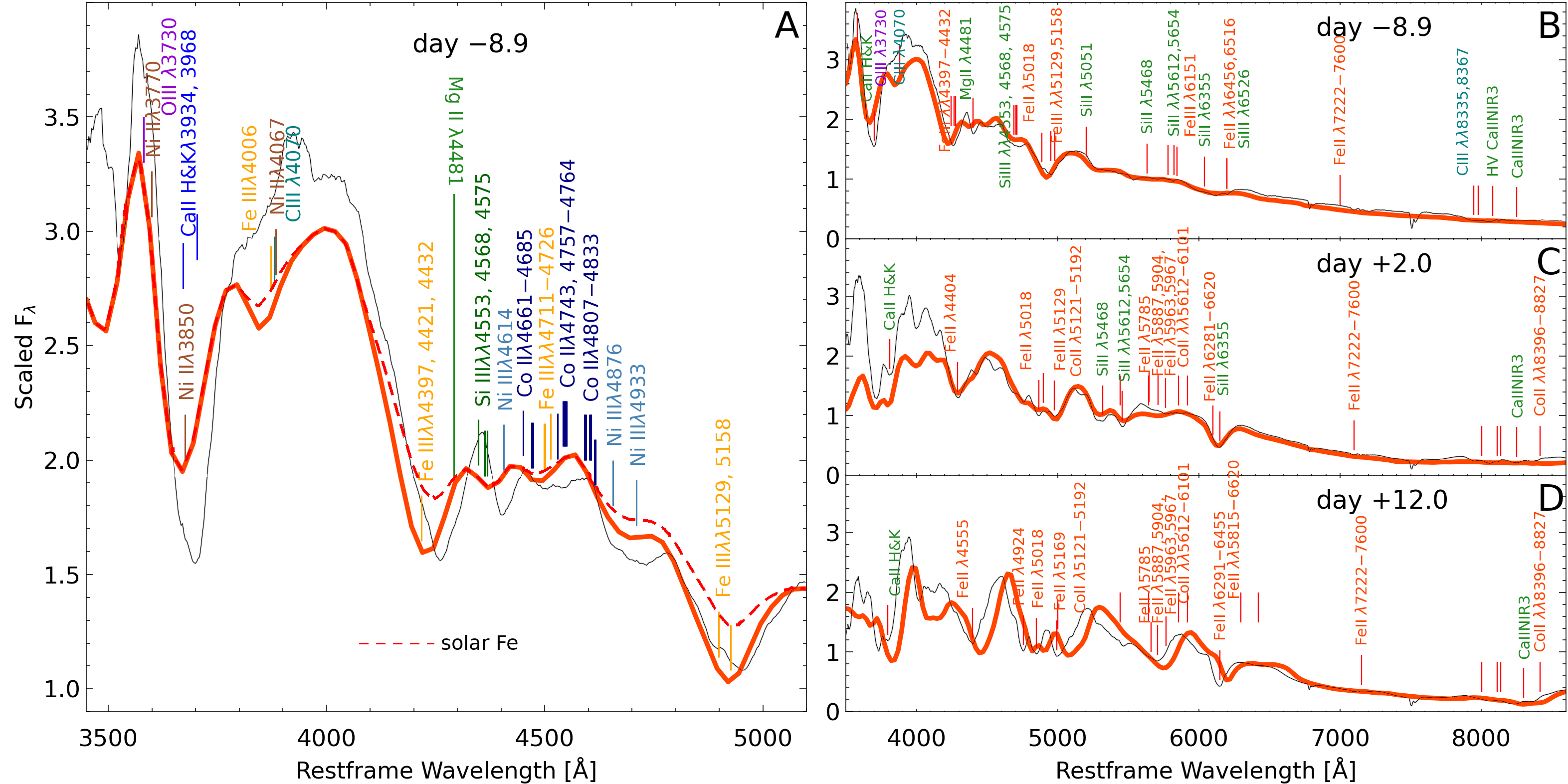}
\caption{\textbf{Synthetic optical spectra (red) of SN\,2019vrq compared to VLT observations (black).} Line identifications are based on Tab.\ref{tab:lines} produced from the simulations. For the mass ranges probed, see Fig. \ref{fig:mass}.
Panel A shows the blue region of the day\,$-9$ spectrum, where a plethora of blended weak NSE lines form the underlying quasi-continuum. Spectral line identifications and Doppler shifts are based on the reference model.
Panels B--D compare the observed and model spectra from $\approx$days\,$-$9 to $+$12 over the entire optical wavelength range. Overall, the model reproduces the observed spectral features, including Ca{\sc\,ii}, singly- and doubly-ionized Si, Fe, Co, and Ni. For demonstration in Panel A,
the reference model (Sect. \ref{sec:refM}) is shown for solar Fe (dashed line) at day $-$9.
In the days\,$+$2 and $+$12 model spectra, the 8000--8200\,\AA\ features are due to freshly synthesized Ca at the interface between incomplete oxygen burning and NSE (Fig. \ref{fig:Model}). Note that, at day\,$+$12, the expansion velocities are too slow by $\approx$1000\,km\,s$^{-1}$.}~\label{fig:model_spec}
\end{figure*}

The adopted PDD model reproduces both the elements and the ionization stages seen 
in the early spectroscopic time series of SN\,2019vrq (Fig.~\ref{fig:model_spec}).
 Throughout the observed epochs, the spectra are dominated by the products of 
 explosive oxygen burning and burning to NSE, superposed on a quasi-continuum 
 formed in the optical by many thousands of overlapping IGE lines, mostly singly 
 ionized Fe and Co (see, e.g., Fig.~1 in \citealp{1993A&A...268..570H}). The spectral 
 evolution is governed by the recombination of the ejecta, that is, by the transition 
 from doubly to singly ionized species, which is mostly complete by day\,$+$2. 
 Because of the large $^{56}$Ni mass and the correspondingly small mass of IMEs, all 
 features related to explosive oxygen burning (Si/S) are strongly blended with IGE 
 lines, which affects the interpretation of both the flux and the polarization spectra (see below).

The spectra probe successively deeper layers of the ejecta. Using the relations 
between velocity and enclosed mass shown in Figs.~\ref{fig:Model} and~\ref{fig:mass}, 
the spectra of SN\,2019vrq at days\,$-$9, $+$2, and $+$12 sample the material above velocities of the inner photosphere   
$v_{\rm ph}\approx$16,000, 12,000, and 9,000\,km\,s$^{-1}$, corresponding to the outer
 0.1, 0.25, and 0.4\,M$_{\rm Ch}$, respectively. Qualitatively, both the strong and 
 the intermediate-strength features are matched by their synthetic counterparts 
 (Fig.~\ref{fig:model_spec}A), which allows a physically founded understanding
 of the progenitor structure and of the explosion physics by means of spectral 
 diagnostics.

\paragraph{Day\,$-$9: composition of the outer ejecta.}
As shown in Figs.~\ref{fig:model_spec}A and B, a Doppler shift $\gtrsim 32,000 {\rm km/s}$ of the wings  of the prominent Fe{\sc\,iii}\,$\lambda$4397 feature extend well into the unburned C/O 
shell of $\approx$0.11\,M$_{\odot}$ (Figs.\ref{fig:mass} \& \ref{fig:Model}), where primordial iron is present. The models 
require a twice-solar metallicity, corresponding to a mass fraction of $\approx$0.04, 
to reproduce the observed strengths of the Fe lines. Primordially, Ni and Co are lower 
than Fe by factors of 16 and 300, respectively~\citep{2007SSRv..130..105G}, so that the 
presence of both Ni and Co lines implies some outward mixing of freshly produced NSE 
material, likely driven by turbulence associated with detonation cells or RT instabilities. In the model, the Co 
and Ni mass ratios relative to Fe are $\approx$5\%, with Co enhanced relative to Ni 
near the photosphere at $\approx 20,000$\,km\,s$^{-1}$ (Fig.~\ref{fig:Model}). Consistently, a flat $^{56}$Ni tail at 
the $\approx$4\% level extends from 22,000 to 28,000\,km\,s$^{-1}$ (Fig.~\ref{fig:Model}). 
The remainder of the day\,$-$9 spectrum is formed by freshly synthesized products of 
explosive carbon burning (e.g. Mg), by incomplete burning yielding silicon, and by NSE 
material characterized by a $^{56}$Co/$^{56}$Ni ratio of $\approx$1/6, consistent with 
radioactive decay.

Neither the synthetic nor the observed spectra show strong C or O features. Given the high luminosity 
and the correspondingly high early-time temperatures, C and O are predominantly doubly ionized.
 Although C{\sc\,iii} and O{\sc\,iii} have significant transitions below 
 4200\,\AA\ (Table~\ref{tab:lines}), these are partially drowned out by Ca{\sc\,ii}\,H\&K and by IGE lines.
  For the same reason, the photospheric Ca{\sc\,ii}\,NIR3 features are very weak at this epoch   because Ca{\sc\,iii} dominates.

The low-amplitude pulsation burns most of the ejecta to NSE dominated by $^{56}$Ni
 (Table~\ref{table_abundances}), so that only weak Si{\sc\,ii}/Si{\sc\,iii} features 
 arise from explosive oxygen burning. In particular, the Si{\sc\,ii}\,$\lambda$6355 doublet 
 is relatively weak and strongly blended with the Fe{\sc\,iii} multiplets near $\approx$6150\,\AA\ 
 and with Si{\sc\,iii}\,$\lambda$6388. These defining PDD signatures are present in both the model 
 and the observations. The somewhat weaker Ni and Co features in the observed spectra may indicate 
 a degree of mixing smaller than assumed in the model.

\paragraph{Structure of the Si{\sc\,ii}\,$\lambda$6355 region.}
The time evolution of the synthetic spectra is overall consistent with the observations of SN\,2019vrq. The broad feature covering $\sim$5900--6500\,\AA\ does not require 
the separation of high- and low-velocity components in the distribution of the Si{\sc\,ii} 
opacity suggested in Paper I. This is consistent with recent empirical findings from a sample study 
of the time evolution of 91T/99aa-like SNe\,Ia~\citep{2022ApJ...938...47P}. In the simulations, 
the  high- and low-velocity contributions of Si{\sc\,ii}\,$\lambda$6355 are instead produced 
by a layered ionization structure, namely a Si{\sc\,iii} zone sandwiched between two layers 
dominated by Si{\sc\,ii} and IGE blends. In the outermost layers, the reduced non-thermal excitation 
favors Si{\sc\,ii}, while in the innermost silicon-rich layers the higher density leads 
to large recombination rates; in between, the high luminosity and radiation temperature 
maintain Si{\sc\,iii}.

\paragraph{Model--observation differences at early times.}
While the model reproduces the overall structure of the spectral modulations between 
$\sim$4300 and 4800\,\AA\ (Fig.~\ref{fig:model_spec}A) without fine-tuning, the synthetic 
spectra exhibit slightly higher Doppler shifts than observed and the models may overestimate the adiabatic cooling. As a consequence, the Ni and Co features are somewhat stronger in the model. In particular, a P~Cygni absorption 
component near $\approx$4560\,\AA\ appears in the day\,$-$9 model spectrum but is 
barely discernible in the data. The spectral evolution at these epochs is fast and therefore 
sensitive to the uncertainty in the time since the explosion (see above); the higher velocities 
may reflect the one-day offset suggested by the LCs \citep{PaperI}. Moreover, the location of the interface 
between burned and unburned material depends on the mixing and on the pulsational amplitude, 
neither of which has been fine-tuned. Finally, the evolution of the broad feature between 
7600 and 8000\,\AA\ can be attributed to singly ionized Fe and Co together with photospheric 
Ca{\sc\,ii}\,IR3 (Table~\ref{tab:lines}), consistent with the  photosphere transitioning 
into the NSE layers early on; the Ca{\sc\,ii}\,IR3 contribution remains weak because most of the Ca 
is doubly ionized.

\paragraph{Days\,$-$14 to $-$7: the outer, chemically homogeneous layers and transition to IGE.}
Apart from the evolving Doppler shift corresponding to the outer 
$\sim$(3--10)$\times 10^{-2}$\,M$_{Ch}$ of the ejecta, the spectra show 
little evolution between days\,$-$14 and $-$7 (Fig.~\ref{Fig_spec}). In particular, 
the profiles near $\approx$ 4500\,\AA\ and $\approx$ 4900\,\AA\ remain nearly unchanged over this period. They are 
formed mostly by the blanketing of iron lines, with only a small admixture (1/6 
of the solar Fe value) of burned material (Fig.~\ref{fig:Model} and 
Sect.~\ref{sec:model_construct}). The persistence of these slowly 
evolving Fe-group features implies chemically homogeneous outer layers 
of $\approx$0.1\,M$_{Ch}$ with a primordial, twice-solar abundance 
pattern, which is incompatible with the evolution expected in the `classical' DD 
scenario (Fig.~\ref{fig:model_spec}A). The corresponding mass of unburned material is an order of 
magnitude larger than the $10^{-2}$\,M$_{\odot}$ found in bright classical  
delayed-detonation models~\citep{2017ApJ...846...58H}. This inference is supported 
by the weakness of Mg{\sc\,ii}\,$\lambda$4481 (Fig.~\ref{fig:model_spec}A), a major 
product of explosive carbon burning, which is barely distinguishable because of the 
strong blending in the early spectra. The implication is that the corresponding layers have not undergone 
explosive burning, and the weak Ni lines are due to mixing.

\paragraph{Day\,$+$2: the photosphere within the IGE region.}
By day\,$+$2, the $\sim$6300\,\AA\ feature is dominated by Fe{\sc\,ii}, with a still 
significant contribution from Si{\sc\,ii}. Such a transition from Si{\sc\,ii} to Fe{\sc\,ii} 
dominance is well known from normal SNe\,Ia (e.g., \citealp{1995ApJ...440..821H, 1995ApJ...441L..33N,2012AJ....143..126B}). 
For the 91T/99aa-like SN\,2019vrq, however, the photosphere lies well within the IGE region as early as $\sim$day\,$-$4 
because the Doppler shift of the absorption minimum at 6300 ${\rm \AA }$ is possibly blueshifted (Fig. \ref{Fig_spec}). This is also consistent with the time series of flux spectra (Fig.~\ref{Fig_spec}), 
which show a continuous increase in the strength of the feature between days\,$-$7 and $+$12, together with a blueshift of the 
absorption minimum around day\,$-$7. The overall Doppler shift of the model agrees well with the observations, and the lines 
are formed by NSE material with only a weak QSE contribution, comparable to normal SNe\,Ia.

The weak S{\sc\,ii} features marked at 5200 and 5600\,\AA\ are already dominated by Co{\sc\,ii} and Fe{\sc\,ii} multiplets, 
a consequence of the large $^{56}$Ni mass and the small Si mass that renders the IME lines substantially blended. The 
comparatively weak Ca{\sc\,ii}\,NIR feature strengthens in both the model and the observations, although it remains blended 
with IGE lines; in the model it is shallow because of the overall lower IME abundance. At this epoch, Ca{\sc\,ii}\,IR3 
is significantly weaker in the model than in the observations because the ionization has not yet fully shifted to the 
singly ionized stage, a consequence of the model photosphere receding faster than observed, as also indicated by the LCs 
(Sect.~\ref{sec:lc_model}).

\begin{figure}[ht]
\centering
\includegraphics[width=0.7\linewidth]{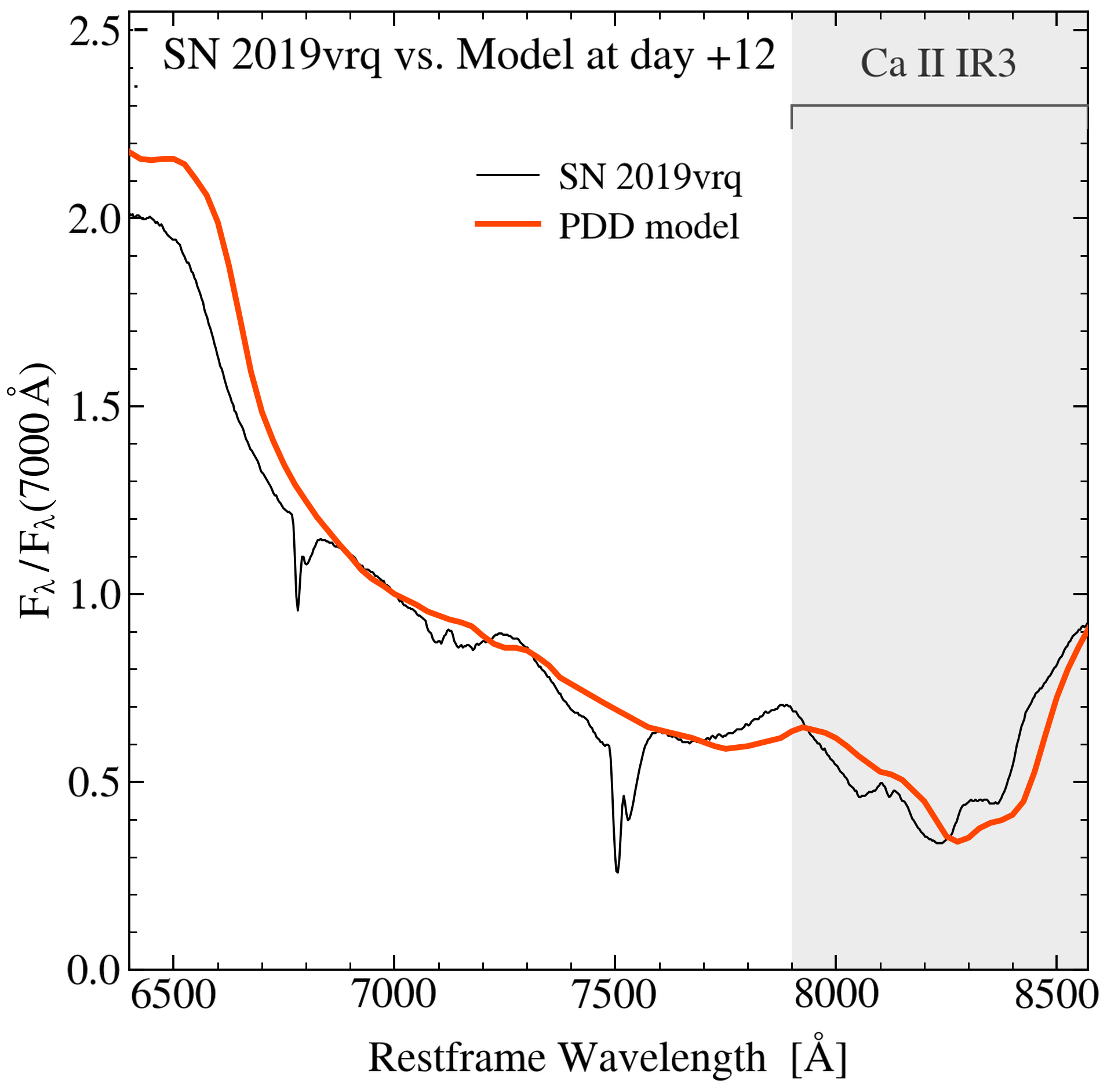}
\caption{\textbf{Profile of the Ca{\sc\,ii}\,NIR3 feature at day\,$+$12.} Same as Panel D of Fig.~\ref{fig:model_spec}, but with the flux normalized at 7000\,\AA\ and zoomed in on the Ca{\sc\,ii}\,NIR3 region. The `blue wing' extends to Doppler shifts of $\approx$33,000\,km\,s$^{-1}$, well beyond the 24,000\,km\,s$^{-1}$ attributed to the HV Ca{\sc\,ii} feature seen at day\,$-$9 (see above and Paper~I), but consistent with the extent of freshly synthesized Ca (Fig.~\ref{fig:Model}). The linear wing is characteristic of features formed in rapidly expanding atmospheres with exponential gradients in density and abundance.
The strongest modulations, at $\approx$8100 and 8350\,\AA\ in both the observations and the model, are produced by singly ionized Fe/Co blends drawn from some 100 moderately strong transitions in the Ca{\sc\,ii}\,IR3 region (Table~\ref{tab:lines}). The features at 6800, 7100, and 7500\,\AA\ are telluric (Fig.~\ref{Fig_spec}).
}~\label{fig:model_ca}
\end{figure}

\paragraph{Day\,$+$12: post-maximum spectra and limitations of the model.}
The day\,$+$12 spectrum of SN\,2019vrq closely resembles that of normal SNe\,Ia, 
as shown in Paper~I, and its evolution reflects the cooling of the envelope. The 
broad absorption between 5200 and 6100\,\AA\ is produced by Co{\sc\,ii} and Fe{\sc\,ii} transitions,
 the strongest contributors being the Fe{\sc\,ii} multiplets with rest wavelengths of 5341.07, 
 5403.50, 5412.88, 5728.14, 5813.28, 5818.31, 5855.80, 5886.64, 5904.45, 5957.34, 5967.27, 
 and 6073.10\,\AA\ (Fig.~\ref{fig:model_spec}D). At this epoch, Ca is 
 fully recombined in the model, and the synthetic and observed profiles agree well 
 (Fig.~\ref{fig:model_ca}).

The model spectrum does, however, exhibit slightly lower Doppler shifts than the 
observations (Fig.~\ref{fig:model_spec}D), indicating that the model photosphere 
recedes too rapidly, by $\approx$1000\,km\,s$^{-1}$. Because the recession of the 
photosphere is governed by the density of the ejecta, this points to a limitation 
of the underlying hydrodynamical model. In PDD models, kinetic energy is redistributed 
from the inner to the outer layers during the pulsation. We treat the deflagration phase 
with spherical models in which the burning rate is calibrated by 3D simulations, which is 
equivalent to a filling factor of unity during the early deflagration phase (Sect.~\ref{sec:model_construct}).
 The result is a local maximum in $\rho$ at $\approx$0.2\,M$_{\rm Ch}$, just outside the 
 deflagration region (Fig.~\ref{fig:Model}). In 3D simulations, the deflagration front reaches 
 larger radii because of a smaller filling factor, which pushes this local density maximum 
 further out. As shown by \citet{2026ApJ..1003L..37S}, the filling factor of burned material  approaches unity 
 in the inner layers owing to a pre-existing turbulent field, whereas in the layers with 
 fully developed RT instabilities it falls below unity by an amount that depends on the 
 size and morphology of the initial magnetic field~\citep{2021ApJ...923..210H} and on 
 further effects such as rotation. 


\begin{figure}
    \centering
    \includegraphics[width=1.0\textwidth]{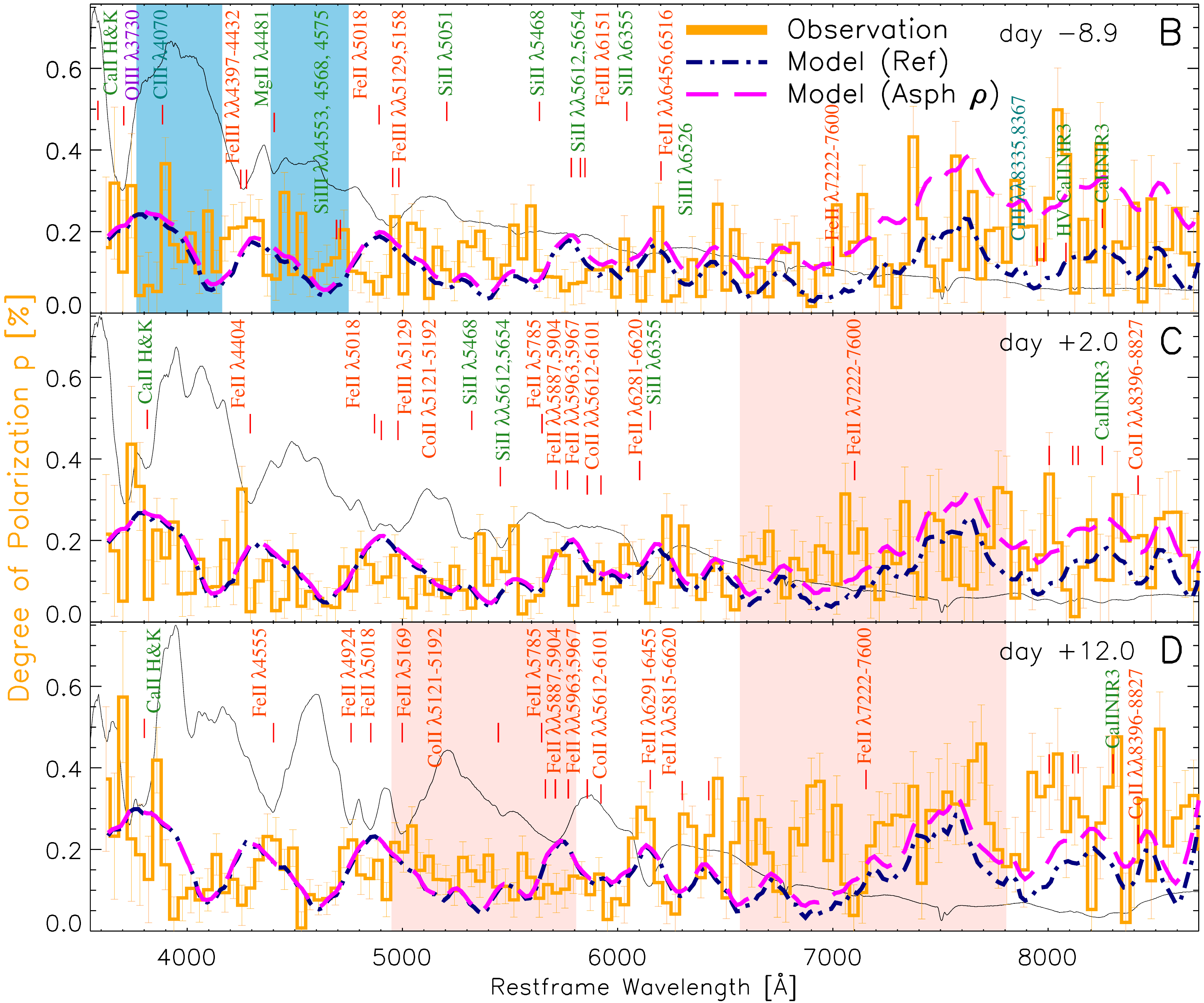}
    \caption{\textbf{Polarization (orange histograms) and scaled flux spectra (gray curves, from Fig.\ref{fig:model_spec}) of SN\,2019vrq.} Polarization spectra were computed after subtracting the interstellar polarization ISP (see Paper I). In each panel, we also overlay the influence of Class I asphericity (see text) on the synthetic polarization for the PDD reference model (navy dotted-dashed curve) and a rotationally symmetric imprint from the WD rotation (magenta long-dashed curve). The model polarization spectra are calculated for the most likely viewing angle of 30$^{\circ}$ determined from a pattern analysis (Sect.~\ref{sec:classes}). Note that the HV Ca{\sc\,ii}  and C{\sc\,iii} at 8335, 8367 and O{\sc\,iii} at 3730 ${\rm \AA}$ hardly show in the flux spectra but can be tentatively detected in polarization as partially resolved features (3 bins) with peaks in P of 0.5 \% in Ca and hints in P peaking at $\approx$ 0.3 \% in C and O. Note that the overall polarization (P) pattern is imprinted in the synthetic P from the thermalization depth at large densities by the quasi-continuum, see e.g. the role of Fe{\sc\,ii} multiplets between 7222-7600 ${\rm \AA}$ (see text). The polarization spectrum of SN\,2019vrq shows many narrow single-bin features with P $\lesssim$ 0.3 \% at all epochs.} 
    \label{fig:pol_spec}
\end{figure}

\section{Limits on Asphericity from Polarimetry and Anisotropic Luminosity of SN\,2019\lowercase{vrq}}\label{sec:asplum}


\subsection{Observations, Classification of Asphericities, and Orientation of the Observer}
\label{sec:classes}
Our simulations rely on the optical flux spectra and their interpretation as a main source of information about the nature of the explosion, but they use polarization to obtain a few valuable pieces of information on the geometry, or better, to ask for an upper limit on the asphericity and directionally dependent luminosity that is compatible with the data. Based on the simulations, we will study the physical origin of the polarization within the framework of low-amplitude PDDs and show that the low polarization is a direct consequence of the physics in our reference model (see Sect. \ref{sec:refM}).
Subsequently, limits for asphericity and directionally dependent luminosity can be calculated.

Observational and synthetic polarization and observed flux spectra are shown  in Fig.~\ref{fig:pol_spec}. We mark some of the prominent features in the flux
as identified in Sect.\ref{sec:spec_model} but show the observed spectra to avoid
the propagation of systematic shifts discussed in Sect. \ref{sec:spec_model}.
The polarization of SN\,2019vrq from days\,$-$9 to $+$12 remained marginal, and all optical features were blended by numerous IGE multiplets.

  Major spectral features that exhibit a moderate level of polarization in our simulations include blends of IGE lines consistently present at $\approx$ 3900, 4800, 6200, 6400, and 7500 ${\rm \AA}$, in particular, the broad feature between 6800 and 7800 ${\rm \AA}$, as well as the Ca{\sc\,ii} $H\&K$ and Ca{\sc\,ii} NIR3. All broad features are related to IGE multiplets or, more precisely, due to  variations in the thermalization depth, which depends on the ratio between electron scattering and quasi-continuum opacities. The contribution of Ca{\sc\,ii} hardly varies with time because, being resonance lines, even a low Ca{\sc\,ii} fraction contributes to the polarization $P$. 
  
  Overall, observations show a similar pattern, sometimes in groups of narrow, one-pixel-wide features. We emphasize that the presence of polarization requires aspherical envelopes.
The same broad pattern produced by Fe{\sc\,ii} and Co{\sc\,ii} multiplets has been observed in SN\,2019np (Fig. 13 in \citet{2023MNRAS.520..560H}), but with smaller amplitude. This pattern is produced by atomic physics and results from the small but variable scattering component in a photosphere dominated by a quasi-continuum of lines. It is one of the components to understand the formation of the polarization spectrum.

Guided by the classification scheme similar to that employed in \citet{2026ApJ...996L..10C}, we consider three classes of asphericity based on their physical origin that are relevant for PDDs: \\
{\bf Class~I: Large-scale, lobed patterns in the abundances at the interfaces between regimes of burning}: During the pulsation or the `rebounce', the DDT can be expected by the Zeldovich mechanism \citep{1970JAMTP..11..264Z}
in the phase and location of maximum deceleration, leading to an off-center DDT at one point \citep{1999ApJ...527L..97L}. The center of mass and the DDT define an axis of symmetry of large-scale lobed abundances at the interface between the QSE layers of explosive oxygen burning (Si/S) and NSE,  the layers of explosive carbon burning (O/Ne/Mg)- and Si/S-rich layers (see e.g. Fig.~11 in \citealp{2003LNP...635..203H}). \\
{\bf Class~II: Small-scale structures in the abundance at the burning interfaces:} This polarization component is due to RT or 'detonation-cell` instabilities of numerous small-scale structures of  $\approx$6,000--7,000\,km\,s$^{-1}$ and less due to the decay of large RT scales with a distribution close to the Kolmogorov relation (Sect.\ref{sec:DDT}). These will produce loop-like patterns in the Stokes $Q-U$ plane because of the picket-fence abundance structures \citep{2026ApJ...996L..10C}, as seen in SN\,2019vrq (Paper I). \textcolor{black}{The amplitude in polarization modulation of $\lesssim 0.05\%$ in the reference model \footnote{This is at the S/N limit of the Monte Carlo noise at  $\approx 0.02 \%$}, thus, is comparable to, but smaller than those found in normal SNe\,Ia~ \citep{2026ApJ...996L..10C} because of the low optical depth of the IME layers of the reference model.} \\
{\bf Class III: A large-scale structure in both density and abundances:} This can be caused by the WD rotation~\citep{1985A&A...146..260E} and is represented here by axisymmetric spheroids. Because our reference model does not include rotation, the effect is presented separately in Sect.~\ref{sec:limits_asymmetry}. \\

Classes~I and III may show different, non-aligned global axes of symmetry. Depending on their respective importance for a specific wavelength range and time, the misalignment will result in scatter in wavelength or a rotation of the dominant axis in the $Q-U$ plane \citep{PaperI}.

Quantitative limits depend on the orientation of the observer.
For SN\,2019vrq, the viewing angles ($\theta,\phi$) of an observer can be estimated by using the broad polarization patterns in the models as they relate to the flux spectrum (Fig.~\ref{fig:pol_spec}).
 The polarization of SN\,2019vrq is marginal and requires averaging over many pixels. Even then, at some epochs, several patterns are grounded in noise. E.g., the 3800 and 4300 ${\rm \AA}$ signatures are present but marginal, and the broad patterns between 6700 and 7600 ${\rm \AA}$ are easy to identify but spurious at earlier times.
 
 As in a previous work \cite{2023MNRAS.520..560H}, we use the broad pattern across the entire wavelength and time spans of the polarization spectra to extract $\theta$ via $\chi^2$ minimization over a range corresponding to a 95\% probability.
In principle, different physical mechanisms may exhibit no symmetry axes or misaligned symmetry axes. Each process has a specific angular dependence (see, e.g., Fig.~5 in \citealp{2026ApJ...996L..10C} and references therein).
Since the observed polarization of SN\,2019vrq remains marginal in all epochs, the possible ranges of $\theta$ are determined separately by finding its most probable value, assuming all polarization is due to one specific process, i.e. Class I, II or III.


For  SN\,2019vrq, the  ranges for $\theta $ (2 $\sigma)$  are found to be $[-10^{\circ},+70^{\circ}]$ for Class~I and $[-60^{\circ},+60^{\circ}]$ for Class~III considered in Sect.~\ref{sec:limits_asymmetry}. For characteristic, class-specific $\theta$ dependencies of $P$,
 see Fig. 5 in \citet{2026ApJ...996L..10C}.
The range allowed for Class~I is related to time evolution or the lack thereof, because the lobe in abundances would be blocked from the view of an observer by the photosphere in the first spectrum observed (see Fig. 11 in \citet{2023MNRAS.520..560H}). Class~III is symmetric around the equator, and the limit is driven by the amplitude of the modulation in the polarization spectra.
Class~II produces about 0.05\% polarization. Since Class~II cannot account for the observed amplitude, the pattern must be dominated by Classes~I and III. Matching their combined amplitudes favors $\theta \approx 30^{\circ}$.
As a reference, we assume an inclination of 30$^{\circ}$ as it is consistent with both Classes~I and III.

Based on our simulation for our reference model and consistent with previous findings for `classical' DD models, the maximum of P is produced for $q\approx 0.8 M_\odot$ \citep{2006NewAR..50..470H,2026ApJ...996L..10C}.  The overall patterns in the polarization spectra at all epochs of SN\,2019vrq are consistent.
More central DDTs, such as those in many normal SNe\,Ia~\citep{2020ApJ...902...46Y, 2023MNRAS.520..560H} are not supported for SN\,2019vrq because the amplitude of the broad features in the synthetic $P_{\rm Si\,II}$ would be smaller by a factor of two, even if seen from the inclination at which $P$ is maximized. \\

\subsection{Physical Origin of the Low  Polarization, and Limits on the Directional Dependent Luminosity}
\label{sec:origin}

The synthetic overall polarization spectrum evolves gradually, with the polarization pattern at amplitudes of $\approx$ 0.1--0.2\% closely following the blends and broad multiplets of IGEs, as identified by the flux spectra, but without strong variations in broad IME features.
In the reference model, the polarization is mostly produced by the off-center DDT, resulting in a small asymmetry in the abundances and directional dependence of $\lesssim 10\%$ in luminosity (Class~I). Note that the persistence over time requires global asphericities (Classes~I/III). Small-scale density clumps (Class~II) imprint a scatter in the $Q-U$ plane at the 0.1$\%$ level corresponding to individual transitions of elements~ within broad blends \citep{2026ApJ...996L..10C}.

The high iron abundance in the photosphere leads to thermalization at a low Thomson-scattering optical depth because the opacity of the quasi-continuum induced by IGE lines dominates Thomson scattering by a factor of 10~\citep{1993A&A...268..570H}.
This results in low polarization, with frequency modulation arising from changes in thermalization depth. The features appear to be broad because they are not formed by individual lines but by multiplets spanning $\approx$100\,\AA.
Note that the frequency resolution is sufficient to resolve the large-scale RT (Class~II), but due to the low mass of IME elements (Table~\ref{table_abundances}), most of the corresponding features are optically thin, with the notable exceptions of Ca{\sc\,ii}\,H$\&$K and Ca{\sc\,ii}\,NIR3.

 Most of the WD  is burned under NSE conditions ($M(\rm NSE)\approx 0.94\,M_{\odot}$) corresponding to an expansion velocity of $\approx$ 19,000 ${\rm km~s^{-1}}$ (Fig.\ref{fig:Model}), and the production of Si is low ($M(\rm Si)\approx$0.1\,M$_{\odot}$, Table~\ref{table_abundances} and Fig.~\ref{fig:Model}). The photospheric velocities are $v_{\rm ph}$(-9d)=16,000,  $v_{\rm ph}$(+2d)=12,000 and $v_{\rm ph}$(+12d)=9,000 km/sec which is well within the NSE layers (Sect. \ref{sec:spec_model}). This prevents the extended phase of
prominent Si{\sc\,ii} polarization frequently seen in normal SNe~Ia (Cikota et al.\ 2026). From the observations, most of the corresponding features predicted by the models are present, though weak, because the transitions of QSE elements are barely optically thick in the co-moving frame.

Within the framework of the PDD model, the low polarization of the Si{\sc\,ii}$\lambda6355$ absorption feature, blended with Fe{\sc\,ii} lines, is a consequence of the significantly lower Si{\sc\,ii} optical depth in 91T/99aa-like events compared to normal SNe~Ia. In addition, already by day $+2$ the photosphere forms within an IGE-rich layer, owing to the much larger amount of IGE relative to normal events. The peak polarization of the broad, blended features
$\approx 0.1\%$ is consistent with an off-center DDT at $0.8\,\ M_\odot$ viewed at an inclination of $\approx 30^{\circ}$, as inferred from the collective pattern analysis above (Sect.~\ref{sec:classes}).

As illustrated by Fig.~\ref{fig:pol_spec}, both observed and model polarization spectra of SN\,2019vrq display little temporal evolution. This can be attributed to the dominance of IGE opacities already early on (day\,$-$9), namely a large number of weak lines forming a quasi-continuum that leads to depolarization, resulting in a low level of polarization in the optical.
Moreover, the density gradients are steep in the outer layers. As a result, even large deviations from spherical symmetry can hardly produce strong polarization across the corresponding spectral line~\citep{1991A&A...246..481H, 2023MNRAS.520..560H}. The evolution lacks the prominent Si{\sc\,II/III} features often seen in normal SNe\,Ia.
However, the size of asphericity ($\approx 10\%)$ is very comparable to (or larger than) that in normal SNe~Ia, with $\Delta L/L_{\rm mean}\approx$10\%~at maximum light in our reference model.

\subsection{Narrow Polarization Features}
\label{sec:narrow}

The narrow nature of the observed IME and some IGE features, along with their offset relative to the rest wavelength, may suggest the presence of small-scale structures. In most instances, these implied structures are comparable to the size of the smallest resolution element in our VLT polarimetric data (Class~II, corresponding to $\approx$700\,km\,s$^{-1}$), which is consistent with the cascades produced by RT instabilities but is much smaller than the largest scales expected in the reference model and classical DD models (Sect. \ref{sec:model}). We note that
many of the labels in Fig. \ref{fig:pol_spec} are not aligned with the absorption
minima of the flux spectra, as may be expected for strong P Cygni-like profiles \citep{2006NewAR..50..470H,2013MNRAS.433L..20M,2024MNRAS.528.3875M}. Due to blending, the correlation may shift. Some peaks attributed to Si{\sc\,ii}\,$\lambda$6355 may in fact be Fe{\sc\,II/III} instead, and will contribute to the scatter identified in the $Q-U$ plane (see Fig. 11 in Paper I).

 The reference model does not show narrow polarization signatures in Si{\sc\,ii} on the resolution scale of the VLT data, although corresponding abundance imprints are present and well resolved in the simulations (Sect.\ref{sec:model}).  Although spatially resolved, for the smallest structures, the Si{\sc\,ii} and most individual IGE lines are optically thin at and outward of the burning interfaces at early times (Sect. \ref{sec:model_construct}). Based on the 3D hydrodynamic models, 
at later times and within the NSE layers any abundance pattern is wiped out by the detonation front~\citep{2005ApJ...623..337G}, and
the scattering fraction of the quasi-continuum becomes very small (e.g. \citet{1993A&A...268..570H,2001ApJ...556..302H,2012A&A...545A...7P,2017hsn..book.1017P,2023MNRAS.520..560H}).

In SN\,2019vrq and, by proxy, all 91T/99aa-like SNe, the IME polarization is expected to be low because the IME features are optically thin, because of the high-ionization, and post-maximum light because the IME layers are well above the photosphere even for asphericities comparable to normal SNe~Ia. By contrast, models for normal-bright SNe~Ia show significant P in Si{\sc\,ii}
because the effective optical depth is $\approx$ 20 to 30 for the entire envelope,
and, consequently, structures down to $\approx $ 1000 km/sec may be optically thick.  According to our reference PDD,  the differences to a normal-bright SN~Ia are that a)
 the single ionization stage dominates throughout the photospheric phase, b) the transition between NSE and QSE layers expands to some 9,000 km/sec, and c) densities in the IME layers are larger by more than an order of magnitude \citep{2023MNRAS.520..560H}.

However, we must also point out limitations.
Our reference model depends on the initial conditions of the simulations. $^{56}$Ni-driven instabilities
or high magnetic fields at the NSE/Ca interface may produce large-scale caustic structures in the core  \citep{2015ApJ...804..140F}, which produce line polarization in a mix of
IGE and Ca within the core region of the SNe~Ia as indicated by late-time
spectropolarimetry \citep{2022ApJ...939...18Y,2023MNRAS.520..560H}.

We detected HV Ca{\sc\,ii} in SN2019vrq in polarization (Fig. \ref{fig:pol_spec} and \citet{PaperI}), but not in the synthetic spectrum.
An important piece of physics may be missing that could produce corresponding scales, e.g., the interaction of the SN ejecta with the circumstellar environment, such as a Roche-lobe due to mass overflow from a companion or with the companion star, as evidenced by the HV Ca{\sc\,ii} discussed below. The polarization data may open up a new, exciting window, or the small, single-bin features observed may be noise.
Note that the $\approx$0.1\% scatter in the $Q$, $U$, and the polarization spectra of SN\,2019vrq, as displayed with a 40-\AA\ binning, may reflect the S/N of our VLT spectropolarimetry (see Paper I).  \\

\subsection{Limits on Global Asphericity in Density and Abundances from Rotation}
\label{sec:limits_asymmetry}


As discussed in Sect. \ref{sec:classes}, polarization can arise from axisymmetric abundance distributions produced by burning physics (Class~I), small-scale local structures (Class~II), and global density and abundance distributions (Class~III, rotation), which we consider in this section. The marginal rise of the continuum polarization towards red wavelengths (Table 3 of Paper I) may suggest the presence of an additional component in the electron scattering or density, previously attributed to rotation~\citep{2001ApJ...556..302H, 2012A&A...545A...7P}.
Moreover, the long-term change in the main axis in the $Q-U$ plane over time (see Paper I) suggests misalignment between the axes of symmetry defined by the rotation and the kinematic center, as well as the location of the DDT, respectively.
Note that the continuum polarization is produced in our PDD model by a Class~III asphericity that peaks at a viewing angle of  $\approx 30^{\circ}$ with an amplitude larger by $\approx$1.5 than at 60$^{\circ}$, consistent with
previous studies \citep{1991A&A...246..481H} because ${\rm P(\theta)\approx sin^2(\theta)}$ \citep{1957lssp.book.....V}.
Some misalignment will change the contribution between Classes I/III, but Class II  will have little impact on the resulting polarization.


In thermonuclear explosions, the released energy depends on local density. The initial structure of a rapidly rotating WD is largely preserved because, to first order, the initial and final structures are self-similar.
We compare the spherical reference model with a deformed density structure by adopting the self-similar solution for a rigidly rotating WD and approximating the asymmetry with ellipsoids (Figure~4 of \citealp{1986A&A...168..130E}). The resulting synthetic polarization spectrum is obtained by mapping the abundance structure and non-LTE level populations of the reference model onto the distorted geometry.


The axis ratio $B/A$ at the photosphere on day\,$-9$ was chosen to be 0.8, with B and A being the minor and major axes, respectively. Using the data in Fig.~\ref{fig:mass}, the corresponding ratios are $\approx0.92$ and $\approx0.98$ at days\,$+$2 and $+$12, respectively. These ratios provide an upper limit on the size of the additional asymmetric component.
The inferred magnitude of the rotational distortion is consistent with values typically found from normal-bright to subluminous SNe\,Ia~\citep{2001ApJ...556..302H,2012A&A...545A...7P,2026ApJ...996L..10C}. What sets 91T/99aa-like SNe apart is not the geometry, but the low production of Si/S. Thus, based on the explosion scenario, we suggest that 91T/99aa-like events represent a separate group of events with a scenario distinguished from 'normal` SNe~Ia but the amount of rotational distortion may be similar.



The axes of the components attributed to Class I and III are likely not aligned, naturally producing loops in the $Q-U$ plane (e.g. \citep{2017hsn..book.1017P}).
In the simulations, structures produced by Class~II would manifest as small-scale instabilities, generating loops on the $Q-U$ plane over short timescales but not contributing significantly to the global directional dependence of the luminosity. Note that the small scales are optically thin in our PDD, so they produce no signal.
Both small scales and the fact that many features are blends of multiple transitions at different rest wavelengths will produce scatter in the $Q-U$ plane (Fig. 9 in Paper I). With blends occurring at different wavelengths, the direct relation between the Doppler shift and the expansion velocity is broken.
For the reference model including rotational distortion, the maximum optical luminosity variation  $\Delta L/L_{\rm mean}$ induced by the asymmetry in the density and abundances is  $\approx$15\%, arising primarily from optically thick layers just below and in the photosphere, consistent with previous estimates for SNe~Ia e.g. \citep{2012A&A...545A...7P}.

\subsection{What was learned from Polarization about SN\,2019vrq and Anisotropic Luminosities?}\label{sec:aniso}
 91T/99aa-like SNe~Ia are different from normal SNe~Ia, and they are all the same. They are different than normal SNe~Ia because P is small (Paper I) due to the quasi-continuum being formed by IGE lines early on. 
Moreover, the density gradients are steep in the outer layers. As a result, even large deviations from spherical symmetry can hardly produce strong polarization across the corresponding spectral line. The evolution of the polarization spectra lack the prominent Si{\sc\,II/III} features often seen in normal SNe\,Ia. 
However, the size of asphericity is suggested to be comparable to (or larger than) that in normal SNe~Ia. All three classes of asphericity are consistent with the $P$ limit of SN\,2019vrq (Sect. \ref{sec:classes}). From the outer layers, the directional
 luminosity dependence is  $\Delta L/L_{\rm mean} \lesssim $15\%~at maximum light in our reference model, making these promising standard candles. The EC capture production in the core (Sect.\ref{sec:model}) currently dominates  and boosts the total diversity in the luminosity of 91T/91aa-like SNe to $\approx \pm 0.35$ mag. The uncertainty in the EC component can be significantly reduced by JWST, thereby enabling the full potential of several key projects with the Roman Space Telescope and the Rubin Observatory.


\subsection{The HV Ca~II NIR3 Feature}~\label{sec:HVCa}
With SN\,2019vrq being representative of the 91T/99aa-like class, the absence of a strong HV
Ca{\sc\,ii}~NIR3 feature at early times is expected and is a direct consequence of the high
luminosity and correspondingly high ionization state at these epochs.\footnote{A detailed
discussion of possible core-region inhomogeneities that could imprint Ca signatures caused by pre-existing velocity fields, off-center DDTs, $^{56}$Ni-decay--driven thermal instabilities, or
caustic structures, and RT instabilities injecting Ca into the core can be found in  \citep{2023MNRAS.520..560H,2026ApJ..1003L..37S}.}
                    
We now examine the HV Ca{\sc\,ii} NIR3 line in the context of our reference PDD model. The reference model does not include shell interaction with circumstellar material (CSM), but produces freshly synthesized Ca between 16,000 and 33,000~km~s$^{-1}$ (Fig.  \ref{fig:Model}).
No CSM or internal physics forming a high-velocity shell has been considered in the reference model.  Deciphering the properties and origin provides a test of explosion physics not included in the current simulation.

Table~\ref{tab:hvca_epochs} summarizes the behavior of the HV and the photospheric Ca{\sc\,ii} NIR3
features in flux and polarization at the three epochs of VLT spectropolarimetry, compared to the predictions of the reference PDD model.
As discussed in Sect.\ref{sec:spec_model}, the absence of strong early Ca{\sc\,ii} NIR3 
in SN\,2019vrq and 91T/99aa-like SNe is caused by high ionization, whereas Ca{\sc\,ii} NIR3 is strong in normal SNe~Ia because singly ionized Ca is the dominant ionization state. Thus, this observational difference does not exclude the same underlying dynamical origin.\\

\noindent 
{\sl The HV Ca{\sc\,ii} NIR3 feature in light of the classical CSM-interaction:}
Flux spectra of almost all normal SNe~Ia show a strong HV Ca{\sc\,ii}~NIR3 feature at Doppler
velocities between 18,000 and 30,000~km~s$^{-1}$ \citep{2013ApJ...777...40M}. It is often attributed to
a dense shell formed by an interaction between the SN ejecta and material bound in the progenitor
system, with an inferred mass of $\approx 10^{-3}$ to $10^{-2}\,M_\odot$ \citep{2004ApJ...607..391G}, rather
than being part of the exploding WD itself. A density shell is required to force recombination to
Ca{\sc\,ii}. This interpretation is supported by the coincidence between the Doppler shift of
the blue wing of Si{\sc\,ii} and that of the absorption minimum of the HV Ca{\sc\,ii} NIR3 feature
\citep{2006ApJ...636..400Q}. It disfavors the alternative explanation of explosive He burning (see
Sect. \ref{sec:alternative}) in a sub-$M_{\rm Ch}$ WD, or of unconsumed He surviving pycnonuclear reactions in a
near-$M_{\rm Ch}$ WD prior to explosion \citep{2019nuco.conf..187H}. Early-time spectropolarimetry of
normal SNe~Ia shows the HV Ca{\sc\,ii} feature to be strongly polarized, with a well-defined
axis in the $Q$--$U$ plane that is \emph{not} aligned with the WD interior
\citep{2020ApJ...902...46Y, 2023MNRAS.520..560H}.

\begin{deluxetable}{llll}
\tablecaption{ Ca{\sc\,ii} NIR3: Observations ({\bf italic}) (Paper I) vs.\ Reference Model ({\bf regular}). Regular font applies to both.
\label{tab:hvca_epochs}}
\tablehead{
\colhead{Epoch} & \colhead{Flux} & \colhead{Polarization} & \colhead{Interpretation/Implication}
}
\startdata
Day $-$9 & {\sl \textbf{Weak HV feature at $\approx 24,000$~km~s$^{-1}$}}  &   {\sl \textbf{Small, resolved HV Ca}} & Ca mostly
doubly ionized, \\
          & {\bf no HV Ca} & on broad $P$ pattern & shell
needed to shift ionization to Ca{\sc\,ii} \\
Day $+2$  & HV Ca component not identifiable; & Consistent with blend of & Recombination
strengthens \\
          & photospheric Ca{\sc\,ii}~NIR3 strong & Ca{\sc\,ii}~NIR3 + Fe/Co{\sc\,ii} 
& photospheric Ca{\sc\,ii} NIR3 and Fe/Co{\sc\,ii}\\
Day $+12$ & Broad wings extend well beyond the & same as day +2 & Early HV Ca \& late-time profile are \\
          & day $-9$ HV component (Fig.~\ref{fig:model_ca}) & & formed by freshly synthesized Ca \\  
\enddata
\tablecomments{Modulations present in $P$ redward of $\approx 6800$~\AA\ at the $\sim 0.1\%$ level
are consistent with a blend of Ca{\sc\,ii}~NIR3 and numerous Fe/Co{\sc\,ii} transitions
(Sect.\ref{sec:spec_model}), though residual instrumental effects such as fringing contribute. All interpretations are model-based.}
\end{deluxetable}

 HYDRA in the version from 2001 for spherical geometry \citep{2001AIPC..586..459H} has previously been used to model the spectral impact of a shell of H-rich CSM
\citep{2004ApJ...607..391G,2006ApJ...636..400Q}, and a high Ca abundance is expected to produce a qualitatively similar
effect. Following \citet{2004ApJ...607..391G,2021ApJ...920..107L,2023MNRAS.520..560H}, we combine momentum conservation with a mass of $10^{-2} ~M_\odot$ and a density contrast
between the shell and its surroundings of $\rho_{\rm sh}/\rho_{\rm sur} = 4$. Note that the mass of the surrounding material may be smaller for an aspherical distribution. During the pulsation, the outer layers of the envelope is $\lesssim 10^9$ km (Fig.~\ref{fig:progenitor}), comfortably within, e.g., the Lagrange point of a
Roche lobe for He- or H-star donors ($0.1$--$1\,R_\odot$) or the size of the binary system
itself ($0.01$--$10\,R_\odot$).  \\


\noindent
{\sl Can a Spherical Shell Illuminated  by an Aspherical PDD Produce the Observed Polarization? }
As a numerical test, we assumed a shell width of $\pm 2000$ km~s$^{-1}$ \citep{2004ApJ...607..391G}.
For the reference model or axisymmetric asphericity  with a spherical shell, the maximum peak polarization reaches only
$P({\rm Ca{\sc\,ii}~NIR3}) \lesssim 0.2$--$0.3\%$, which is a factor of two below the observed value of 0.5 \% (Fig. \ref{fig:pol_spec}). Changing the width of the shell alters the depth of the flux depression but has little effect on $P$, because the expansion velocity greatly exceeds the width of the shell. This result suggests that the shell itself is aspherical. While perhaps unsurprising given current CSM candidates which include stellar and accretion winds, symbiotic binary ejecta including recurrent nova and wind/ISM and wind/wind interactions \citep{Hachisu1999,2004ApJ...607..391G,Mazzali2005,Patat2007,Goobar2008,2023MNRAS.525.1867N,Chiotellis2012,Silverman2013_11kx,PerezTorres2014, TsebrenkoSoker2015,2016ApJ...818...26D,Darnley2019,Graham2019,Kool2023,Hosseinzadeh2022,Siebert2023}, this finding highlights a genuine limitation of one snapshot and opens up Pandora's box.\\

\noindent
{\sl Summary:}
The following picture emerges within the PDD framework. Ca originates from incomplete explosive oxygen burning.
Due to the smooth outer density structure and high expansion velocities, the newly synthesized Ca contributes to the broad features. Due to the low densities exposed at day -9, Ca{\sc\,iii} dominates. Recombination to Ca{\sc\,ii} requires a thin shell with a density enhancement of an order of magnitude. The shell is asymmetric related to a physical multi-dimensional mechanism.
It may require an interaction with a disk, Roche lobe, or companion star, as suggested above, or hydrodynamical instabilities in the outer layers produced, e.g., by WD rotation or higher pulsation modes. The result shows a pathway, but higher-cadence observations are needed.

\section{Alternative Scenarios for SN~2019\lowercase{vrq}}~\label{sec:alternative}
In the following, we compare the basic properties from our analysis of SN\,2019vrq with other scenarios. For the translation between the time evolution of the spectra and layers in the explosion models, we make use of Fig.~\ref{fig:mass} in lieu of detailed simulations. This approach can be justified because, to first order, the density structures can be approximated as polytropes and, for every scenario for 91T/99aa-like SNe\,Ia, the total mass is sufficiently large to contain $\approx 0.86 M_\odot$ of $^{56}$Ni (Sects.~\ref{sec:model} and~\ref{sec:lc_model}).
\begin{itemize}
    \item 
    One alternative explosion framework involves an initial detonation of a thin helium layer on the surface of a sub-$M_{\mathrm{Ch}}$ C/O WD, which sends a shock wave inward and generates compressional heating that triggers a secondary detonation in the CO-core~\citep{1980tsup.work...96W, 1982ApJ...253..798N, 1982ApJ...257..780N, 1990ApJ...354L..53L, 1994ApJ...423..371W, 1996ApJ...457..500H, 2010A&A...514A..53F, 2010ApJ...715..767S, 2015ApJ...805L...6S}. 
Advanced simulations have allowed the mass of the He surface layer to be reduced, or its presence to be neglected entirely~\citep{2018MNRAS.474.3931B, 2019ApJ...873...84P, 2018ApJ...854...52S, 2021ApJ...909L..18S}. 
While providing reasonable fits to the LCs and spectra of normal and subluminous SNe\,Ia,
the helium shell-triggered sub-M$_{\rm Ch}$ double-detonation framework encounters difficulties in accounting for overluminous events (e.g., \citealp{2018ApJ...854...52S, 2021ApJ...906...65P}). 
Despite possibly being consistent with the abundance asymmetries~ in the surface layers \citep{2016MNRAS.462.1039B}, they are inconsistent with the unburned layer of $\gtrsim 0.1\,M_\odot$ required for SN\,2019vrq (Sect. \ref{sec:spec_model}). 
Note that, after reducing the mass of the He-shell, this scenario produces HV Ca in the outermost layers as a natural product of the low-mass He-shell ash~\citep{2010A&A...514A..53F, 2010ApJ...719.1067K, 2013ApJ...776...97M, 2019ApJ...873...84P, 2019ApJ...878L..38T}, as invoked for SN 2018byg and SN 2016hnk~\citep{2019ApJ...873L..18D, 2020ApJ...896..165J}, and thus, unlike virtually all other scenarios, does not require invoking CSM to produce a HV Ca{\sc\,ii} feature. 
However, the supersolar iron abundance and the lack of a significant amount of 'explosive burning` products in the early spectra of SN\,2019vrq do not support this scenario. Moreover, the Ca{\sc\,ii} feature at day\,$+$12 indicates extended instabilities in the deeper ejecta, which is difficult to reconcile with a scenario in which the burning proceeds in a pure detonation mode. \\
\item Dynamical and violent mergers are unlikely because they are inconsistent with the low overall continuum polarization. The lower limit on the axis ratio of 0.8 (Class III, Sect.\ref{sec:limits_asymmetry}) is incompatible with the remarkably aspherical configuration predicted by the dynamical merger of binary WDs~\citep{1984ApJS...54..335I, 1984ApJ...277..355W, 1990ApJ...348..647B, 2010Natur.463...61P, 2016MNRAS.455.1060B}, unless seen close to `pole-on'.
Moreover, one may expect $\gtrsim $ 0.2 $M_\odot$ of unburned C/O (see the references above). 
\item As discussed in Sect.~\ref{sec:model_explosion}, classical  delayed-detonation models  may provide a reasonable first-order agreement with typical 91T/99aa-like SNe but, for SN\,2019vrq, show too little unburned material ($\lesssim 10^{-2}$\,M$_{\odot}$)~\citep{2017ApJ...846...58H, 2026arXiv260521575P}. 
\item Large-amplitude pulsating delayed-detonation models with high-density WD progenitors ~\citep{1993A&A...270..223K,1996ApJ...457..500H} produce too much unburned material, in contrast to low-amplitude pulsating delayed detonation models that are consistent with the observations. However, the assumption of radial pulsations without rotation makes them parameterized, not first-principles models. 
Note that previous studies of the early deflagration phase and light curve shapes \citep{2024ApJ...969...80C} suggest low-density WDs with a production of EC elements reduced to that of `normal' SNe\,Ia ($\approx 0.06-0.12$\,M$_{\odot}$)~\citep{2023ApJ...945L...2D, 2024ApJ...961..187D, 2024ApJ...975..203A}. 
\end{itemize}

Note that all alternative scenarios have serious shortcomings compared to the scenario presented in this work.

\section{Discussion and Summary}~\label{summary}
 High-quality flux and polarization spectra of the 91T/99aa-like SN\,2019vrq  have been obtained by the VLT from days $-9$ to $+12$ (Paper I), complemented by LCs and spectra from the literature. Based on detailed radiation-transport simulations with the 3D code HYDRA, the data were analyzed within the framework of low-amplitude PDD with an off-center DDT, including the line identification (Tab. \ref{tab:lines}). Combined with the spectra and LC similarity within this group, our results suggest an explosion scenario different from bright, 'normal' SNe~Ia\footnote{Though PDDs may be a subgroup among all SNe~Ia, in particular the underluminous SNe~Ia \citep{1995ApJ...444..831H}.}. Possible answers to the questions (Q1...6) as posed in Sect. \ref{sec:intro} have been presented in the framework of PDDs, but our results also suggest pathways for future directions.

\begin{itemize}


\item
Our conclusions are driven by quantitative flux spectroscopy. Identification of elements responsible for individual weak spectral features required a combination of detailed simulations and high S/N spectra. The polarization signal is marginal and used to provide limits on the asphericity. 

\item
The observed LCs and the evolution of the flux spectra are well reproduced by a low-amplitude PDD model with WD mass near $M_{\rm Ch}$ that leaves an outer, unburned layer of $\approx$ 0.1 $M_{\odot}$ and produces 0.86 $M_{\odot}$ of $^{56}$Ni. 
The mass of the unburned layers is significantly larger than expected within the `classical' delayed-detonation by a factor of $\geq 10$ and smaller than expected by $\geq 2$ for merger scenarios. 
The model WD derives from a 7\,M$_{\odot}$ progenitor of solar metallicity, has a central density of $\rho_{c} = 10^{9}$\,g\,cm$^{-3}$ with a WD mass of 1.356\,M$_{\odot}$ at the time of the explosion, has a slow deflagration phase with Euler number of 0.14, deflagration-burned mass of $M \approx 0.2$\,M$_{\odot}$, and an off-center DDT at M(DDT)=0.8\,M$_{\odot}$. Twice solar metallicity has been adopted for the outermost layers (Section~\ref{sec:model_construct}), which is possibly due to a failed runaway many years prior to the explosion. Note a possible link of PDD progenitors to some SNe~Iax from $M_{\rm Ch}$ which leave a massive remnant behind \citep{2012ApJ...761L..23J,2014MNRAS.438.1762F}.

\item
We suggest that 91T/99aa-like objects form a class distinct from the majority of SNe\,Ia. A pulsation requires a low C/O ratio during the deflagration phase, which is characteristic of progenitors near the upper limit in M$_{\rm MS}$ (Sect.~\ref{sec:model}). Because of their short evolutionary time relative to the WD stage, PDDs are strongly tied to recent star formation, consistent with observations. 

\item
A slow deflagration phase is required to produce the low-amplitude pulsation. `Classical' DD and PDD models are related and form a quasi-continuum between a WD that remains bound and one that is unbound at the end of the deflagration phase~\citep{2017ApJ...846...58H}. 
They are, however, physically distinct in the mechanism that drives the DDT: RT-driven turbulence from nuclear burning in the classical case, versus mixing from the interaction between the expanding inner layers and the fallback of the outer layers in the pulsational case. This distinction makes 91T-like SNe a distinct class of explosions (Sect.~\ref{sec:model_construct}). 

\item
The low-amplitude pulsation and low deflagration speed limit the WD's pre-expansion, leading to high ejecta density at the time of detonation. As a result, 0.94\,M$_{\odot}$ undergo burning to NSE, with 0.16\,M$_{\odot}$ of explosive oxygen products, and 0.11\,M$_{\odot}$ remaining unburned (Sect.~\ref{sec:model_explosion}). 
The Si layers are not split into high- and low-velocity components, consistent with the findings of \citet{2024ApJS..273...16P}. Specifically, the high-velocity and photospheric Si{\sc\,ii} are a result of the ionization structure, and the high-velocity component is a measure of the extent of the Si region (Sect.~\ref{sec:spec_model}).

\item 
The weak \textcolor{black}{Ca{\sc\,ii}\,H\&K} and Ca{\sc\,ii}\,NIR3 in the early-time spectrum can be understood as an ionization effect (Sect. \ref{sec:spec_model}). Because of the similarity 
of SN\,2019vrq to other 91T/99aa-like SNe, it can be understood why early HV   and photospheric features are commonly weak \citep{2024ApJS..273...16P}.

\item
 The HV Ca{\sc\,ii} indicates a density shell, but the linear line wing of the photospheric Ca{\sc\,ii} strongly suggests that  Ca is produced in explosive incomplete oxygen burning, i.e., layers between Si/S and NSE.
 We find that the maximum asphericity allowed in models for SN\,2019vrq ejecta is insufficient to produce the line polarization, and we require an aspherical density shell. Possible origins include an aspherical CSM that punches through the envelope or an internal origin, such as a rotating WD (Sect. \ref{sec:HVCa}).

\item
Pulsation is needed to produce a substantial outer layer that shows no evidence for significant nuclear burning. There are two pieces of evidence for this. The first is that the abundance of Fe in the outer layers is consistent with unburned material of twice- solar abundance (Sect.~\ref{sec:spec_model}). The second is the slow rise and decline of the monochromatic and bolometric LCs. An unburned shell reduces the explosion energy, gains energy on impact, but further reduces the expansion rate of the inner layer responsible for the diffusion timescales (Sect.~\ref{sec:lc_model}). 

\item
 A recent analysis of classical  DD models with DDT triggered by RT-driven small-scale turbulence predicts $\approx$0.2--0.3\,M$_{\odot}$ of $^{58}$Ni~\citep{2026arXiv260521575P}. Our PDD model produces $\approx1/3$ to $1/2$ of the stable $^{58}$Ni found in models of normal-bright SNe\,Ia (Sect.~\ref{sec:model_abundance}). The low $^{58}$Ni mass in low-amplitude PDDs  
is supported by previous analyses of the LCs, which include 91T/99aa-like SNe~\citep{2024ApJ...969...80C}. The $^{58}$Ni mass thus provides a crucial test for deciphering the origin of 91T/99aa-like SNe, a question that has implications for 91T-based cosmology and can be addressed with JWST MIR observations because seeing all ionization states is required to measure the $^{58}$Ni mass.

\item
For SN\,2019vrq, the degrees of both the continuum and the line polarization are close to the observational limit due to the S/N ratio of our data. Even before maximum light, the marginal individual polarization features are blends rather than single features, with the blends often morphing with time between different elements and ionization stages.

\item
Within the PDD model, the low polarization of the absorption feature of Si{\sc\,ii}\,$\lambda$6355 which is blended with Fe{\sc\,ii} lines is a natural result of the significantly lower Si{\sc\,ii} optical depth in the 91T/99aa-like events compared to normal SNe\,Ia. 
In addition, by day $+$2, the photosphere forms in an IGE-rich layer due to the much larger amount of IGE than in normal events. 
The peak polarization of the broad, blended features, $\approx0.1\%$, is consistent with an off-center DDT at 0.8\,M$_{\odot}$ viewed at an inclination of $\approx 30^{\circ}$ inferred from the collective patterns (Sect.~\ref{sec:classes}).

\item
Prior to maximum light, the continuum polarization is consistent with a possible rise to the red of $\approx$ 6800 ${\rm \AA}$ (Paper I). Note that a non-rotating, off-center pulsating DD model hardly shows the rise (Fig. \ref{fig:pol_spec}). We use a rotationally symmetric (ellipsoidal) density distribution as a proxy for a rotating WD. For demonstration, we use snapshots with the occupation numbers fixed at each epoch and an axis ratio of $R = B/A = 0.8$ as a lower limit~\citep{2001ApJ...556..302H, 2026ApJ...996L..10C}. 
This axis ratio is larger than expected for the dynamical merging of two WDs. The symmetry axis of this rotational ellipsoid may differ from the axis of the off-center DDT formed during the pulsation, introducing one more free parameter than the DDT axis alone and a scatter of up to $\approx$0.1\% in the $Q-U$ plane (Sect.~\ref{sec:model}). 

\item
The polarization is consistent with the typical, broad pattern in wavelength imprinted by atomic lines and multiplets of IGE, very similar in size and time evolution as found in normal and underluminous events, suggesting common origins: large-scale chemical distributions (Class~I), a large-scale axisymmetry indicated by a decreasing continuum component (Class~III), and small-scale inhomogeneities (Class~II). 

\item
The evolution of the individual polarization features is consistent with an off-center chemical distribution. However, in SN\,2019vrq, the low silicon mass reduces the line polarization by an order of magnitude relative to typical SNe\,Ia, a direct consequence of the low IME mass. 

\item
The inferred limit on the asphericity corresponds to a total, directional luminosity variation of $\Delta L/\bar{L} \approx 0.15$ (Sect.\ref{sec:aniso}) by the outer layers. The similarity in LCs and spectra suggests that it applies to the entire  sample of 91T/99aa-like events. This implies a magnitude dispersion small enough to make these events potentially good standard candles.

\item
 Based on our model, the possible variance in brightness of 91T/99aa is currently dominated by the inner layers through the production of electron-capture elements. The difference in $^{58}$Ni between PDD and classical  DD models for SN\,2019vrq is $\approx 0.2 M_\odot$, and the total dispersion in L may reach $\approx \pm 0.35$\,mag. Though LC shapes favor low central density $\rho_c$ in the initial WDs (Sect.\ref{sec:model_construct}), measuring the $^{58}$Ni mass in 91T/99aa with JWST is essential to quantify the total amount of $^{58}$Ni, and to potentially make 91T/99aa-like SNe true standard candles for cosmology. 
 Moreover, within the PDD scenario, 99aa-like SNe may form a continuum approaching 91T-like SNe. The distribution may be ruled by the EC production.

\item
\textcolor{black}{We note that the class of 91T/99aa-like SNe can be identified by their LC characteristics, namely blue color, slow rise and decline. Another clue may be the potentially increasing fraction among SNe\,Ia because they are correlated with periods of starbursts \citep{2022ApJ...938...47P}, and they originate from WDs at the upper end of the main-sequence mass range \citep{2024ApJ...969...80C}, both of which indicate a short time between star formation and explosion.} 
\end{itemize}

Finally, we note some limitations of this analysis regarding 91T/99aa-like SNe. For the initial WD we assumed $\rho_c = 10^9 {\rm g/cm^3}$ for being the most likely value though, based on LC shapes $\rho_c$ may be as high as $\rho_c = 3~10^9 {\rm g/cm^3}$ for SN1999aa \citep{2024ApJ...969...80C}, opening up the intriguing possibility that $\rho_c$ dominates the difference for 99aa and 91T-like SNe with implications for high-precision cosmology. Testing, though, requires JWST. All the common geometric components (classes~I-III) appear to be present, but the limited temporal coverage of the observations prevents their separate identification and linking of observations to the progenitor system. One limitation is the mixing during the deflagration and pulsational phases. The former depends sensitively on the initial conditions of the WD, such as pre-existing turbulence and magnetic fields \citep{2004PhRvL..92u1102G,2026ApJ..1003L..37S,2026arXiv260813432S}, and possibly on WD rotation. 
The outer layers of PDDs hardly affect the luminosity relevant for cosmology with 91T/99aa-like events. However, the amount of electron-capture elements has been identified in our previous works as the key to potential variations in brightness; late-time MIR observations with JWST are required to constrain the abundances of electron-capture elements. Full 3D MHD simulations are underway to study the physics of the deflagration with and without pulsation. 
Radial pulsation can be expected as the main pulsational  mode, though higher-order modes may contribute in rotating WDs or in the presence of high magnetic fields. 
Owing to the limited cadence and S/N, SN\,2019vrq observations severely limit the polarization-based analysis of the asphericity and, in particular, the interpretation and nature of the feature attributed to the weak HV Ca{\sc\,ii} because one epoch does not allow to derive the distribution in the relevant layers \citep{2023MNRAS.520..560H}.
Owing to the large parameter space of initial conditions, first-principles simulations remain unrealistic, and progress will continue to be guided by future observations.

\section*{Acknowledgments}
All observations in the paper are present in Paper I, and/or have been published previously.
P.H. acknowledges the support of the National Science Foundation (NSF) awards AST-1715133 and  AST-2306395 for supporting students (E.F, T.M.), the postdoc (S.S.), and the development of the methods and code HYDRA used for the simulations. J.C.W is supported by NSF grant AST1813825. 
The work of A.C. is supported by NOIRLab, which is managed by the Association of Universities for Research in Astronomy (AURA) under a cooperative agreement with the NSF. Y.Y.'s research is partially supported by the Tsinghua University Dushi Program. 
The research of Y.Y. has been supported through a Benoziyo Prize Postdoctoral Fellowship and the Bengier-Winslow-Robertson Fellowship. 
L.G. acknowledges financial support from MCIN and AEI 10.13039/501100011033 under projects PID2023-151307NB-I00, CEX2020-001058-M, and by the MaX-CSIC Excellence Award MaX4-SOMMA-ICE.
All VLT spectropolarimetry data are based on observations collected at the European Organization for Astronomical Research in the Southern Hemisphere under ESO program 0104.D-0109 (PI Y.\,Yang) and related programs (PI A.\,Cikota) as part of the SPECPOL collaboration and can be accessed via: \url{https://archive.eso.org/cms.html}. The simulation data presented are available as complementary material on the publisher's website. This research has made heavy use of NASA's Astrophysics Data System Bibliographic Services, the SIMBAD database, operated by CDS in Strasbourg, France, and the NASA/IPAC Extragalactic Database (NED), operated by the Jet Propulsion Laboratory, California Institute of Technology, under contract with NASA. The paper has been produced on Overleaf with the tools Grammarly to check for typos and grammar, and Claude as a research tool to find additional observations and to format Table 2.  The simulations have been performed on the Astrophysics Group's computer cluster at Florida State University.

\bibliographystyle{aasjournal}

\begin{thebibliography}{}
\expandafter\ifx\csname natexlab\endcsname\relax\def\natexlab#1{#1}\fi
\providecommand{\url}[1]{\href{#1}{#1}}
\providecommand{\dodoi}[1]{doi:~\href{http://doi.org/#1}{\nolinkurl{#1}}}
\providecommand{\doeprint}[1]{\href{http://ascl.net/#1}{\nolinkurl{http://ascl.net/#1}}}
\providecommand{\doarXiv}[1]{\href{https://arxiv.org/abs/#1}{\nolinkurl{https://arxiv.org/abs/#1}}}

\bibitem[{{Aldoroty} {et~al.}(2023){Aldoroty}, {Wang}, {Hoeflich}, {Yang}, {Suntzeff}, {Aldering}, {Antilogus}, {Aragon}, {Bailey}, {Baltay}, {Bongard}, {Boone}, {Buton}, {Copin}, {Dixon}, {Fouchez}, {Gangler}, {Gupta}, {Hayden}, {Karmen}, {Kim}, {Kowalski}, {K{\"u}sters}, {L{\'e}get}, {Mondon}, {Nordin}, {Pain}, {Pecontal}, {Pereira}, {Perlmutter}, {Ponder}, {Rabinowitz}, {Rigault}, {Rubin}, {Runge}, {Saunders}, {Smadja}, {Suzuki}, {Tao}, {Thomas}, \& {Vincenzi}}]{2023ApJ...948...10A}
{Aldoroty}, L., {Wang}, L., {Hoeflich}, P., {et~al.} 2023, \apj, 948, 10, \dodoi{10.3847/1538-4357/acad78}

\bibitem[{{Arnett} \& {Livne}(1994)}]{1994ApJ...427..330A}
{Arnett}, D., \& {Livne}, E. 1994, \apj, 427, 330, \dodoi{10.1086/174143}

\bibitem[{{Arnett}(1982)}]{1982ApJ...253..785A}
{Arnett}, W.~D. 1982, \apj, 253, 785, \dodoi{10.1086/159681}

\bibitem[{{Ashall} {et~al.}(2024){Ashall}, {Hoeflich}, {Baron}, {Shahbandeh}, {DerKacy}, {Medler}, {Shappee}, {Tucker}, {Fereidouni}, {Mera}, {Andrews}, {Baade}, {Bostroem}, {Brown}, {Burns}, {Burrow}, {Cikota}, {de Jaeger}, {Do}, {Dong}, {Dominguez}, {Fox}, {Galbany}, {Hsiao}, {Krisciunas}, {Khaghani}, {Kumar}, {Lu}, {Maund}, {Mazzali}, {Morrell}, {Patat}, {Pfeffer}, {Phillips}, {Schmidt}, {Stangl}, {Stevens}, {Stritzinger}, {Suntzeff}, {Telesco}, {Wang}, \& {Yang}}]{2024ApJ...975..203A}
{Ashall}, C., {Hoeflich}, P., {Baron}, E., {et~al.} 2024, \apj, 975, 203, \dodoi{10.3847/1538-4357/ad6608}

\bibitem[{{Benz} {et~al.}(1990){Benz}, {Bowers}, {Cameron}, \& {Press}}]{1990ApJ...348..647B}
{Benz}, W., {Bowers}, R.~L., {Cameron}, A.~G.~W., \& {Press}, W.~H.~. 1990, \apj, 348, 647, \dodoi{10.1086/168273}

\bibitem[{{Blondin} {et~al.}(2018){Blondin}, {Dessart}, \& {Hillier}}]{2018MNRAS.474.3931B}
{Blondin}, S., {Dessart}, L., \& {Hillier}, D.~J. 2018, \mnras, 474, 3931, \dodoi{10.1093/mnras/stx3058}

\bibitem[{{Blondin} {et~al.}(2012){Blondin}, {Matheson}, {Kirshner}, {Mandel}, {Berlind}, {Calkins}, {Challis}, {Garnavich}, {Jha}, {Modjaz}, {Riess}, \& {Schmidt}}]{2012AJ....143..126B}
{Blondin}, S., {Matheson}, T., {Kirshner}, R.~P., {et~al.} 2012, \aj, 143, 126, \dodoi{10.1088/0004-6256/143/5/126}

\bibitem[{{Blondin} {et~al.}(2022){Blondin}, {Blinnikov}, {Callan}, {Collins}, {Dessart}, {Even}, {Fl{\"o}rs}, {Fullard}, {Hillier}, {Jerkstrand}, {Kasen}, {Katz}, {Kerzendorf}, {Kozyreva}, {O'Brien}, {P{\'a}ssaro}, {Roth}, {Shen}, {Shingles}, {Sim}, {Singhal}, {Smith}, {Sorokina}, {Utrobin}, {Vogl}, {Williamson}, {Wollaeger}, {Woosley}, \& {Wygoda}}]{2022A&A...668A.163B}
{Blondin}, S., {Blinnikov}, S., {Callan}, F.~P., {et~al.} 2022, \aap, 668, A163, \dodoi{10.1051/0004-6361/202244134}

\bibitem[{{Branch} \& {Wheeler}(2017)}]{2017suex.book.....B}
{Branch}, D., \& {Wheeler}, J.~C. 2017, {Supernova Explosions} (Springer-Verlag Berlin and Heidelberg GmbH \& Co. KG, Germany), \dodoi{10.1007/978-3-662-55054-0}

\bibitem[{{Bravo} \& {Garc{\'\i}a-Senz}(2009)}]{2009ApJ...695.1244B}
{Bravo}, E., \& {Garc{\'\i}a-Senz}, D. 2009, \apj, 695, 1244, \dodoi{10.1088/0004-637X/695/2/1244}

\bibitem[{{Bulla} {et~al.}(2016{\natexlab{a}}){Bulla}, {Sim}, {Pakmor}, {Kromer}, {Taubenberger}, {R{\"o}pke}, {Hillebrandt}, \& {Seitenzahl}}]{2016MNRAS.455.1060B}
{Bulla}, M., {Sim}, S.~A., {Pakmor}, R., {et~al.} 2016{\natexlab{a}}, \mnras, 455, 1060, \dodoi{10.1093/mnras/stv2402}

\bibitem[{{Bulla} {et~al.}(2016{\natexlab{b}}){Bulla}, {Sim}, {Kromer}, {Seitenzahl}, {Fink}, {Ciaraldi-Schoolmann}, {R{\"o}pke}, {Hillebrandt}, {Pakmor}, {Ruiter}, \& {Taubenberger}}]{2016MNRAS.462.1039B}
{Bulla}, M., {Sim}, S.~A., {Kromer}, M., {et~al.} 2016{\natexlab{b}}, \mnras, 462, 1039, \dodoi{10.1093/mnras/stw1733}

\bibitem[{{Burns} {et~al.}(2014){Burns}, {Stritzinger}, {Phillips}, {Hsiao}, {Contreras}, {Persson}, {Folatelli}, {Boldt}, {Campillay}, {Castell{\'o}n}, {Freedman}, {Madore}, {Morrell}, {Salgado}, \& {Suntzeff}}]{2014ApJ...789...32B}
{Burns}, C.~R., {Stritzinger}, M., {Phillips}, M.~M., {et~al.} 2014, \apj, 789, 32, \dodoi{10.1088/0004-637X/789/1/32}

\bibitem[{{Chakraborty} {et~al.}(2024){Chakraborty}, {Sadler}, {Hoeflich}, {Hsiao}, {Phillips}, {Burns}, {Diamond}, {Dominguez}, {Galbany}, {Uddin}, {Ashall}, {Krisciunas}, {Kumar}, {Mera}, {Morrell}, {Baron}, {Contreras}, {Stritzinger}, \& {Suntzeff}}]{2024ApJ...969...80C}
{Chakraborty}, S., {Sadler}, B., {Hoeflich}, P., {et~al.} 2024, \apj, 969, 80, \dodoi{10.3847/1538-4357/ad4702}

\bibitem[{{Chandrasekhar}(1931)}]{1931ApJ....74...81C}
{Chandrasekhar}, S. 1931, Astrophysical Journal, 74, 81, \dodoi{10.1086/143324}

\bibitem[{{Chandrasekhar}(1961)}]{Chandrasekhar1961}
---. 1961, Hydrodynamic and Hydromagnetic Stability (Oxford Univ. Press)

\bibitem[{{Chiotellis} {et~al.}(2012){Chiotellis}, {Schure}, \& {Vink}}]{Chiotellis2012}
{Chiotellis}, A., {Schure}, K.~M., \& {Vink}, J. 2012, \aap, 537, A139

\bibitem[{{Cikota} {et~al.}(2026){Cikota}, {Hoeflich}, {Baade}, {Patat}, {Wang}, {Wheeler}, {Yang}, {Fereidouni}, \& {Mishra}}]{2026ApJ...996L..10C}
{Cikota}, A., {Hoeflich}, P., {Baade}, D., {et~al.} 2026, \apjl, 996, L10, \dodoi{10.3847/2041-8213/ae27c8}

\bibitem[{{Cyburt} {et~al.}(2010){Cyburt}, {Amthor}, {Ferguson}, {Meisel}, {Smith}, {Warren}, {Heger}, {Hoffman}, {Rauscher}, {Sakharuk}, {Schatz}, {Thielemann}, \& {Wiescher}}]{cy10}
{Cyburt}, R.~H., {Amthor}, A.~M., {Ferguson}, R., {et~al.} 2010, \apjs, 189, 240

\bibitem[{{Darnley} {et~al.}(2019)}]{Darnley2019}
{Darnley}, M.~J., {et~al.} 2019, \nat, 565, 460

\bibitem[{{De} {et~al.}(2019){De}, {Kasliwal}, {Polin}, {Nugent}, {Bildsten}, {Adams}, {Bellm}, {Blagorodnova}, {Cenko}, {Fremling}, {Graham}, {Ho}, {Kulkarni}, {Laher}, {Masci}, {Miller}, {Sollerman}, \& et~al.}]{2019ApJ...873L..18D}
{De}, K., {Kasliwal}, M.~M., {Polin}, A., {et~al.} 2019, \apjl, 873, L18, \dodoi{10.3847/2041-8213/ab0aec}

\bibitem[{{DerKacy} {et~al.}(2023){DerKacy}, {Ashall}, {Hoeflich}, {Baron}, {Shappee}, {Baade}, {Andrews}, {Bostroem}, {Brown}, {Burns}, {Burrow}, {Cikota}, {de Jaeger}, {Do}, {Dong}, {Dominguez}, {Galbany}, {Hsiao}, {Karamehmetoglu}, {Krisciunas}, {Kumar}, {Lu}, {Evans}, {Maund}, {Mazzali}, {Medler}, {Morrell}, {Patat}, {Phillips}, {Shahbandeh}, {Stangl}, {Stevens}, {Stritzinger}, {Suntzeff}, {Telesco}, {Tucker}, {Valenti}, {Wang}, {Yang}, {Jha}, \& {Kwok}}]{2023ApJ...945L...2D}
{DerKacy}, J.~M., {Ashall}, C., {Hoeflich}, P., {et~al.} 2023, \apjl, 945, L2, \dodoi{10.3847/2041-8213/acb8a8}

\bibitem[{{DerKacy} {et~al.}(2024){DerKacy}, {Ashall}, {Hoeflich}, {Baron}, {Shahbandeh}, {Shappee}, {Andrews}, {Baade}, {Balangan}, {Bostroem}, {Brown}, {Burns}, {Burrow}, {Cikota}, {de Jaeger}, {Do}, {Dong}, {Dominguez}, {Fox}, {Galbany}, {Hoang}, {Hsiao}, {Janzen}, {Jencson}, {Krisciunas}, {Kumar}, {Lu}, {Lundquist}, {Mera Evans}, {Maund}, {Mazzali}, {Medler}, {Meza Retamal}, {Morrell}, {Patat}, {Pearson}, {Phillips}, {Shrestha}, {Stangl}, {Stevens}, {Stritzinger}, {Suntzeff}, {Telesco}, {Tucker}, {Valenti}, {Wang}, \& {Yang}}]{2024ApJ...961..187D}
---. 2024, \apj, 961, 187, \dodoi{10.3847/1538-4357/ad0b7b}

\bibitem[{{DESI Collaboration} {et~al.}(2025{\natexlab{a}}){DESI Collaboration}, Karim, Aguilar, Ahlen, Alam, Allen, Prieto, Alves, Anand, Andrade, Armengaud, Aviles, Bailey, Brieden, Forero-Romero, Eisenstein, Levi, Dey, Schlegel, Weinberg, White, {et~al.}}]{AbdulKarim2025DESIDR2II}
{DESI Collaboration}, Karim, M.~A., Aguilar, J., {et~al.} 2025{\natexlab{a}}, Physical Review D, \dodoi{10.1103/tr6y-kpc6}

\bibitem[{{DESI Collaboration} {et~al.}(2025{\natexlab{b}}){DESI Collaboration}, Karim, Aguilar, Ahlen, Prieto, Alves, Anand, Andrade, Armengaud, Aviles, Bailey, Brieden, Forero-Romero, Font-Ribera, Eisenstein, Levi, Percival, Weinberg, White, {et~al.}}]{AbdulKarim2025DESIDR2LyA}
---. 2025{\natexlab{b}}, Physical Review D, \dodoi{10.1103/2wwn-xjm5}

\bibitem[{{Diamond} {et~al.}(2018){Diamond}, {Hoeflich}, {Hsiao}, {Sand}, {Sonneborn}, {Phillips}, {Hristov}, {Collins}, {Ashall}, {Marion}, {Stritzinger}, {Morrell}, {Gerardy}, \& {Penney}}]{2018ApJ...861..119D}
{Diamond}, T.~R., {Hoeflich}, P., {Hsiao}, E.~Y., {et~al.} 2018, \apj, 861, 119, \dodoi{10.3847/1538-4357/aac434}

\bibitem[{{Dom{\'\i}nguez} \& {H{\"o}flich}(2000)}]{2000ApJ...528..854D}
{Dom{\'\i}nguez}, I., \& {H{\"o}flich}, P. 2000, \apj, 528, 854, \dodoi{10.1086/308223}

\bibitem[{{Dom{\'\i}nguez} {et~al.}(2001){Dom{\'\i}nguez}, {H{\"o}flich}, \& {Straniero}}]{2001ApJ...557..279D}
{Dom{\'\i}nguez}, I., {H{\"o}flich}, P., \& {Straniero}, O. 2001, \apj, 557, 279, \dodoi{10.1086/321661}

\bibitem[{{Dragulin} \& {Hoeflich}(2016)}]{2016ApJ...818...26D}
{Dragulin}, P., \& {Hoeflich}, P. 2016, \apj, 818, 26, \dodoi{10.3847/0004-637X/818/1/26}

\bibitem[{{Eriguchi} \& {Mueller}(1985)}]{1985A&A...146..260E}
{Eriguchi}, Y., \& {Mueller}, E. 1985, \aap, 146, 260

\bibitem[{{Eriguchi} {et~al.}(1986){Eriguchi}, {Mueller}, \& {Hachisu}}]{1986A&A...168..130E}
{Eriguchi}, Y., {Mueller}, E., \& {Hachisu}, I. 1986, \aap, 168, 130

\bibitem[{{Fesen} {et~al.}(2015){Fesen}, {H{\"o}flich}, \& {Hamilton}}]{2015ApJ...804..140F}
{Fesen}, R.~A., {H{\"o}flich}, P.~A., \& {Hamilton}, A. J.~S. 2015, \apj, 804, 140, \dodoi{10.1088/0004-637X/804/2/140}

\bibitem[{{Filippenko} {et~al.}(1992){Filippenko}, {Richmond}, {Matheson}, {Shields}, {Burbidge}, {Cohen}, {Dickinson}, {Malkan}, {Nelson}, {Pietz}, {Schlegel}, {Schmeer}, {Spinrad}, {Steidel}, {Tran}, \& {Wren}}]{1992ApJ...384L..15F}
{Filippenko}, A.~V., {Richmond}, M.~W., {Matheson}, T., {et~al.} 1992, \apjl, 384, L15, \dodoi{10.1086/186252}

\bibitem[{{Fink} {et~al.}(2010){Fink}, {R{\"o}pke}, {Hillebrandt}, {Seitenzahl}, {Sim}, \& {Kromer}}]{2010A&A...514A..53F}
{Fink}, M., {R{\"o}pke}, F.~K., {Hillebrandt}, W., {et~al.} 2010, \aap, 514, A53, \dodoi{10.1051/0004-6361/200913892}

\bibitem[{{Fink} {et~al.}(2014){Fink}, {Kromer}, {Seitenzahl}, {Ciaraldi-Schoolmann}, {R{\"o}pke}, {Sim}, {Pakmor}, {Ruiter}, \& {Hillebrandt}}]{2014MNRAS.438.1762F}
{Fink}, M., {Kromer}, M., {Seitenzahl}, I.~R., {et~al.} 2014, \mnras, 438, 1762, \dodoi{10.1093/mnras/stt2315}

\bibitem[{{Gall} {et~al.}(2018){Gall}, {Stritzinger}, {Ashall}, {Baron}, {Burns}, {Hoeflich}, {Hsiao}, {Mazzali}, {Phillips}, {Filippenko}, {Anderson}, {Benetti}, {Brown}, {Campillay}, {Challis}, {Contreras}, {Elias de la Rosa}, {Folatelli}, {Foley}, {Fraser}, {Holmbo}, {Marion}, {Morrell}, {Pan}, {Pignata}, {Suntzeff}, {Taddia}, {Torres Robledo}, \& {Valenti}}]{2018A&A...611A..58G}
{Gall}, C., {Stritzinger}, M.~D., {Ashall}, C., {et~al.} 2018, \aap, 611, A58, \dodoi{10.1051/0004-6361/201730886}

\bibitem[{{Gamezo} {et~al.}(2004){Gamezo}, {Khokhlov}, \& {Oran}}]{2004PhRvL..92u1102G}
{Gamezo}, V.~N., {Khokhlov}, A.~M., \& {Oran}, E.~S. 2004, \prl, 92, 211102, \dodoi{10.1103/PhysRevLett.92.211102}

\bibitem[{{Gamezo} {et~al.}(2005){Gamezo}, {Khokhlov}, \& {Oran}}]{2005ApJ...623..337G}
---. 2005, \apj, 623, 337, \dodoi{10.1086/428767}

\bibitem[{{Gamezo} {et~al.}(2003){Gamezo}, {Khokhlov}, {Oran}, {Chtchelkanova}, \& {Rosenberg}}]{2003Sci...299...77G}
{Gamezo}, V.~N., {Khokhlov}, A.~M., {Oran}, E.~S., {Chtchelkanova}, A.~Y., \& {Rosenberg}, R.~O. 2003, Science, 299, 77, \dodoi{10.1126/science.299.5603.77}

\bibitem[{{Ganeshalingam} {et~al.}(2011){Ganeshalingam}, {Li}, \& {Filippenko}}]{2011MNRAS.416.2607G}
{Ganeshalingam}, M., {Li}, W., \& {Filippenko}, A.~V. 2011, \mnras, 416, 2607, \dodoi{10.1111/j.1365-2966.2011.19213.x}

\bibitem[{{Gerardy} {et~al.}(2004){Gerardy}, {H{\"o}flich}, {Fesen}, {Marion}, {Nomoto}, {Quimby}, {Schaefer}, {Wang}, \& {Wheeler}}]{2004ApJ...607..391G}
{Gerardy}, C.~L., {H{\"o}flich}, P., {Fesen}, R.~A., {et~al.} 2004, \apj, 607, 391, \dodoi{10.1086/383488}

\bibitem[{{Goobar}(2008)}]{Goobar2008}
{Goobar}, A. 2008, \apjl, 686, L103

\bibitem[{{Graboske} {et~al.}(1973){Graboske}, {Dewitt}, {Grossman}, \& {Cooper}}]{graboske73}
{Graboske}, H.~C., {Dewitt}, H.~E., {Grossman}, A.~S., \& {Cooper}, M.~S. 1973, \apj, 181, 457, \dodoi{10.1086/152062}

\bibitem[{{Graham} {et~al.}(2019)}]{Graham2019}
{Graham}, M.~L., {et~al.} 2019, \apj, 871, 62

\bibitem[{{Grevesse} {et~al.}(2007){Grevesse}, {Asplund}, \& {Sauval}}]{2007SSRv..130..105G}
{Grevesse}, N., {Asplund}, M., \& {Sauval}, A.~J. 2007, \ssr, 130, 105, \dodoi{10.1007/s11214-007-9173-7}

\bibitem[{{Hachisu} {et~al.}(1999){Hachisu}, {Kato}, \& {Nomoto}}]{Hachisu1999}
{Hachisu}, I., {Kato}, M., \& {Nomoto}, K. 1999, \apj, 522, 487

\bibitem[{{Hamuy} {et~al.}(2000){Hamuy}, {Trager}, {Pinto}, {Phillips}, {Schommer}, {Ivanov}, \& {Suntzeff}}]{2000AJ....120.1479H}
{Hamuy}, M., {Trager}, S.~C., {Pinto}, P.~A., {et~al.} 2000, \aj, 120, 1479, \dodoi{10.1086/301527}

\bibitem[{{Hillebrandt} {et~al.}(2013){Hillebrandt}, {Kromer}, {R{\"o}pke}, \& {Ruiter}}]{2013FrPhy...8..116H}
{Hillebrandt}, W., {Kromer}, M., {R{\"o}pke}, F.~K., \& {Ruiter}, A.~J. 2013, Frontiers of Physics, 8, 116, \dodoi{10.1007/s11467-013-0303-2}

\bibitem[{{Hillier} \& {Dessart}(2012)}]{2012MNRAS.424..252H}
{Hillier}, D.~J., \& {Dessart}, L. 2012, \mnras, 424, 252, \dodoi{10.1111/j.1365-2966.2012.21192.x}

\bibitem[{{Hoeflich}(2017)}]{2017hsn..book.1151H}
{Hoeflich}, P. 2017, in Handbook of Supernovae, ed. A.~W. {Alsabti} \& P.~{Murdin} (Springer Cham), 1151, \dodoi{10.1007/978-3-319-21846-5_56}

\bibitem[{{Hoeflich} {et~al.}(2024){Hoeflich}, {Fereidouni}, \& {Mera}}]{2024JPhCS2742a2024H}
{Hoeflich}, P., {Fereidouni}, E., \& {Mera}, T.~B. 2024, {Type Ia supernovae in the age of JWST: Finding the 'right' questions and the path to answers},  IOP, \dodoi{10.1088/1742-6596/2742/1/012024}

\bibitem[{{Hoeflich} \& {Khokhlov}(1996)}]{1996ApJ...457..500H}
{Hoeflich}, P., \& {Khokhlov}, A. 1996, \apj, 457, 500, \dodoi{10.1086/176748}

\bibitem[{{Hoeflich} {et~al.}(1992){Hoeflich}, {Khokhlov}, \& {Mueller}}]{1992A&A...259..549H}
{Hoeflich}, P., {Khokhlov}, A., \& {Mueller}, E. 1992, \aap, 259, 549

\bibitem[{{Hoeflich} {et~al.}(1994){Hoeflich}, {Khokhlov}, \& {Mueller}}]{1994ApJS...92..501H}
---. 1994, \apjs, 92, 501, \dodoi{10.1086/192004}

\bibitem[{{Hoeflich} {et~al.}(1995){Hoeflich}, {Khokhlov}, \& {Wheeler}}]{1995ApJ...444..831H}
{Hoeflich}, P., {Khokhlov}, A.~M., \& {Wheeler}, J.~C. 1995, \apj, 444, 831, \dodoi{10.1086/175656}

\bibitem[{{Hoeflich} {et~al.}(1993{\natexlab{a}}){Hoeflich}, {Mueller}, \& {Khokhlov}}]{1993A&AS...97..221H}
{Hoeflich}, P., {Mueller}, E., \& {Khokhlov}, A. 1993{\natexlab{a}}, \aaps, 97, 221

\bibitem[{{Hoeflich} {et~al.}(1993{\natexlab{b}}){Hoeflich}, {Mueller}, \& {Khokhlov}}]{1993A&A...268..570H}
---. 1993{\natexlab{b}}, \aap, 268, 570

\bibitem[{{Hoeflich} {et~al.}(2017){Hoeflich}, {Hsiao}, {Ashall}, {Burns}, {Diamond}, {Phillips}, {Sand}, {Stritzinger}, {Suntzeff}, {Contreras}, {Krisciunas}, {Morrell}, \& {Wang}}]{2017ApJ...846...58H}
{Hoeflich}, P., {Hsiao}, E.~Y., {Ashall}, C., {et~al.} 2017, \apj, 846, 58, \dodoi{10.3847/1538-4357/aa84b2}

\bibitem[{{Hoeflich} {et~al.}(2019){Hoeflich}, {Ashall}, {Fisher}, {Hristov}, {Collins}, {Hsiao}, {Wiedenhoever}, {Chakraborty}, \& {Diamond}}]{2019nuco.conf..187H}
{Hoeflich}, P., {Ashall}, C., {Fisher}, A., {et~al.} 2019, {Thermonuclear Supernovae: Prospecting in the Age of Time-Domain and Multi-Wavelength Astronomy}, \dodoi{10.1007/978-3-030-13876-9_31}

\bibitem[{{Hoeflich} {et~al.}(2021){Hoeflich}, {Ashall}, {Bose}, {Baron}, {Stritzinger}, {Davis}, {Shahbandeh}, {Anand}, {Baade}, {Burns}, {Collins}, {Diamond}, {Fisher}, {Galbany}, {Hristov}, {Hsiao}, {Phillips}, {Shappee}, {Suntzeff}, \& {Tucker}}]{2021ApJ...922..186H}
{Hoeflich}, P., {Ashall}, C., {Bose}, S., {et~al.} 2021, \apj, 922, 186, \dodoi{10.3847/1538-4357/ac250d}

\bibitem[{{Hoeflich} {et~al.}(2023){Hoeflich}, {Yang}, {Baade}, {Cikota}, {Maund}, {Mishra}, {Patat}, {Patra}, {Wang}, {Wheeler}, {Filippenko}, {Gal-Yam}, \& {Schulze}}]{2023MNRAS.520..560H}
{Hoeflich}, P., {Yang}, Y., {Baade}, D., {et~al.} 2023, \mnras, 520, 560, \dodoi{10.1093/mnras/stad172}

\bibitem[{{Hoeflich} {et~al.}(2025){Hoeflich}, {Fereidouni}, {Fisher}, {Mera}, {Ashall}, {Brown}, {Baron}, {DerKacy}, {Diamond}, {Shabandeh}, \& {Stritzinger}}]{2025arXiv250107654H}
{Hoeflich}, P., {Fereidouni}, E., {Fisher}, A., {et~al.} 2025, arXiv e-prints, arXiv:2501.07654, \dodoi{10.48550/arXiv.2501.07654}

\bibitem[{{Hoflich}(1991)}]{1991A&A...246..481H}
{Hoflich}, P. 1991, \aap, 246, 481

\bibitem[{{H{\"o}flich}(1995)}]{1995ApJ...440..821H}
{H{\"o}flich}, P. 1995, \apj, 440, 821, \dodoi{10.1086/175317}

\bibitem[{{H{\"o}flich}(2003)}]{2003ASPC..288..185H}
{H{\"o}flich}, P. 2003, in Astronomical Society of the Pacific Conference Series, Vol. 288, Stellar Atmosphere Modeling, ed. I.~{Hubeny}, D.~{Mihalas}, \& K.~{Werner}, 185

\bibitem[{{H{\"o}flich}(2006)}]{2006NuPhA.777..579H}
---. 2006, \nphysa, 777, 579, \dodoi{10.1016/j.nuclphysa.2004.12.038}

\bibitem[{{H{\"o}flich} {et~al.}(2003){H{\"o}flich}, {Gerardy}, {Linder}, \& {et al.}}]{2003LNP...635..203H}
{H{\"o}flich}, P., {Gerardy}, C., {Linder}, E., \& {et al.} 2003, {Models for Type Ia Supernovae and Cosmology}, \dodoi{10.1007/978-3-540-39882-0_11}

\bibitem[{{H{\"o}flich} {et~al.}(2006){H{\"o}flich}, {Gerardy}, {Marion}, \& {Quimby}}]{2006NewAR..50..470H}
{H{\"o}flich}, P., {Gerardy}, C.~L., {Marion}, H., \& {Quimby}, R. 2006, \nar, 50, 470, \dodoi{10.1016/j.newar.2006.06.074}

\bibitem[{{H{\"o}flich} {et~al.}(2001){H{\"o}flich}, {Khokhlov}, \& {Wang}}]{2001AIPC..586..459H}
{H{\"o}flich}, P., {Khokhlov}, A., \& {Wang}, L. 2001, in American Institute of Physics Conference Series, Vol. 586, 20th Texas Symposium on relativistic astrophysics, ed. J.~C. {Wheeler} \& H.~{Martel} (AIP), 459--471, \dodoi{10.1063/1.1419593}

\bibitem[{{H{\"o}flich} \& {Stein}(2002)}]{2002ApJ...568..779H}
{H{\"o}flich}, P., \& {Stein}, J. 2002, \apj, 568, 779, \dodoi{10.1086/338981}

\bibitem[{{H{\"o}flich} {et~al.}(1998){H{\"o}flich}, {Wheeler}, \& {Thielemann}}]{1998ApJ...495..617H}
{H{\"o}flich}, P., {Wheeler}, J.~C., \& {Thielemann}, F.~K. 1998, \apj, 495, 617, \dodoi{10.1086/305327}

\bibitem[{{Hosseinzadeh} {et~al.}(2022)}]{Hosseinzadeh2022}
{Hosseinzadeh}, G., {et~al.} 2022, \apjl, 933, L45

\bibitem[{{Howell}(2011)}]{2011NatCo...2..350H}
{Howell}, D.~A. 2011, Nature Communications, 2, 350, \dodoi{10.1038/ncomms1344}

\bibitem[{{Howell} {et~al.}(2001){Howell}, {H{\"o}flich}, {Wang}, \& {Wheeler}}]{2001ApJ...556..302H}
{Howell}, D.~A., {H{\"o}flich}, P., {Wang}, L., \& {Wheeler}, J.~C. 2001, \apj, 556, 302, \dodoi{10.1086/321584}

\bibitem[{{Hristov} {et~al.}(2021){Hristov}, {Hoeflich}, \& {Collins}}]{2021ApJ...923..210H}
{Hristov}, B., {Hoeflich}, P., \& {Collins}, D.~C. 2021, \apj, 923, 210, \dodoi{10.3847/1538-4357/ac0ef8}

\bibitem[{{Iben} \& {Tutukov}(1984)}]{1984ApJS...54..335I}
{Iben}, I., J., \& {Tutukov}, A.~V. 1984, \apjs, 54, 335, \dodoi{10.1086/190932}

\bibitem[{{Ihanec} {et~al.}(2019){Ihanec}, {Wevers}, {Callis}, {Gromadzki}, \& {Yaron}}]{2019TNSCR2483....1I}
{Ihanec}, N., {Wevers}, T., {Callis}, E., {Gromadzki}, M., \& {Yaron}, O. 2019, Transient Name Server Classification Report, 2019-2483, 1

\bibitem[{{Itoh} {et~al.}(1979){Itoh}, {Totsuji}, {Ichimaru}, \& {Dewitt}}]{itoh79}
{Itoh}, N., {Totsuji}, H., {Ichimaru}, S., \& {Dewitt}, H.~E. 1979, \apj, 234, 1079, \dodoi{10.1086/157590}

\bibitem[{{Jacobson-Gal{\'a}n} {et~al.}(2020){Jacobson-Gal{\'a}n}, {Polin}, {Foley}, {Dimitriadis}, {Kilpatrick}, {Swift}, \& et~al.}]{2020ApJ...896..165J}
{Jacobson-Gal{\'a}n}, W.~V., {Polin}, A., {Foley}, R.~J., {et~al.} 2020, \apj, 896, 165, \dodoi{10.3847/1538-4357/ab94b8}

\bibitem[{{Jeffery} {et~al.}(1992){Jeffery}, {Leibundgut}, {Kirshner}, {Benetti}, {Branch}, \& {Sonneborn}}]{1992ApJ...397..304J}
{Jeffery}, D.~J., {Leibundgut}, B., {Kirshner}, R.~P., {et~al.} 1992, \apj, 397, 304, \dodoi{10.1086/171787}

\bibitem[{{Jones} {et~al.}(2009){Jones}, {Read}, {Saunders}, {Colless}, {Jarrett}, {Parker}, {Fairall}, {Mauch}, {Sadler}, {Watson}, {Burton}, {Campbell}, {Cass}, {Croom}, {Dawe}, {Fiegert}, {Frankcombe}, {Hartley}, {Huchra}, {James}, {Kirby}, {Lahav}, {Lucey}, {Mamon}, {Moore}, {Peterson}, {Prior}, {Proust}, {Russell}, {Safouris}, {Wakamatsu}, {Westra}, \& {Williams}}]{2009MNRAS.399..683J}
{Jones}, D.~H., {Read}, M.~A., {Saunders}, W., {et~al.} 2009, \mnras, 399, 683, \dodoi{10.1111/j.1365-2966.2009.15338.x}

\bibitem[{{Jordan} {et~al.}(2012){Jordan}, {Perets}, {Fisher}, \& {van Rossum}}]{2012ApJ...761L..23J}
{Jordan}, IV, G.~C., {Perets}, H.~B., {Fisher}, R.~T., \& {van Rossum}, D.~R. 2012, \apjl, 761, L23, \dodoi{10.1088/2041-8205/761/2/L23}

\bibitem[{{Khokhlov}(1993)}]{1993ApJ...419L..77K}
{Khokhlov}, A. 1993, \apjl, 419, L77, \dodoi{10.1086/187141}

\bibitem[{{Khokhlov}(1998)}]{1998JCoPh.143..519K}
---. 1998, Journal of Computational Physics, 143, 519, \dodoi{10.1006/jcph.1998.9998}

\bibitem[{{Khokhlov} \& {H{\"o}flich}(2001)}]{2001AIPC..556..301K}
{Khokhlov}, A., \& {H{\"o}flich}, P. 2001, in American Institute of Physics Conference Series, Vol. 556, Explosive Phenomena in Astrophysical Compact Objects, ed. H.-Y. {Chang}, C.-H. {Lee}, M.~{Rho}, \& I.~{Yi} (AIP), 301--312, \dodoi{10.1063/1.1368287}

\bibitem[{{Khokhlov} {et~al.}(1992){Khokhlov}, {Mueller}, \& {Hoeflich}}]{1992A&A...253L...9K}
{Khokhlov}, A., {Mueller}, E., \& {Hoeflich}, P. 1992, \aap, 253, L9

\bibitem[{{Khokhlov} {et~al.}(1993){Khokhlov}, {Mueller}, \& {Hoeflich}}]{1993A&A...270..223K}
---. 1993, \aap, 270, 223

\bibitem[{{Khokhlov} {et~al.}(2025){Khokhlov}, {Dom{\'\i}nguez}, {Chtchelkanova}, {Hoeflich}, {Baron}, {Krisciunas}, {Phillips}, {Suntzeff}, \& {Wang}}]{2025ApJ...982..204K}
{Khokhlov}, A., {Dom{\'\i}nguez}, I., {Chtchelkanova}, A.~Y., {et~al.} 2025, \apj, 982, 204, \dodoi{10.3847/1538-4357/adb0c1}

\bibitem[{{Khokhlov}(1989)}]{khok89}
{Khokhlov}, A.~M. 1989, \mnras, 239, 785

\bibitem[{{Khokhlov}(1991)}]{1991A&A...245..114K}
---. 1991, \aap, 245, 114

\bibitem[{{Khokhlov}(1995)}]{1995ApJ...449..695K}
---. 1995, \apj, 449, 695, \dodoi{10.1086/176091}

\bibitem[{{Khokhlov}(2000)}]{2000astro.ph..8463K}
---. 2000, arXiv e-prints, astro, \dodoi{10.48550/arXiv.astro-ph/0008463}

\bibitem[{{Kool} {et~al.}(2023)}]{Kool2023}
{Kool}, E.~C., {et~al.} 2023, \nat, 617, 477

\bibitem[{{Kromer} {et~al.}(2010){Kromer}, {Sim}, {Fink}, {R{\"o}pke}, {Seitenzahl}, \& {Hillebrandt}}]{2010ApJ...719.1067K}
{Kromer}, M., {Sim}, S.~A., {Fink}, M., {et~al.} 2010, \apj, 719, 1067, \dodoi{10.1088/0004-637X/719/2/1067}

\bibitem[{{Li} {et~al.}(2001){Li}, {Filippenko}, {Treffers}, {Riess}, {Hu}, \& {Qiu}}]{2001ApJ...546..734L}
{Li}, W., {Filippenko}, A.~V., {Treffers}, R.~R., {et~al.} 2001, \apj, 546, 734, \dodoi{10.1086/318299}

\bibitem[{{Liu} {et~al.}(2023){Liu}, {R{\"o}pke}, \& {Han}}]{2023RAA....23h2001L}
{Liu}, Z.-W., {R{\"o}pke}, F.~K., \& {Han}, Z. 2023, Research in Astronomy and Astrophysics, 23, 082001, \dodoi{10.1088/1674-4527/acd89e}

\bibitem[{{Livne}(1990)}]{1990ApJ...354L..53L}
{Livne}, E. 1990, \apjl, 354, L53, \dodoi{10.1086/185721}

\bibitem[{{Livne}(1999)}]{1999ApJ...527L..97L}
---. 1999, \apjl, 527, L97, \dodoi{10.1086/312405}

\bibitem[{{Livne} {et~al.}(2005){Livne}, {Asida}, \& {H{\"o}flich}}]{2005ApJ...632..443L}
{Livne}, E., {Asida}, S.~M., \& {H{\"o}flich}, P. 2005, \apj, 632, 443, \dodoi{10.1086/432975}

\bibitem[{{Lu} {et~al.}(2021){Lu}, {Ashall}, {Hsiao}, {Hoeflich}, {Galbany}, {Baron}, {Phillips}, {Contreras}, {Burns}, {Suntzeff}, {Stritzinger}, {Anais}, {Anderson}, {Brown}, {Busta}, {Castell{\'o}n}, {Davis}, {Diamond}, {Falco}, {Gonzalez}, {Hamuy}, {Holmbo}, {Holoien}, {Krisciunas}, {Kirshner}, {Kumar}, {Kuncarayakti}, {Marion}, {Morrell}, {Persson}, {Piro}, {Prieto}, {Sand}, {Shahbandeh}, {Shappee}, \& {Taddia}}]{2021ApJ...920..107L}
{Lu}, J., {Ashall}, C., {Hsiao}, E.~Y., {et~al.} 2021, \apj, 920, 107, \dodoi{10.3847/1538-4357/ac1606}

\bibitem[{{Maoz} {et~al.}(2014){Maoz}, {Mannucci}, \& {Nelemans}}]{2014ARA&A..52..107M}
{Maoz}, D., {Mannucci}, F., \& {Nelemans}, G. 2014, \araa, 52, 107, \dodoi{10.1146/annurev-astro-082812-141031}

\bibitem[{{Marion} {et~al.}(2013){Marion}, {Vinko}, {Wheeler}, {Foley}, {Hsiao}, {Brown}, {Challis}, {Filippenko}, {Garnavich}, {Kirshner}, {Landsman}, {Parrent}, {Pritchard}, {Roming}, {Silverman}, \& {Wang}}]{2013ApJ...777...40M}
{Marion}, G.~H., {Vinko}, J., {Wheeler}, J.~C., {et~al.} 2013, \apj, 777, 40, \dodoi{10.1088/0004-637X/777/1/40}

\bibitem[{{Maund}(2024)}]{2024MNRAS.528.3875M}
{Maund}, J.~R. 2024, \mnras, 528, 3875, \dodoi{10.1093/mnras/stad2572}

\bibitem[{{Maund} {et~al.}(2010){Maund}, {H{\"o}flich}, {Patat}, {Wheeler}, {Zelaya}, {Baade}, {Wang}, {Clocchiatti}, \& {Quinn}}]{2010ApJ...725L.167M}
{Maund}, J.~R., {H{\"o}flich}, P., {Patat}, F., {et~al.} 2010, \apjl, 725, L167, \dodoi{10.1088/2041-8205/725/2/L167}

\bibitem[{{Maund} {et~al.}(2013){Maund}, {Spyromilio}, {Hoflich}, {Wheeler}, {Baade}, {Clocchiatti}, {Patat}, {Reilly}, {Wang}, \& {Zelaya}}]{2013MNRAS.433L..20M}
{Maund}, J.~R., {Spyromilio}, J., {Hoflich}, P.~A., {et~al.} 2013, \mnras, 433, L20, \dodoi{10.1093/mnrasl/slt050}

\bibitem[{{Mazzali} {et~al.}(1995){Mazzali}, {Danziger}, \& {Turatto}}]{1995A&A...297..509M}
{Mazzali}, P.~A., {Danziger}, I.~J., \& {Turatto}, M. 1995, \aap, 297, 509

\bibitem[{{Mazzali} {et~al.}(2005){Mazzali}, {Benetti}, {Altavilla}, {Blanc}, {Cappellaro}, {Elias-Rosa}, {Garavini}, {Goobar}, {Harutyunyan}, {Kotak}, {Leibundgut}, {Lundqvist}, {Mattila}, {Mendez}, {Nobili}, {Pain}, {Pastorello}, {Patat}, {Pignata}, {Podsiadlowski}, {Ruiz-Lapuente}, {Salvo}, {Schmidt}, {Sollerman}, {Stanishev}, {Stehle}, {Tout}, {Turatto}, \& {Hillebrandt}}]{Mazzali2005}
{Mazzali}, P.~A., {Benetti}, S., {Altavilla}, G., {et~al.} 2005, \apjl, 623, L37, \dodoi{10.1086/429874}

\bibitem[{{Moore} {et~al.}(2013){Moore}, {Townsley}, \& {Bildsten}}]{2013ApJ...776...97M}
{Moore}, K., {Townsley}, D.~M., \& {Bildsten}, L. 2013, \apj, 776, 97, \dodoi{10.1088/0004-637X/776/2/97}

\bibitem[{{Mueller} \& {Arnett}(1986)}]{1986ApJ...307..619M}
{Mueller}, E., \& {Arnett}, W.~D. 1986, \apj, 307, 619, \dodoi{10.1086/164448}

\bibitem[{{Niemeyer}(1994)}]{1994MsT..........1N}
{Niemeyer}, J.~C. 1994, Master's thesis, -

\bibitem[{{Niemeyer} \& {Hillebrandt}(1995)}]{1995ApJ...452..779N}
{Niemeyer}, J.~C., \& {Hillebrandt}, W. 1995, \apj, 452, 779, \dodoi{10.1086/176346}

\bibitem[{{Noirot} {et~al.}(2023){Noirot}, {Desprez}, {Asada}, {Sawicki}, {Estrada-Carpenter}, {Martis}, {Sarrouh}, {Strait}, {Abraham}, {Brada{\v{c}}}, {Brammer}, {Iyer}, {MacFarland}, {Matharu}, {Mowla}, {Muzzin}, {Pacifici}, {Ravindranath}, {Willott}, {Albert}, {Doyon}, {Hutchings}, \& {Rowlands}}]{2023MNRAS.525.1867N}
{Noirot}, G., {Desprez}, G., {Asada}, Y., {et~al.} 2023, \mnras, 525, 1867, \dodoi{10.1093/mnras/stad1019}

\bibitem[{{Nomoto}(1982{\natexlab{a}})}]{1982ApJ...253..798N}
{Nomoto}, K. 1982{\natexlab{a}}, \apj, 253, 798, \dodoi{10.1086/159682}

\bibitem[{{Nomoto}(1982{\natexlab{b}})}]{1982ApJ...257..780N}
---. 1982{\natexlab{b}}, \apj, 257, 780, \dodoi{10.1086/160031}

\bibitem[{{Nugent} {et~al.}(1995){Nugent}, {Baron}, {Hauschildt}, \& {Branch}}]{1995ApJ...441L..33N}
{Nugent}, P., {Baron}, E., {Hauschildt}, P.~H., \& {Branch}, D. 1995, \apjl, 441, L33, \dodoi{10.1086/187782}

\bibitem[{{Pakmor} {et~al.}(2010){Pakmor}, {Kromer}, {R{\"o}pke}, {Sim}, {Ruiter}, \& {Hillebrandt}}]{2010Natur.463...61P}
{Pakmor}, R., {Kromer}, M., {R{\"o}pke}, F.~K., {et~al.} 2010, \nat, 463, 61, \dodoi{10.1038/nature08642}

\bibitem[{{Pakmor} {et~al.}(2024){Pakmor}, {Seitenzahl}, {Ruiter}, {Sim}, {R{\"o}pke}, {Taubenberger}, {Bieri}, \& {Blondin}}]{2024A&A...686A.227P}
{Pakmor}, R., {Seitenzahl}, I.~R., {Ruiter}, A.~J., {et~al.} 2024, \aap, 686, A227, \dodoi{10.1051/0004-6361/202449637}

\bibitem[{{Patat}(2017)}]{2017hsn..book.1017P}
{Patat}, F. 2017, in Handbook of Supernovae, ed. A.~W. {Alsabti} \& P.~{Murdin} (Springer Cham), 1017, \dodoi{10.1007/978-3-319-21846-5_110}

\bibitem[{{Patat} {et~al.}(2012){Patat}, {H{\"o}flich}, {Baade}, {Maund}, {Wang}, \& {Wheeler}}]{2012A&A...545A...7P}
{Patat}, F., {H{\"o}flich}, P., {Baade}, D., {et~al.} 2012, \aap, 545, A7, \dodoi{10.1051/0004-6361/201219146}

\bibitem[{{Patat} {et~al.}(2007)}]{Patat2007}
{Patat}, F., {et~al.} 2007, Science, 317, 924

\bibitem[{{Patel} {et~al.}(2026){Patel}, {Dongre}, {Fisher}, {Poludnenko}, {Gamezo}, {Ugalino}, \& {Byrohl}}]{2026arXiv260521575P}
{Patel}, K., {Dongre}, A., {Fisher}, R., {et~al.} 2026, arXiv e-prints, arXiv:2605.21575, \dodoi{10.48550/arXiv.2605.21575}

\bibitem[{{Penney} \& {Hoeflich}(2014)}]{2014ApJ...795...84P}
{Penney}, R., \& {Hoeflich}, P. 2014, \apj, 795, 84, \dodoi{10.1088/0004-637X/795/1/84}

\bibitem[{{P{\'e}rez-Torres} {et~al.}(2014)}]{PerezTorres2014}
{P{\'e}rez-Torres}, M.~A., {et~al.} 2014, \apj, 792, 38

\bibitem[{{Perlmutter} {et~al.}(1999){Perlmutter}, {Aldering}, {Goldhaber}, {Knop}, {Nugent}, {Castro}, {Deustua}, {Fabbro}, {Goobar}, {Groom}, {Hook}, {Kim}, {Kim}, {Lee}, {Nunes}, {Pain}, {Pennypacker}, {Quimby}, {Lidman}, {Ellis}, {Irwin}, {McMahon}, {Ruiz-Lapuente}, {Walton}, {Schaefer}, {Boyle}, {Filippenko}, {Matheson}, {Fruchter}, {Panagia}, {Newberg}, {Couch}, \& {Project}}]{1999ApJ...517..565P}
{Perlmutter}, S., {Aldering}, G., {Goldhaber}, G., {et~al.} 1999, \apj, 517, 565, \dodoi{10.1086/307221}

\bibitem[{{Phillips} {et~al.}(1992){Phillips}, {Wells}, {Suntzeff}, {Hamuy}, {Leibundgut}, {Kirshner}, \& {Foltz}}]{1992AJ....103.1632P}
{Phillips}, M.~M., {Wells}, L.~A., {Suntzeff}, N.~B., {et~al.} 1992, \aj, 103, 1632, \dodoi{10.1086/116177}

\bibitem[{{Phillips} {et~al.}(2022){Phillips}, {Ashall}, {Burns}, {Contreras}, {Galbany}, {Hoeflich}, {Hsiao}, {Morrell}, {Nugent}, {Uddin}, {Baron}, {Freedman}, {Harris}, {Krisciunas}, {Kumar}, {Lu}, {Persson}, {Piro}, {Polin}, {Shahbandeh}, {Stritzinger}, \& {Suntzeff}}]{2022ApJ...938...47P}
{Phillips}, M.~M., {Ashall}, C., {Burns}, C.~R., {et~al.} 2022, \apj, 938, 47, \dodoi{10.3847/1538-4357/ac9305}

\bibitem[{{Phillips} {et~al.}(2024){Phillips}, {Ashall}, {Brown}, {Galbany}, {Tucker}, {Burns}, {Contreras}, {Hoeflich}, {Hsiao}, {Kumar}, {Morrell}, {Uddin}, {Baron}, {Freedman}, {Krisciunas}, {Persson}, {Piro}, {Shappee}, {Stritzinger}, {Suntzeff}, {Chakraborty}, {Kirshner}, {Lu}, {Marion}, {Polin}, \& {Shahbandeh}}]{2024ApJS..273...16P}
{Phillips}, M.~M., {Ashall}, C., {Brown}, P.~J., {et~al.} 2024, \apjs, 273, 16, \dodoi{10.3847/1538-4365/ad4f7e}

\bibitem[{{Polin} {et~al.}(2019){Polin}, {Nugent}, \& {Kasen}}]{2019ApJ...873...84P}
{Polin}, A., {Nugent}, P., \& {Kasen}, D. 2019, \apj, 873, 84, \dodoi{10.3847/1538-4357/aafb6a}

\bibitem[{{Polin} {et~al.}(2021){Polin}, {Nugent}, \& {Kasen}}]{2021ApJ...906...65P}
---. 2021, \apj, 906, 65, \dodoi{10.3847/1538-4357/abcccc}

\bibitem[{{Poludnenko}(2016)}]{poludnenko2016b}
{Poludnenko}, A. 2016, in {Handbook of Supernovae}, ed. P.~{Murdin} (Berlin: Springer), 999

\bibitem[{{Poludnenko} {et~al.}(2019){Poludnenko}, {Chambers}, {Ahmed}, {Gamezo}, \& {Taylor}}]{2019AAS...23311307P}
{Poludnenko}, A., {Chambers}, J., {Ahmed}, K., {Gamezo}, V., \& {Taylor}, B. 2019, in American Astronomical Society Meeting Abstracts, Vol. 233, American Astronomical Society Meeting Abstracts \#233, ed. A.~A. Society (American Astronomical Society), 113.07

\bibitem[{{Poludnenko} {et~al.}(2011){Poludnenko}, {Gardiner}, \& {Oran}}]{2011PhRvL.107e4501P}
{Poludnenko}, A.~Y., {Gardiner}, T.~A., \& {Oran}, E.~S. 2011, \prl, 107, 054501, \dodoi{10.1103/PhysRevLett.107.054501}

\bibitem[{{Quimby} {et~al.}(2006){Quimby}, {H{\"o}flich}, {Kannappan}, {Rykoff}, {Rujopakarn}, {Akerlof}, {Gerardy}, \& {Wheeler}}]{2006ApJ...636..400Q}
{Quimby}, R., {H{\"o}flich}, P., {Kannappan}, S.~J., {et~al.} 2006, \apj, 636, 400, \dodoi{10.1086/498014}

\bibitem[{{Quimby} {et~al.}(2007){Quimby}, {H{\"o}flich}, \& {Wheeler}}]{2007ApJ...666.1083Q}
{Quimby}, R., {H{\"o}flich}, P., \& {Wheeler}, J.~C. 2007, \apj, 666, 1083, \dodoi{10.1086/520527}

\bibitem[{{Riess} {et~al.}(1998){Riess}, {Filippenko}, {Challis}, {Clocchiatti}, {Diercks}, {Garnavich}, {Gilliland}, {Hogan}, {Jha}, {Kirshner}, {Leibundgut}, {Phillips}, {Reiss}, {Schmidt}, {Schommer}, {Smith}, {Spyromilio}, {Stubbs}, {Suntzeff}, \& {Tonry}}]{1998AJ....116.1009R}
{Riess}, A.~G., {Filippenko}, A.~V., {Challis}, P., {et~al.} 1998, \aj, 116, 1009, \dodoi{10.1086/300499}

\bibitem[{{Riess} {et~al.}(2016){Riess}, {Macri}, {Hoffmann}, {Scolnic}, {Casertano}, {Filippenko}, {Tucker}, {Reid}, {Jones}, {Silverman}, {Chornock}, {Challis}, {Yuan}, {Brown}, \& {Foley}}]{2016ApJ...826...56R}
{Riess}, A.~G., {Macri}, L.~M., {Hoffmann}, S.~L., {et~al.} 2016, \apj, 826, 56, \dodoi{10.3847/0004-637X/826/1/56}

\bibitem[{{Riess} {et~al.}(2022){Riess}, {Yuan}, {Macri}, {Scolnic}, {Brout}, {Casertano}, {Jones}, {Murakami}, {Anand}, {Breuval}, {Brink}, {Filippenko}, {Hoffmann}, {Jha}, {D'arcy Kenworthy}, {Mackenty}, {Stahl}, \& {Zheng}}]{2022ApJ...934L...7R}
{Riess}, A.~G., {Yuan}, W., {Macri}, L.~M., {et~al.} 2022, \apjl, 934, L7, \dodoi{10.3847/2041-8213/ac5c5b}

\bibitem[{{R{\"o}pke} {et~al.}(2007){R{\"o}pke}, {Hillebrandt}, {Schmidt}, {Niemeyer}, {Blinnikov}, \& {Mazzali}}]{2007ApJ...668.1132R}
{R{\"o}pke}, F.~K., {Hillebrandt}, W., {Schmidt}, W., {et~al.} 2007, \apj, 668, 1132, \dodoi{10.1086/521347}

\bibitem[{{R{\"o}pke} {et~al.}(2012){R{\"o}pke}, {Kromer}, {Seitenzahl}, {Pakmor}, {Sim}, {Taubenberger}, {Ciaraldi-Schoolmann}, {Hillebrandt}, {Aldering}, {Antilogus}, {Baltay}, {Benitez-Herrera}, {Bongard}, {Buton}, {Canto}, {Cellier-Holzem}, {Childress}, {Chotard}, {Copin}, {Fakhouri}, {Fink}, {Fouchez}, {Gangler}, {Guy}, {Hachinger}, {Hsiao}, {Chen}, {Kerschhaggl}, {Kowalski}, {Nugent}, {Paech}, {Pain}, {Pecontal}, {Pereira}, {Perlmutter}, {Rabinowitz}, {Rigault}, {Runge}, {Saunders}, {Smadja}, {Suzuki}, {Tao}, {Thomas}, {Tilquin}, \& {Wu}}]{2012ApJ...750L..19R}
{R{\"o}pke}, F.~K., {Kromer}, M., {Seitenzahl}, I.~R., {et~al.} 2012, \apjl, 750, L19, \dodoi{10.1088/2041-8205/750/1/L19}

\bibitem[{{Ruiz-Lapuente} {et~al.}(1992){Ruiz-Lapuente}, {Cappellaro}, {Turatto}, {Gouiffes}, {Danziger}, {della Valle}, \& {Lucy}}]{1992ApJ...387L..33R}
{Ruiz-Lapuente}, P., {Cappellaro}, E., {Turatto}, M., {et~al.} 1992, \apjl, 387, L33, \dodoi{10.1086/186299}

\bibitem[{{Sasdelli} {et~al.}(2014){Sasdelli}, {Mazzali}, {Pian}, {Nomoto}, {Hachinger}, {Cappellaro}, \& {Benetti}}]{2014MNRAS.445..711S}
{Sasdelli}, M., {Mazzali}, P.~A., {Pian}, E., {et~al.} 2014, \mnras, 445, 711, \dodoi{10.1093/mnras/stu1777}

\bibitem[{{Shen}(2015)}]{2015ApJ...805L...6S}
{Shen}, K.~J. 2015, \apjl, 805, L6, \dodoi{10.1088/2041-8205/805/1/L6}

\bibitem[{{Shen} {et~al.}(2021){Shen}, {Blondin}, {Kasen}, {Dessart}, {Townsley}, {Boos}, \& {Hillier}}]{2021ApJ...909L..18S}
{Shen}, K.~J., {Blondin}, S., {Kasen}, D., {et~al.} 2021, \apjl, 909, L18, \dodoi{10.3847/2041-8213/abe69b}

\bibitem[{{Shen} {et~al.}(2018){Shen}, {Kasen}, {Miles}, \& {Townsley}}]{2018ApJ...854...52S}
{Shen}, K.~J., {Kasen}, D., {Miles}, B.~J., \& {Townsley}, D.~M. 2018, \apj, 854, 52, \dodoi{10.3847/1538-4357/aaa8de}

\bibitem[{{Shen} {et~al.}(2010){Shen}, {Kasen}, {Weinberg}, {Bildsten}, \& {Scannapieco}}]{2010ApJ...715..767S}
{Shen}, K.~J., {Kasen}, D., {Weinberg}, N.~N., {Bildsten}, L., \& {Scannapieco}, E. 2010, \apj, 715, 767, \dodoi{10.1088/0004-637X/715/2/767}

\bibitem[{{Shiber} {et~al.}(2026{\natexlab{a}}){Shiber}, {Hoeflich}, {Mera}, {Fereidouni}, {Levy}, {Maci}, {Ashall}, {Baron}, {Shahbandeh}, {Medler}, {Hoogendam}, \& {Pfeffer}}]{2026ApJ..1003L..37S}
{Shiber}, S., {Hoeflich}, P., {Mera}, T., {et~al.} 2026{\natexlab{a}}, \apjl, 1003, L37, \dodoi{10.3847/2041-8213/ae664a}

\bibitem[{{Shiber} {et~al.}(2026{\natexlab{b}}){Shiber}, {Hoeflich}, {Mera}, {Fereidouni}, {Levy}, {Maci}, {Ashall}, {Medler}, {DerKacy}, {Baron}, {Shahbandeh}, \& {Pfeffer}}]{2026arXiv260813432S}
---. 2026{\natexlab{b}}, arXiv e-prints, arXiv:2608.13432, \dodoi{10.48550/arXiv.2608.13432}

\bibitem[{{Siebert} {et~al.}(2023)}]{Siebert2023}
{Siebert}, M.~R., {et~al.} 2023, \apj, 958, 173

\bibitem[{{Silverman} {et~al.}(2013)}]{Silverman2013_11kx}
{Silverman}, J.~M., {et~al.} 2013, \apj, 772, 125

\bibitem[{{Slattery} {et~al.}(1982){Slattery}, {Doolen}, \& {Dewitt}}]{slattery82}
{Slattery}, W.~L., {Doolen}, G.~D., \& {Dewitt}, H.~E. 1982, \pra, 26, 2255, \dodoi{10.1103/PhysRevA.26.2255}

\bibitem[{{Taubenberger}(2017)}]{2017hsn..book..317T}
{Taubenberger}, S. 2017, in Handbook of Supernovae, ed. A.~W. {Alsabti} \& P.~{Murdin} (Springer Cham), 317, \dodoi{10.1007/978-3-319-21846-5_37}

\bibitem[{{Taylor}(1950)}]{taylor1950}
{Taylor}, G. 1950, Proc. R. Soc. London A, 201, 192

\bibitem[{{Telesco} {et~al.}(2015){Telesco}, {H{\"o}flich}, {Li}, {{\'A}lvarez}, {Wright}, {Barnes}, {Fern{\'a}ndez}, {Hough}, {Levenson}, {Mari{\~n}as}, {Packham}, {Pantin}, {Rebolo}, {Roche}, \& {Zhang}}]{2015ApJ...798...93T}
{Telesco}, C.~M., {H{\"o}flich}, P., {Li}, D., {et~al.} 2015, \apj, 798, 93, \dodoi{10.1088/0004-637X/798/2/93}

\bibitem[{{Townsley} {et~al.}(2019){Townsley}, {Miles}, {Shen}, \& {Kasen}}]{2019ApJ...878L..38T}
{Townsley}, D.~M., {Miles}, B.~J., {Shen}, K.~J., \& {Kasen}, D. 2019, \apjl, 878, L38, \dodoi{10.3847/2041-8213/ab27cd}

\bibitem[{{Tsebrenko} \& {Soker}(2015)}]{TsebrenkoSoker2015}
{Tsebrenko}, D., \& {Soker}, N. 2015, \mnras, 447, 2568

\bibitem[{{van de Hulst}(1957)}]{1957lssp.book.....V}
{van de Hulst}, H.~C. 1957, {Light Scattering by Small Particles}

\bibitem[{{Van Horn}(1969)}]{vanHorn69}
{Van Horn}, H.~M. 1969, Physics Letters A, 28, 706, \dodoi{10.1016/0375-9601(69)90699-9}

\bibitem[{{Webbink}(1984)}]{1984ApJ...277..355W}
{Webbink}, R.~F. 1984, \apj, 277, 355, \dodoi{10.1086/161701}

\bibitem[{{Woosley} \& {Weaver}(1994)}]{1994ApJ...423..371W}
{Woosley}, S.~E., \& {Weaver}, T.~A. 1994, \apj, 423, 371, \dodoi{10.1086/173813}

\bibitem[{{Woosley} {et~al.}(1980){Woosley}, {Weaver}, \& {Taam}}]{1980tsup.work...96W}
{Woosley}, S.~E., {Weaver}, T.~A., \& {Taam}, R.~E. 1980, in Texas Workshop on Type I Supernovae, ed. J.~C. {Wheeler}, 96--112

\bibitem[{{Yang} {et~al.}(2022{\natexlab{a}}){Yang}, {Wang}, {Suntzeff}, {Hu}, {Aldoroty}, {Brown}, {Krisciunas}, {Arcavi}, {Burke}, {Galbany}, {Hiramatsu}, {Hosseinzadeh}, {Howell}, {McCully}, {Pellegrino}, \& {Valenti}}]{2022ApJ...938...83Y}
{Yang}, J., {Wang}, L., {Suntzeff}, N., {et~al.} 2022{\natexlab{a}}, \apj, 938, 83, \dodoi{10.3847/1538-4357/ac8c97}

\bibitem[{{Yang} {et~al.}(2018){Yang}, {Wang}, {Baade}, {Brown}, {Cikota}, {Cracraft}, {H{\"o}flich}, {Maund}, {Patat}, {Sparks}, {Spyromilio}, {Stevance}, {Wang}, \& {Wheeler}}]{2018ApJ...854...55Y}
{Yang}, Y., {Wang}, L., {Baade}, D., {et~al.} 2018, \apj, 854, 55, \dodoi{10.3847/1538-4357/aaa76a}

\bibitem[{{Yang} {et~al.}(2020){Yang}, {Hoeflich}, {Baade}, {Maund}, {Wang}, {Brown}, {Stevance}, {Arcavi}, {Burke}, {Cikota}, {Clocchiatti}, {Gal-Yam}, {Graham}, {Hiramatsu}, {Hosseinzadeh}, {Howell}, {Jha}, {McCully}, {Patat}, {Sand}, {Schulze}, {Spyromilio}, {Valenti}, {Vink{\'o}}, {Wang}, {Wheeler}, {Yaron}, \& {Zhang}}]{2020ApJ...902...46Y}
{Yang}, Y., {Hoeflich}, P., {Baade}, D., {et~al.} 2020, \apj, 902, 46, \dodoi{10.3847/1538-4357/aba759}

\bibitem[{{Yang} {et~al.}(2022{\natexlab{b}}){Yang}, {Yan}, {Wang}, {Wheeler}, {Baade}, {Isaacson}, {Cikota}, {Maund}, {Hoeflich}, {Patat}, {Giacalone}, {Rice}, {Tyler}, {Mishra}, {Ashall}, {Brink}, {Filippenko}, {Galbany}, {Patra}, {Shahbandeh}, {Vasylyev}, \& {Vink{\'o}}}]{2022ApJ...939...18Y}
{Yang}, Y., {Yan}, H., {Wang}, L., {et~al.} 2022{\natexlab{b}}, \apj, 939, 18, \dodoi{10.3847/1538-4357/ac8d5f}

\bibitem[{{Yang} {et~al.}(2026){Yang}, {Hoeflich}, {Wheeler}, {Baade}, {Wang}, {Cikota}, {Howell}, {McCully}, {Patat}, {Ashall}, {Bulla}, {Gal-Yam}, \& {Schulze}}]{PaperI}
{Yang}, Y., {Hoeflich}, P., {Wheeler}, C., {et~al.} 2026, ApJ, Submitted, \dodoi{10.3847/2041-8213/ae664a}

\bibitem[{{Zel'dovich} {et~al.}(1970){Zel'dovich}, {Librovich}, {Makhviladze}, \& {Sivashinskil}}]{1970JAMTP..11..264Z}
{Zel'dovich}, Y.~B., {Librovich}, V.~B., {Makhviladze}, G.~M., \& {Sivashinskil}, G.~I. 1970, Journal of Applied Mechanics and Technical Physics, 11, 264, \dodoi{10.1007/BF00908106}

\bibitem[{{Zhang} {et~al.}(2019){Zhang}, {Xu}, \& {Wang}}]{2019TNSCR2898....1Z}
{Zhang}, J., {Xu}, L., \& {Wang}, X. 2019, Transient Name Server Classification Report, 2019-2898, 1

\bibitem[{{Zingale} {et~al.}(2011){Zingale}, {Nonaka}, {Almgren}, {Bell}, {Malone}, \& {Woosley}}]{2011ApJ...740....8Z}
{Zingale}, M., {Nonaka}, A., {Almgren}, A.~S., {et~al.} 2011, \apj, 740, 8, \dodoi{10.1088/0004-637X/740/1/8}

\end{thebibliography}

\vfill\eject
\subsection*{Strongest Optical Lines - relevant for the model spectra}~\label{sec:lines}
\providecommand{\colhead}[1]{\multicolumn{1}{c}{#1}}
\providecommand{\mkfont}{\tiny}
\providecommand{\mko}{{\mkfont\#}}       
\providecommand{\mkt}{{\mkfont\#\#}}     
\providecommand{\mkh}{{\mkfont\#\#\#}}   
\begingroup
\setlength{\tabcolsep}{1.2pt}
\renewcommand{\arraystretch}{1.05}
\footnotesize
\begin{longtable}{@{}c r l@{\hspace{1.4pt}}|@{\hspace{1.4pt}}c r l@{\hspace{1.4pt}}|@{\hspace{1.4pt}}c r l@{\hspace{1.4pt}}|@{\hspace{1.4pt}}c r l@{\hspace{1.4pt}}|@{\hspace{1.4pt}}c r l@{\hspace{1.4pt}}|@{\hspace{1.4pt}}c r l@{\hspace{1.4pt}}|@{\hspace{1.4pt}}c r l@{}}
\caption{Strong allowed line transitions between 3200 and 9300 ${\rm \AA}$ in vacuum (out of $\approx 50,000$) based on the total optical depth {$\tau_l$} above the photosphere defined by the thermalization optical depth being one. The markers (blank, \mko, \mkt, \mkh) correspond to  {$\tau_l \geq 0.05, 0.2, 1.0, 4.0$} at day -9. 
For days +2 and +12, the marks are to be reduced by one and two steps, respectively, whereas the marks for singly ionized Co and Fe are to be increased by about one step because the Fe/Co abundance increases by an order of magnitude and the ionization balance shifts. Due to recombination after day -9, Ca{\sc\,II} becomes very strong. Note that strong spectral features arise from many overlapping weaker transitions. Thus, individual components of IGE element multiplets may be weak, but strong spectral features can be formed by combining dozens of blends. Note that we give vacuum wavelengths in the table because it is the relevant parameter for model atoms.  }\label{tab:lines}\\
\hline\hline
\colhead{Mk} & \colhead{$\lambda$\,(\AA)} & \colhead{Ion} & \colhead{Mk} & \colhead{$\lambda$\,(\AA)} & \colhead{Ion} & \colhead{Mk} & \colhead{$\lambda$\,(\AA)} & \colhead{Ion} & \colhead{Mk} & \colhead{$\lambda$\,(\AA)} & \colhead{Ion} & \colhead{Mk} & \colhead{$\lambda$\,(\AA)} & \colhead{Ion} & \colhead{Mk} & \colhead{$\lambda$\,(\AA)} & \colhead{Ion} & \colhead{Mk} & \colhead{$\lambda$\,(\AA)} & \colhead{Ion} \\
\hline
\endfirsthead
\multicolumn{21}{l}{\footnotesize\emph{Table~\ref{tab:lines} --- continued}}\\
\hline\hline
\colhead{Mk} & \colhead{$\lambda$\,(\AA)} & \colhead{Ion} & \colhead{Mk} & \colhead{$\lambda$\,(\AA)} & \colhead{Ion} & \colhead{Mk} & \colhead{$\lambda$\,(\AA)} & \colhead{Ion} & \colhead{Mk} & \colhead{$\lambda$\,(\AA)} & \colhead{Ion} & \colhead{Mk} & \colhead{$\lambda$\,(\AA)} & \colhead{Ion} & \colhead{Mk} & \colhead{$\lambda$\,(\AA)} & \colhead{Ion} & \colhead{Mk} & \colhead{$\lambda$\,(\AA)} & \colhead{Ion} \\
\hline
\endhead
\hline
\multicolumn{21}{r}{\footnotesize\emph{continued on next page}}\\
\endfoot
\hline
\multicolumn{21}{@{}l@{}}{\footnotesize $^{}$ .}\\
\endlastfoot
& 3203.5 & Ti\,{\sc ii} & \mko & 3204.8 & Fe\,{\sc ii} & \mko & 3204.8 & Si\,{\sc ii} & \mko & 3211.0 & Si\,{\sc ii} & \mko & 3211.0 & Fe\,{\sc ii} &  & 3211.0 & Si\,{\sc ii} & \mko & 3211.4 & Fe\,{\sc ii}  \\
& 3211.5 & Si\,{\sc iii} &  & 3211.6 & Si\,{\sc iii} & \mko & 3214.2 & Fe\,{\sc ii} & \mko & 3215.5 & Fe\,{\sc ii} &  & 3218.0 & Ti\,{\sc ii} & \mko & 3218.9 & Fe\,{\sc ii} &  & 3219.2 & Ti\,{\sc ii}  \\
\mkt & 3219.3 & Fe\,{\sc iii} &  & 3223.8 & Ti\,{\sc ii} & \mko & 3223.9 & Fe\,{\sc ii} & \mko & 3224.3 & Fe\,{\sc ii} &  & 3225.2 & Ti\,{\sc ii} & \mko & 3225.5 & Fe\,{\sc ii} & \mko & 3227.3 & Fe\,{\sc ii}  \\
\mko & 3228.7 & Fe\,{\sc ii} &  & 3229.5 & Ti\,{\sc ii} &  & 3229.8 & Ti\,{\sc iii} &  & 3230.0 & Co\,{\sc ii} &  & 3230.4 & Ti\,{\sc ii} &  & 3231.4 & Si\,{\sc iii} &  & 3231.5 & Co\,{\sc ii}  \\
& 3232.9 & Co\,{\sc ii} &  & 3233.2 & Ti\,{\sc ii} & \mko & 3233.3 & Fe\,{\sc ii} & \mkt & 3233.7 & Fe\,{\sc ii} &  & 3234.9 & Si\,{\sc iii} &  & 3235.5 & Ti\,{\sc ii} & \mko & 3235.5 & Co\,{\sc ii}  \\
& 3237.1 & Ti\,{\sc ii} &  & 3237.5 & Ti\,{\sc ii} & \mko & 3237.8 & Fe\,{\sc ii} & \mko & 3238.1 & Fe\,{\sc iii} & \mkt & 3238.3 & Fe\,{\sc ii} & \mkt & 3238.8 & Fe\,{\sc ii} &  & 3240.0 & Ti\,{\sc ii}  \\
& 3240.6 & Ti\,{\sc ii} &  & 3242.6 & Si\,{\sc iii} &  & 3242.9 & Ti\,{\sc ii} & \mkt & 3244.7 & Fe\,{\sc ii} &  & 3246.7 & Fe\,{\sc ii} & \mkt & 3248.1 & Fe\,{\sc ii} & \mko & 3249.5 & Ti\,{\sc ii}  \\
& 3253.9 & Ti\,{\sc ii} &  & 3253.9 & Ti\,{\sc ii} & \mkt & 3259.7 & Fe\,{\sc ii} & \mkt & 3260.0 & Fe\,{\sc ii} &  & 3260.3 & Co\,{\sc ii} &  & 3262.5 & Ti\,{\sc ii} &  & 3262.6 & Ti\,{\sc ii}  \\
\end{longtable}
\endgroup

\end{document}